\documentclass[submission,copyright]{eptcs}

\usepackage[utf8]{inputenc}
\usepackage[english]{babel}
\usepackage[T1]{fontenc}
\providecommand{\event}{QPL 2022} 

\usepackage{graphicx}
\usepackage{appendix}
\usepackage{color}
\usepackage[dvipsnames]{xcolor}
\usepackage{hyperref}
\hypersetup{colorlinks=true, linkcolor=blue, citecolor=ForestGreen, urlcolor=DarkOrchid}
\usepackage{xparse}
\usepackage{dirtytalk}
\usepackage[colorinlistoftodos]{todonotes}
\usepackage{xspace}
\usepackage{eqparbox}
\usepackage{pgfmath}
\usepackage{tikz}
\usetikzlibrary{calc,
    babel,
    arrows,
    shapes,
    graphs,
	fadings,
	positioning,
	decorations.markings,
	intersections,
	backgrounds,
	fit,
	quantikz,
    circuits.logic.US,
    tikzmark,
    zx-calculus,
    external
}
\usetikzlibrary{zx-calculus}
\newcommand{\externalize}[2]{%
  {#2}%
}
\newcommand{\externalizezx}[2]{%
  {\tikz{\node[inner sep=0pt, outer sep=0pt]{%
          #2
  };}}%
}

\tikzexternaldisable

\definecolor{colorZxZ}{gray}{1.0}
\definecolor{colorZxX}{gray}{0.6}
\definecolor{colorZxH}{gray}{1.0}

\tikzset{%
    zxstyle/.style={
        rounded rectangle,
        anchor=center, 
        line width=0.4pt,
        execute at begin node={\thinmuskip=0mu\medmuskip=0mu\thickmuskip=0mu}, 
        minimum size=5mm,
        font={\fontsize{8}{10}\selectfont\boldmath},
        inner sep=0.0mm,
        inner xsep=1.2mm,
        scale=0.8,
    },
    zxstyletight/.style={
        zxstyle,
        minimum size=4mm,
        inner ysep=0.2mm,
        inner xsep=0.8mm,
    },
    spider/.style={%
        circle,
        font=\small,
        draw=black,
        inner sep=1pt,
    },
    coordinate/.style={%
        inner sep=0.0pt,
        outer sep=0.0pt,
        anchor=center
    },
    leftsummator/.style={%
        circuit logic US,
        buffer gate,
        draw=black,thin,
        inner sep=0.0ex,
        label={center:\small $\Sigma$},
        anchor=west,
        node contents={}
    },
    rightsummator/.style={%
        circuit logic US,
        buffer gate,
        draw=black,thin,
        rotate=180,
        inner sep=0.0ex,
        label={[rotate=180]center:\small $\Sigma$},
        anchor=west,
        node contents={}
    },
    bubble/.style={%
        ellipse,
        draw=black,thin, 
        inner sep=0.1ex
    },
    xspider/.style={%
        spider,
        fill=gray!70,
        inner sep=#1
    },
    xspider/.default={1pt},
    zspider/.style={%
        spider,
        fill=white,
        inner sep=#1
    },
    zspider/.default={1pt},
    hadamard/.style={%
        rectangle,
        fill=white,
        draw=black,
        inner sep=1pt,
        node contents={}
    },
    invisible/.style={%
        inner sep=0.0pt,
        outer sep=0.0pt,
        minimum width=0pt,
        minimum height=0pt,
        draw=none,
        anchor=center,
        node contents={}
    },
    gamma/.style={%
        zspider,
        zxstyle,
        inner sep=0.3pt,
        anchor=center,
        node contents={#1}
    },
    gamma/.default={$\gamma$},
    gammatimes2/.style={%
        zspider,
        zxstyle,
        inner sep=0.2pt,
        anchor=center,
        node contents={$2\gamma$}
    },
    gammainv/.style={%
        zspider,
        zxstyle,
        inner sep=0.2pt,
        anchor=center,
        minimum width=0.4em,
        node contents={$\zxMinus$#1}
    },
    gammainv/.default={$\gamma$},
    gammainvtimes2/.style={%
        zspider,
        zxstyle,
        inner sep=0.2pt,
        anchor=center,
        node contents={$\zxMinus2\gamma$}
    },
    beta/.style={%
        xspider,
        zxstyle,
        inner sep=0.3pt,
        anchor=center,
        node contents={#1}
    },
    beta/.default={$\beta$},
    betainv/.style={%
        xspider,
        zxstyle,
        inner sep=0.2pt,
        anchor=center,
        node contents={$\zxMinus$#1}
    },
    betainv/.default={$\beta$},
    betainvtimes2/.style={%
        xspider,
        zxstyle,
        inner sep=0.2pt,
        anchor=center,
        node contents={$\zxMinus2\beta$}
    },
    betatimes2/.style={%
        xspider,
        zxstyle,
        inner sep=0.2pt,
        anchor=center,
        node contents={$2\beta$}
    },
    zpi/.style={%
        zspider,
        zxstyle,
        inner sep=0.4pt,
        anchor=center,
        node contents={$\pi$}
    },
    xpi/.style={%
        xspider,
        zxstyle,
        inner sep=0.4pt,
        anchor=center,
        node contents={$\pi$}
    }
}

\NewDocumentCommand{\identity}{O{2.0em}O{0.4em}}{%
\begin{tikzpicture}
    \node (left) {};
    \node[right=#1 of left] (right) {};
    \draw[thin] (left) -- (right);
    \path ($(left.center)!0.5!(right.center)$) circle [radius=#2];
\end{tikzpicture}
}

\NewDocumentCommand{\xpi}{O{2.0em}O{0.4em}}{%
\begin{tikzpicture}
    \node (left) {};
    \node[right=#1 of left] (right) {};
    \node at ($(left.center)!0.5!(right.center)$) [name=center, xpi, zxstyletight] ;
    \draw[thin] (left) -- (center);
    \draw[thin] (right) -- (center);
    \path ($(left.center)!0.5!(right.center)$) circle [radius=#2];
\end{tikzpicture}
}

\NewDocumentCommand{\zpi}{O{2.0em}O{0.4em}}{%
\begin{tikzpicture}
    \node (left) {};
    \node[right=#1 of left] (right) {};
    \node at ($(left.center)!0.5!(right.center)$) [name=center, zpi, zxstyletight] ;
    \draw[thin] (left) -- (center);
    \draw[thin] (right) -- (center);
    \path ($(left.center)!0.5!(right.center)$) circle [radius=#2];
\end{tikzpicture}
}

\NewDocumentCommand{\zxsumtwo}{m m O{0.0em} O{0.5ex} O{0.7} o o t{r} t{c}}{%
    \tikzmath{
        \xshift = #3;
        \ytop = #4;
        \ybottom = -#4;
    }
    \IfBooleanTF{#9}{%
        \IfBooleanTF{#8}{%
            \coordinate (coordbubble) at (sumleft.center);
            }{%
            \coordinate (coordbubble) at (sumright.center);
        }
    }{%
        \coordinate (coordbubble) at ($(sumleft.center)!0.5!(sumright.center)$); 
    }

    \node[draw=none, scale=#5, inner sep=0pt, anchor=center, xshift=\xshift, yshift=\ytop] (contenttop) %
    at (coordbubble) {%
        #1
    };
    \node[bubble, fit=(contenttop)] (bubbletop) {};
    \node[draw=none, scale=#5, inner sep=0pt, anchor=center, xshift=\xshift, yshift=\ybottom] (contentbottom) %
    at (coordbubble) {%
        #2
    };
    \node[bubble, fit=(contentbottom)] (bubblebottom) {};
    \IfBooleanTF{#8}
    {%
        \IfNoValueTF{#6}
        {%
            \draw[thin] (sumright) -- (bubbletop.east);
        }
        {%
            \draw[thin] (sumright) -- (bubbletop.east) node[fill=white, midway, inner sep=1.0pt, anchor=center] {$#6$};
        }
        \IfNoValueTF {#7}
        {%
            \draw[thin] (sumright) -- (bubblebottom.east);
        }
        {%
            \draw[thin] (sumright) -- (bubblebottom.east) node[fill=white, midway, inner sep=1.0pt, anchor=center] {$#7$};
        }
        \IfBooleanF {#9}
        {%
            \draw[thin] (sumleft) -- (bubblebottom.west);
            \draw[thin] (sumleft) -- (bubbletop.west);
        }
    }
    {%
        \IfNoValueTF {#6}
        {%
            \draw[thin] (sumleft) -- (bubbletop.west);
        }
        {%
            \draw[thin] (sumleft) -- (bubbletop.west) node[fill=white, midway, inner sep=1.0pt, anchor=center] {$#6$};
        }
        \IfNoValueTF {#7}
        {%
            \draw[thin] (sumleft) -- (bubblebottom.west);
        }
        {%
            \draw[thin] (sumleft) -- (bubblebottom.west) node[fill=white, midway, inner sep=1.0pt, anchor=center] {$#7$};
        }
        \IfBooleanF {#9}
        {%
            \draw[thin] (sumright) -- (bubbletop.east);
            \draw[thin] (sumright) -- (bubblebottom.east);
        }
    }
}

\NewDocumentCommand{\zxsumconfig}{D(){0,0} O{0.3em} O{0.5ex} O{0.7}}{%
    \tikzmath{
        \xshift = #2;
        \ytop = 3.0 * #3;
        \ymidtop = #3;
        \ymidbottom = -#3;
        \ybottom = -3.0 * #3;
    }
    \def\origincoordinate{#1}
    \def\bubblescale{#4}
}
\NewDocumentCommand{\zxsumfour}{m m m m o o o o}{%
    \node[draw=none, inner sep=0pt, scale=\bubblescale, anchor=center, xshift=\xshift, yshift=\ytop] (contenttop) %
    at ($(sumleft.center)!0.5!(sumright.center)$){%
        #1
    };
    \node[bubble, fit=(contenttop)] (bubbletop) {};

    \node[draw=none, inner sep=0pt, scale=\bubblescale, anchor=center, xshift=\xshift, yshift=\ymidtop] (contentmidtop) %
    at ($(sumleft.center)!0.5!(sumright.center)$){%
        #2
    };
    \node[bubble, fit=(contentmidtop)] (bubblemidtop) {};

    \node[draw=none, inner sep=0pt, scale=\bubblescale, anchor=center, xshift=\xshift, yshift=\ymidbottom] (contentmidbottom) %
    at ($(sumleft.center)!0.5!(sumright.center)$){%
        #3
    };
    \node[bubble, fit=(contentmidbottom)] (bubblemidbottom) {};

    \node[draw=none, inner sep=0pt, scale=\bubblescale, anchor=center, xshift=\xshift, yshift=\ybottom] (contentbottom) %
    at ($(sumleft.center)!0.5!(sumright.center)$){%
        #4
    };
    \node[bubble, fit=(contentbottom)] (bubblebottom) {};

    \IfNoValueTF {#5}
    {%
        \draw[thin] (sumleft) -- (bubbletop.west);
    }
    {%
        \draw[thin] (sumleft) -- (bubbletop.west) node[fill=white, midway, anchor=center, inner sep=1.0pt] {$#5$};
    }
    \IfNoValueTF {#6}
    {%
        \draw[thin] (sumleft) -- (bubblemidtop.west);
    }
    {%
        \draw[thin] (sumleft) -- (bubblemidtop.west) node[fill=white, midway, anchor=center, inner sep=1.0pt] {$#6$};
    }
    \IfNoValueTF {#7}
    {%
        \draw[thin] (sumleft) -- (bubblemidbottom.west);
    }
    {%
        \draw[thin] (sumleft) -- (bubblemidbottom.west) node[fill=white, midway, anchor=center, inner sep=1.0pt] {$#7$};
    }
    \IfNoValueTF {#8}
    {%
        \draw[thin] (sumleft) -- (bubblebottom.west);
    }
    {%
        \draw[thin] (sumleft) -- (bubblebottom.west) node[fill=white, circle, midway, anchor=center, inner sep=1.0pt] {$#8$};
    }
    \draw[thin] (sumright) -- (bubbletop.east);
    \draw[thin] (sumright) -- (bubblemidtop.east);
    \draw[thin] (sumright) -- (bubblemidbottom.east);
    \draw[thin] (sumright) -- (bubblebottom.east);
}

\NewDocumentCommand{\diagplacehold}{m m m O{1.5em} O{5ex} O{1.0em} O{0.4ex} O{0.20em}}{%
\begin{tikzpicture}[node distance=#5 and #4,
                    label/.style={outer sep=0pt, inner sep=0pt}]
    \node[name=bl, invisible];
    \node[on grid, name=bm, right=of bl, invisible];
    \node[on grid, name=br, right=of bm, invisible];
    \node[on grid, name=tl, above=of bl, invisible];
    \node[on grid, name=tm, right=of tl, invisible];
    \node[on grid, name=tr, right=of tm, invisible];

    \node[rectangle, 
          draw=black,thin, 
          fill=white,
          fit=(tm) (bm), 
          inner xsep=#6, 
          inner ysep=#7] (box) {};
    \node[left=#8 of box , label] (dots) {\rvdots};
    \node[left=#8 of dots, label] {#1};
    \node[right=#8 of box, label] (dots) {\rvdots};
    \node[right=#8 of dots, label] {#3};
    \node[fill=white, inner sep=0pt, anchor=center] at ($(box.north west)!0.5!(box.south east)$) {#2};

    \begin{scope}[on background layer]
        \draw (bl) -- (br);
        \draw (tl) -- (tr);
    \end{scope}
\end{tikzpicture}
}

\NewDocumentCommand{\diagplaceholdnoinput}{m m O{1.5em} O{5ex} O{1.0em} O{0.4ex} O{0.20em}}{%
\begin{tikzpicture}[node distance=#4 and #3,
                    label/.style={outer sep=0pt, inner sep=0pt}]
    \node[name=bl, invisible];
    \node[on grid, name=bm, right=of bl, invisible];
    \node[on grid, name=br, right=of bm, invisible];
    \node[on grid, name=tl, above=of bl, invisible];
    \node[on grid, name=tm, right=of tl, invisible];
    \node[on grid, name=tr, right=of tm, invisible];

    \node[rectangle, 
          draw=black,thin, 
          fill=white,
          fit=(tm) (bm), 
          inner xsep=#5, 
          inner ysep=#6] (box) {};
    \node[right=#7 of box, label] (dots) {\rvdots};
    \node[right=#7 of dots, label] {#2};
    \node[fill=white, inner sep=0pt, anchor=center] at ($(box.north west)!0.5!(box.south east)$) {#1};

    \begin{scope}[on background layer]
        \draw (bm) -- (br);
        \draw (tm) -- (tr);
    \end{scope}
\end{tikzpicture}
}

\NewDocumentCommand{\diagplaceholdnooutput}{m m O{1.5em} O{5ex} O{1.0em} O{0.4ex} O{0.20em}}{%
\begin{tikzpicture}[node distance=#4 and #3,
                    label/.style={outer sep=0pt, inner sep=0pt}]
    \node[name=bl, invisible];
    \node[on grid, name=bm, right=of bl, invisible];
    \node[on grid, name=br, right=of bm, invisible];
    \node[on grid, name=tl, above=of bl, invisible];
    \node[on grid, name=tm, right=of tl, invisible];
    \node[on grid, name=tr, right=of tm, invisible];

    \node[rectangle, 
          draw=black,thin, 
          fill=white,
          fit=(tm) (bm), 
          inner xsep=#5, 
          inner ysep=#6] (box) {};
    \node[left=#7 of box , label] (dots) {\rvdots};
    \node[left=#7 of dots, label] {#1};
    \node[fill=white, inner sep=0pt, anchor=center] at ($(box.north west)!0.5!(box.south east)$) {#2};

    \begin{scope}[on background layer]
        \draw (bl) -- (bm);
        \draw (tl) -- (tm);
    \end{scope}
\end{tikzpicture}
}

\NewDocumentCommand{\diagplaceholdproduct}{m m m m O{1.5em} O{5ex} O{1.0em} O{0.4ex} O{0.20em}}{%
\begin{tikzpicture}[node distance=#6 and #5,
                    label/.style={outer sep=0pt, inner sep=0pt}]
    \node[name=bl, invisible];
    \node[on grid, name=bml, right=of bl, invisible];
    \node[on grid, name=bm, right=of bml, invisible];
    \node[on grid, name=bmr, right=of bm, invisible];
    \node[on grid, name=br, right=of bmr, invisible];
    \node[on grid, name=tl, above=of bl, invisible];
    \node[on grid, name=tml, right=of tl, invisible];
    \node[on grid, name=tm, right=of tml, invisible];
    \node[on grid, name=tmr, right=of tm, invisible];
    \node[on grid, name=tr, right=of tmr, invisible];

    \node[rectangle, 
          draw=black,thin, 
          fill=white,
          fit=(tml) (bml), 
          inner xsep=#7, 
          inner ysep=#8] (leftbox) {};
    \node[left=#9 of leftbox , label] (dots) {\rvdots};
    \node[left=#9 of dots, label] {#1};
    \node[fill=white, inner sep=0pt, anchor=center] at ($(leftbox.north west)!0.5!(leftbox.south east)$) {#2};

    \node[anchor=center] at ($(tm.center)!0.5!(bm.center)$) (dots) {\rvdots};

    \node[rectangle, 
          draw=black,thin, 
          fill=white,
          fit=(tmr) (bmr), 
          inner xsep=#7, 
          inner ysep=#8] (rightbox) {};
    \node[right=#9 of rightbox, label] (dots) {\rvdots};
    \node[right=#9 of dots, label] {#4};
    \node[fill=white, inner sep=0pt, anchor=center] at ($(rightbox.north west)!0.5!(rightbox.south east)$) {#3};
    \begin{scope}[on background layer]
        \draw (bl) -- (br);
        \draw (tl) -- (tr);
    \end{scope}
\end{tikzpicture}
}

\NewDocumentCommand{\combineddiagplaceholdconfig}{O{1.5em} O{5ex} O{0.8em} O{0.4ex} O{0.20em} O{2.00em}}{%
    \def\xnode{#1}
    \def\ynode{#2}
    \def\xsep{#3}
    \def\ysep{#4}
    \def\ldist{#5}
    \def\mdist{#6}
}
\NewDocumentCommand{\combineddiagplacehold}{m O{$m$} O{$n$} O{$\ell$} O{$p$} O{$k$}}{%
    \begin{tikzpicture}[node distance=\ynode and \xnode,
                        label/.style={outer sep=0pt, inner sep=0pt}]
    \def\labelmbox{#1}
    \def\labelmboxin{#2}
    \def\labelmboxout{#3}
    \def\labelleft{#4}
    \def\labelmiddle{#5}
    \def\labelright{#6}

    \node[name=1-lb, invisible];
    \node[on grid, name=2-lb, above=of 1-lb, invisible];
    \node[on grid, name=3-lb, above=of 2-lb, invisible];
    \node[on grid, name=4-lb, above=of 3-lb, invisible];

    \node[on grid, name=1-l, left=of 1-lb, invisible];
    \node[on grid, name=4-l, left=of 4-lb, invisible];

    \node[rectangle, 
          draw=black,thin, 
          fill=white,
          fit=(1-lb) (4-lb), 
          inner xsep=\xsep, 
          inner ysep=\ysep] (leftbox) {};
    \node[left=\ldist of leftbox , label] (dots) {\rvdots};
    \node[left=\ldist of dots, label] {\labelleft};

    \node[on grid, right=of 1-lb, name=1-ll, invisible];
    \node[right=\mdist of 1-ll, name=1-m, invisible];
    \node[on grid, right=of 2-lb, name=2-ll, invisible];
    \node[right=\mdist of 2-ll, name=2-m, invisible];
    \node[right=\mdist of 1-m, name=1-rl, invisible];
    \node[right=\mdist of 2-m, name=2-rl, invisible];
    \node[rectangle, 
          draw=black,thin, 
          fill=white,
          fit=(1-m) (2-m), 
          inner xsep=\xsep, 
          inner ysep=\ysep] (middlebox) {};
    \node[right=\ldist of middlebox , label] (dots) {\rvdots};
    \node[right=\ldist of dots, label] {\labelmboxout};
    \node[left=\ldist of middlebox , label] (dots) {\rvdots};
    \node[left=\ldist of dots, label] {\labelmboxin};

    \node[on grid, name=1-rb, right=of 1-rl, invisible];
    \node[on grid, name=2-rb, above=of 1-rb, invisible];
    \node[on grid, name=3-rb, above=of 2-rb, invisible];
    \node[on grid, name=4-rb, above=of 3-rb, invisible];
    \node[on grid, name=4-r, left=of 4-rb, invisible];

    \node[rectangle, 
          draw=black,thin, 
          fill=white,
          fit=(1-rb) (4-rb), 
          inner xsep=\xsep, 
          inner ysep=\ysep] (rightbox) {};
    \node[right=\ldist of rightbox , label] (dots) {\rvdots};
    \node[right=\ldist of dots, label] {\labelright};
    \node[fill=white, inner sep=0pt, anchor=center] at ($(middlebox.north west)!0.5!(middlebox.south east)$) {\labelmbox};

    \node[anchor=center] at ($(3-lb.center)!0.5!(4-rb.center)$) (dots) {\rvdots};
    \node[left=\ldist of dots, label] {$p$};

    \node[on grid, name=1-r, right=of 1-rb, invisible];
    \node[on grid, name=4-r, right=of 4-rb, invisible];

    \begin{scope}[on background layer]
        \draw (1-l) -- (1-r);
        \draw (2-lb) -- (2-rb);
        \draw (3-lb) -- (3-rb);
        \draw (4-l) -- (4-r);
    \end{scope}
    \end{tikzpicture}
}

\NewDocumentCommand{\pslayer}{m m m m m}{%
    \node[rectangle, 
          anchor=center, 
          fill=white,
          draw=black,thin, 
          name=box,
          fit=#3,
          #4 
          ] {};
    \node[rectangle, 
          anchor=center, 
          fill=white,
          draw=black,thin, 
          name=lowerbox,
          fit=#2,
          #4 
          ] {};
          
    \draw (lowerbox.north west) -- (lowerbox.south east);
    \draw (lowerbox.north east) -- (lowerbox.south west);
    \node[fill=white, #5, anchor=center] at ($(lowerbox.north west)!0.5!(lowerbox.south east)$) {#1};
}

\NewDocumentCommand{\mixlayer}{m m m m}{%
    \node[rectangle, 
          anchor=center, 
          fill=gray!70,
          draw=black,thin, 
          name=box,
          fit=#2,
          #3 
          ] {};
          
    \draw (box.west) -- (box.east);
    \node[fill=gray!70, anchor=center, #4] at ($(box.north west)!0.5!(box.south east)$) {#1};
}

\usepackage{amsmath}
\usepackage{amsthm}
\usepackage{amssymb}
\allowdisplaybreaks[1]
\usepackage{mathtools}
\mathtoolsset{showonlyrefs=true}	
\definecolor{sbblue}{rgb}{0.2823529411764706, 0.47058823529411764, 0.8156862745098039}
\definecolor{sborange}{rgb}{0.9333333333333333, 0.5215686274509804, 0.2901960784313726}
\definecolor{sbgreen}{rgb}{0.41568627450980394, 0.8, 0.39215686274509803}
\definecolor{sbred}{rgb}{0.8392156862745098, 0.37254901960784315, 0.37254901960784315}
\definecolor{sbpurple}{rgb}{0.5843137254901961, 0.4235294117647059, 0.7058823529411765}
\definecolor{sbbrown}{rgb}{0.5490196078431373, 0.3803921568627451, 0.23529411764705882}
\definecolor{sbmagenta}{rgb}{0.8627450980392157, 0.49411764705882355, 0.7529411764705882}
\definecolor{sbgray}{rgb}{0.4745098039215686, 0.4745098039215686, 0.4745098039215686}
\definecolor{sbocca}{rgb}{0.8352941176470589, 0.7333333333333333, 0.403921568627451}
\definecolor{sblightblue}{rgb}{0.5098039215686274, 0.7764705882352941, 0.8862745098039215}

\newcommand{\ii}{\mathrm{i}}
\newcommand{\ee}{\mathrm{e}}

\theoremstyle{definition}
\newtheorem{defn}{Definition}[section]

\newcommand{\SpiderRule}{\ensuremath{(\hyperref[fig:zx_rules]{\bm f})}\xspace}

\newcommand{\PiRule}{\ensuremath{(\hyperref[fig:zx_rules]{\bm \pi})}\xspace}
\newcommand{\CopyRule}{\ensuremath{(\hyperref[fig:zx_rules]{\bm c})}\xspace}
\newcommand{\HadamardRule}{\ensuremath{(\hyperref[fig:zx_rules]{\bm{h}})}\xspace}
\newcommand{\IdRule}{\ensuremath{(\hyperref[fig:zx_rules]{\bm{id}})}\xspace}
\newcommand{\HHRule}{\ensuremath{(\hyperref[fig:zx_rules]{\bm{hh}})}\xspace}
\newcommand{\BialgRule}{\ensuremath{(\hyperref[fig:zx_rules]{\bm b})}\xspace}

\newcommand{\HopfRule}{\ensuremath{(\hyperref[fig:zx_rules]{\bm{hopf}})}\xspace}

\newcommand{\bialgrule}{{\footnotesize\BialgRule}}
\newcommand{\idrule}{{\footnotesize\IdRule}}
\newcommand{\hadamardrule}{{\footnotesize\HadamardRule}}
\newcommand{\hhrule}{\footnotesize\HHRule}
\newcommand{\copyrule}{\footnotesize\CopyRule}
\newcommand{\pirule}{\footnotesize\PiRule}
\newcommand{\spiderrule}{\footnotesize\SpiderRule}

\newcommand{\hopfrule}{\footnotesize\HopfRule}

\makeatletter
\DeclareRobustCommand{\rvdots}{%
  \vbox{
    \baselineskip4\p@\lineskiplimit\z@
    \kern-\p@
    \hbox{.}\hbox{.}\hbox{.}
  }}
\makeatother

\makeatletter
\newcommand{\oset}[3][0ex]{%
  \mathrel{\mathop{#3}\limits^{
    \vbox to#1{\kern-2\ex@
    \hbox{$\scriptstyle#2$}\vss}}}}
\makeatother

\title{Diagrammatic Analysis for Parameterized Quantum Circuits}
\author{Tobias Stollenwerk
\institute{Institute for Quantum Computing Analytics (PGI-12), J\"ulich Research Centre, Wilhelm-Johnen-Stra\ss e, 52428 J\"ulich, Germany} 
\institute{German Aerospace Center (DLR), Linder H\"ohe, 51147 Cologne, Germany}
\email{to.stollenwerk@fz-juelich.de}
\and
Stuart Hadfield
\institute{Quantum Artificial Intelligence Lab (QuAIL), NASA Ames Research Center, Moffett Field, CA 94035, USA}
\institute{USRA Research Institute for Advanced Computer Science (RIACS), Mountain View, CA 94043, USA}
\email{stuart.hadfield@nasa.gov}
}

\def\titlerunning{Diagrammatic Analysis for Parameterized Quantum Circuits}
\def\authorrunning{T. Stollenwerk \& S. Hadfield}

\usetikzlibrary{external}
\tikzexternaldisable
\begin{document}

\maketitle

\begin{abstract}
Diagrammatic representations of quantum algorithms and circuits offer novel approaches to their design and analysis. 
In this work, we describe extensions of the ZX-calculus especially suitable for parameterized quantum circuits, in particular for computing observable expectation values as functions of or for fixed parameters, 
which are important algorithmic quantities in a variety of applications ranging from combinatorial optimization to quantum chemistry. 
We provide several new ZX-diagram rewrite rules and generalizations for this setting. 
In particular, we give formal rules for dealing with linear combinations of ZX-diagrams,
where the relative complex-valued scale factors of each diagram must be kept track of, 
in contrast to most previously studied single-diagram realizations 
where these coefficients can be effectively ignored. 
This allows us to directly import a number useful relations from the operator analysis to ZX-calculus setting, 
including causal cone and quantum gate commutation rules.
We demonstrate that the diagrammatic approach offers
useful insights into algorithm structure and performance by considering 
several ans\"atze from the literature including realizations of hardware-efficient ans\"atze and QAOA. 
We find that by employing a diagrammatic representation, 
calculations across different ans\"atze can become more intuitive and potentially easier to 
approach systematically than by alternative means. 
Finally, we outline how diagrammatic approaches may aid in the design and study of new and more effective quantum circuit ans\"atze. 
\end{abstract}

\section{Introduction}
\label{sec:introduction}

Diagrammatic approaches to quantum mechanics~\cite{coecke2008interacting,coecke2017picturing,coecke2021kindergarden} 
have gained much attention in recent years as an advantageous alternative approach to analyzing and understanding quantum systems, 
providing simpler intuition and in some cases improved algorithmic approaches. 
These  methods provide straightforward rules for representing, manipulating, 
and simplifying quantum objects, while at the same time are underpinned by sophisticated mathematical ideas 
(in particular, category theory~\cite{abramsky2004categorical,van2020zx}). 
An important example is the ZX-calculus~\cite{coecke2008interacting,coecke2011interacting,van2020zx} and its closely related  variants~\cite{ranchin2014depicting,wang2014qutrit,hadzihasanovic2015diagrammatic,jeandel2017calculus,hadzihasanovic2018diagrammatic,backens2018zh,east2022aklt} which have seen a number of successful applications in quantum computing, 
ranging from circuit optimization~\cite{duncan2020graph,kissinger2020reducing,backens2021there,gorard2021zx} 
and synthesis~\cite{cowtan2019phase,de2020architecture}, 
to algorithm analysis~\cite{carette2021quantum,townsend2021classifying}, 
natural language processing~\cite{coecke2020foundations} and 
machine learning~\cite{yeung2020diagrammatic,toumi2021diagrammatic,zhao2021analyzing}, among others. 

In this paper we show how the ZX-calculus is also useful for analyzing 
algorithms based on 
parameterized quantum circuits (PQCs), 
such as variational quantum algorithms, 
in particular for 
calculating important derived quantities such as expectation values of quantum observables, or their gradients. Such quantities may be computed as functions of the circuit parameters, in which case the parameters are symbolically carried through subsequent ZX-diagrams, or as numbers for the case of fixed parameters of interest. 
To enable this, we present several new ZX-rules generalizing the standard ones appearing in the literature; in particular, 
we present rules and notation for explicitly handling linear combinations of ZX-diagrams which naturally arise, 
for example, when incorporating commutation rules for unitary operators which are used for instance in computing 
expectation values. 
For linear combination of diagrams, clearly, it is critical to keep track of the scalar multiplier of each diagram, 
whereas in previous single-diagram applications such global phases or normalization constants can typically be ignored. 
In our application these multipliers will typically be complex-valued functions of the quantum circuit parameters. 
Furthermore, our formalism then allows direct importation of a number of useful relations from the operator analysis to ZX calculus setting, such as causal cone and operator commutation rules, among others.

After stating the new rules we demonstrate their efficacy with
several prototypical examples of parameterized quantum circuits in the context of combinatorial optimization, 
including straightforward derivation of some new and existing results concerning 
example circuits drawn from the literature. 
While for computing expectation values of relatively shallow circuits we are able to show most of the key diagram reduction steps explicitly, 
for deeper circuits our approach can be aided by integration with software implementations of the ZX-calculus (e.g., \cite{kissinger2015quantomatic,pyzx2019}). 
Though we focus on the common task of analyzing quantum circuit expectation values, important in particular for assessing algorithm performance, our proposed rules are general and may find much broader application in future work. 
For instance, toward analyzing phenomena related to parameter setting, expectation value gradients may be obtained either by 
differentiating directly~\cite{toumi2021diagrammatic}, or by 
reducing the calculation to that of computing further circuit expectation values as in
parameter shift rules~\cite{crooks2019gradients,Wierichs2022generalparameter}. 
We emphasize that our approach may be applied to a wide variety of application problems and related quantum circuits beyond those explicitly considered in our examples, 
and further 
ZX results and 
generalizations from the literature may be leveraged, 
including extensions to qudits~\cite{wang2014qutrit} or fermions~\cite{hadzihasanovic2018diagrammatic,cowtan2020generic}, among others. 


\section{Preliminaries}%
\subsection{ZX-Calculus}
\begin{figure}[htpb]
    \centering
        \begin{tikzpicture}
            \input{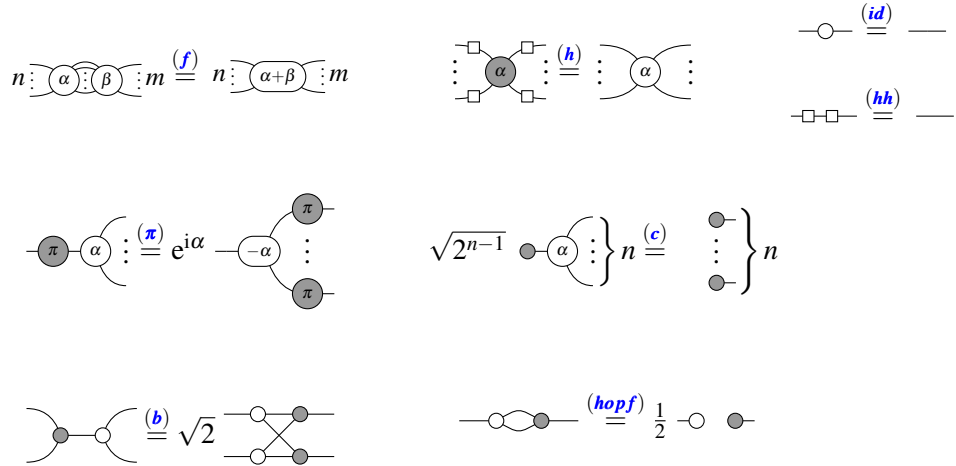}
        \end{tikzpicture}
    \caption{The ZX-diagram rewrite rules (cf.\ for example~\cite{van2020zx} or~\cite{zhao2021analyzing}). Note 
    the explicit 
    scalar factors.}%
    \label{fig:zx_rules}
\end{figure}

We refer the reader to~\cite{vilmart2018near,van2020zx} and the references therein for comprehensive introductions, including complete sets of graphical rewrite rules 
as well as their mathematical details. 
A number of the most important ZX-diagram rewrite rules are displayed in Figure~\ref{fig:zx_rules}. 
We use the label attached to each equation to reference these rules when we apply them in the examples we consider below. 

\subsection{Parameterized Quantum Circuits}
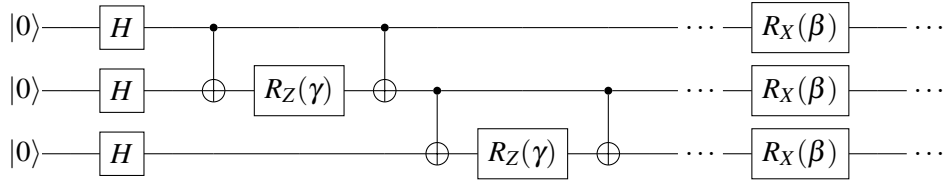
\begin{figure}[htpb]
    \centering
    \begin{quantikz}[thin lines, row sep=1ex, column sep=1em]
          \ket{0} & \qw & \gate{H} & \qw & \ctrl{1} & \qw                
        & \ctrl{1}  & \qw      &    \qw             &  \qw     & \qw 
        & \ \ldots\ \qw & \gate{R_X(\beta)} & \qw & \ \ldots\  \qw \\
         \ket{0} & \qw & \gate{H} & \qw & \targ{}  
        & \gate{R_Z(\gamma)} & \targ{}   & \ctrl{1}  &    \qw             & \ctrl{1} & \qw 
        & \ \ldots\ \qw & \gate{R_X(\beta)} & \qw & \ \ldots\  \qw \\
        \ket{0} & \qw & \gate{H} & \qw & \qw      & \qw               
        &   \qw   & \targ{}  & \gate{R_Z(\gamma)} & \targ{}    & \qw 
        & \ \ldots\ \qw & \gate{R_X(\beta)} & \qw & \ \ldots\  \qw 
    \end{quantikz}
    \caption{Example of a parameterized quantum circuit: QAOA on 3 qubits. Here the phase and mixing operators as well as initial state preparation have been compiled to basic quantum gates.}%
    \label{fig:example_pqc}
\end{figure}

Parameterized quantum circuits (PQC) have gained much attention in recent years, in particular as heuristic approaches suitable for NISQ~\cite{preskill2018quantum} era devices that are classically optimized  (often variationally) as part of a hybrid 
protocol, though we emphasize they are by no means restricted to this setting; see \cite{cerezo2021variational,bharti2022noisy} for reviews of recent developments.  
Two particular approaches of interest are the QAOA (quantum alternating operator ansatz~\cite{hadfield2019quantum}, which generalizes the quantum approximate optimization algorithm~\cite{farhi2014quantum}) and VQE (variational quantum eigensolver~\cite{peruzzo2014variational,mcclean2016theory}) paradigms, 
as well as a number of more recent variants of these approaches. 
Here we briefly review the original QAOA paradigm and its application to combinatorial optimization, though our results to follow may be applied more generally to a variety of problems and algorithms. 
In QAOA we are given a cost function $c(x)$ and corresponding classical Hamiltonian $C$ (i.e., diagonal in the computational basis,  $C\ket{x}=c(x)\ket{x}$) we seek to optimize over bit strings $x\in\{0,1\}^n$. 
A QAOA$_p$ circuit 
consists of $2p$ alternating layers 
specified by $2p$ angles $\gamma_i,\beta_i$ in some domain (e.g.~$[-\pi,\pi]$) to create the state 
\begin{equation}
    \ket{\boldsymbol{\gamma \beta}}= U_M(\beta_p)U_P(\gamma_p)\dots U_M(\beta_1)U_P(\gamma_1)\ket{s} \, ,
\end{equation}
for phase operator $U_P(\gamma)=\exp(-\ii \gamma C)$, (transverse-field) mixing operator $U_M(\beta)=\exp(-\ii \beta B)$ where $B=\sum_{i=1}^n X_i$, and standard initial product state $\ket{s}=\ket{+}^{\otimes n}$. The state is then measured in the computational basis 
which returns some $y\in\{0,1\}^n$ achieving cost $c(y)$. 
Figure~\ref{fig:example_pqc} shows a simple example of a QAOA circuit.
Repeated state preparation and measurement gives further samples which may be used to estimate the cost expectation $\langle C\rangle_p$ or other important quantities. 
These quantities may be used to update or search for better circuit parameters if desired; we emphasize that in different cases parameters may be found through analytic~\cite{wang2018quantum}, numeric~\cite{farhi2014quantum}, or average-case~\cite{streif2019training} techniques, or, distinctly, searched for empirically 
(e.g., variationally). 
After a set number of runs overall, or when other suitable termination criteria has been reached, the best solution found is returned. 

A fundamentally important quantity for QAOA as well as related approaches is the cost expectation value $\langle C\rangle$, 
which  may be computed  for a single instance or over a suitable class, 
and can be used to bound the expected approximation ratio achieved~\cite{farhi2014quantum,hadfield2018quantum,hadfield2019quantum,hadfield2021analytical} for the given problem. 
Importantly, we are often given a decomposition of the cost Hamiltonian such as $C=\sum_j C_j$ which we may exploit in computing $\langle C\rangle = \sum_j \langle C_j\rangle$ as a sum of terms (typically, a linear combination of Pauli $Z$ operators~\cite{hadfield2021representation}), which directly motivates the 
rules we introduce for accommodating linear combinations of ZX-diagrams. 
For combinatorial optimization the $C_j$ terms mutually commute which leads to further simplifications, whereas this may not be true for more general problems and applications such as quantum chemistry (though linearity of expectation still applies).
In general, many quantities of interest for PQCs can be expressed as expectation values and are hence amenable to similar analysis via diagrammatic techniques as we explore below. 

\subsection{Related Work}
Several 
recent papers 
provide 
related but distinct results towards applying the ZX calculus in the PQC setting. 
In particular, three
papers  \cite{toumi2021diagrammatic,wang2022differentiating,jeandel2022addition} 
which appeared during 
preparation of this work that consider  differentiation and addition of ZX-diagrams. 
These papers introduce diagrammatic extensions 
complementary to our results. 
However they do not deal with expectation values explicitly which is the focus of this work. 
%
In terms of previous applications to variational quantum algorithms, 
a recent paper~\cite{zhao2021analyzing} considers using the ZX-calculus for computing and analyzing expectation values of derivatives of the cost expectation for particular classes of random parameterized quantum circuits built from particular gate sets (see in particular~\cite[Assumption 1]{zhao2021analyzing}), 
in the context of detecting possible barren plateaus~\cite{mcclean2018barren}. 
Our work differs in that we consider expectation values of the cost function themselves, and make no similar assumption of randomly selected parameters. 
A particular similarity with~\cite{zhao2021analyzing} is both their application and ours require explicit accounting of scalar factors associated to ZX-diagrams (see Section~\ref{sec:ZXforPQC}). 
However, while it is observed  in~\cite[Eq. 7]{zhao2021analyzing} that quantum expectation values may represented with the ZX-calculus in~\cite[Eq. 7]{zhao2021analyzing}, the authors do not apply the decomposition $C=\sum_j C_j$, 
which we exploit to derive novel ZX rules and analysis. 
Our approach and results are complementary to those of~\cite{zhao2021analyzing}.  
Another work~\cite{fontana2020optimizing} applied ZX-calculus in analysis of symmetries in the parameter landscape of the cost function expectation. 
We note that a different diagrammatic approach to constructing parameterized quantum circuits is considered in~\cite{herasymenko2021diagrammatic}.
Concepts related to linear combinations of ZX-diagrams have been discussed in the framework of category theory for example in~\cite{comfort2020sheet,duncan2009generalised}.

\section{ZX-Calculus for Parameterized Quantum Circuits} \label{sec:ZXforPQC}
\label{sec:zx_for_pqc}
In this section we extend the ZX-calculus to 
accommodate linear combinations of (conventional) ZX-diagrams.
Then, toward its application to parameterized quantum circuits we derive a collection of general rules and useful identities within the new framework. 
We will apply these rules to several concrete quantum circuit examples in Section~\ref{sec:example_applications} and the Appendices.

\subsection{Diagrammatic Rules for Linear Combinations}
Here we define linear combinations of diagrams, in which case diagram constants give the relative weights of the sum. For example, for computing the expectation value of an observable $H=\sum_{j=1}^m a_j H_j$ for some quantum circuit state $\ket{\psi}=U\ket{\psi_0}$ we have $\langle H \rangle_\psi = \sum_{j=1}^m a_j\langle H_j \rangle_\psi$, which hence corresponds to a single ZX-diagram or equivalently to a sum of $m$ weighted diagrams. This idea generalizes in the natural way to sums of linear maps and more general ZX objects. 
We also show new ZX-diagram rules which relate single (sub)diagrams to sums or products of (sub)diagrams, such that the resulting diagram reductions involve differing numbers of diagrams.

As mentioned, we do not use the common convention of considering diagrams equivalent up to scalars or phases; hence we include complex scalar multipliers explicitly in our diagrams and rules to follow, i.e., 
\begin{equation}
    a \, \cdot \, \;
    \vcenter{\hbox{\diagplacehold{$m$}{$A$}{$n$}[2.5em]}}
    \, \neq \, \,\,
    b \, \cdot \, \;
    \vcenter{\hbox{\diagplacehold{$m$}{$A$}{$n$}[2.5em]}} \label{eqn:zxsumrules_scalar}
    \qquad 
    \text{unless } a=b \,,
\end{equation}
where $a, b$ are complex scalar multipliers and the diagram is a placeholder for an arbitrary ZX-diagram with $m$ inputs and $n$ outputs. 
In particular, care must be taken in applying the usual rules of ZX-calculus to account for any implicit constant factors. 
We note that scalar factors are also retained in the distinct application of~\cite{zhao2021analyzing}; see~\cite[Fig. 4]{zhao2021analyzing} for an example list of some ZX-diagram rewrite rules with explicit scalars.

\begin{defn}[Sum notation]
We define novel diagram notation 
for describing arbitrary linear combinations.
The linear combination of two ZX-diagrams with $m$ inputs and $n$ outputs is written 
\begin{equation}
    a \, \cdot \, \;
    \vcenter{\hbox{\diagplacehold{$m$}{$A$}{$n$}[2.5em]}}
    +
    \,\, b \, \cdot \, \;
    \vcenter{\hbox{\diagplacehold{$m$}{$B$}{$n$}[2.5em]}} \\
     =: \vcenter{\hbox{
            \externalize{zxsumrules-sum}{%
                \begin{tikzpicture}[node distance=5ex and 2.5em]
    \def\sumwidth{14em}
    \def\yscale{1.3}
    \node[name=anchorlt, invisible];
    \node[on grid, below=of anchorlt, name=anchorlb, invisible];
    \node[right=\sumwidth of anchorlt.center, name=anchorrt, invisible];
    \node[on grid, below=of anchorrt, name=anchorrb, invisible];
    \node at ($(anchorlt.center)!0.5!(anchorlb.center)$) [yscale=\yscale, name=sumleft, leftsummator] ;
    \node[right=\sumwidth of sumleft.west, yscale=\yscale, name=sumright, rightsummator];
    \zxsumtwo{%
        \diagplacehold{$m$}{$A$}{$n$}
        }{%
        \diagplacehold{$m$}{$B$}{$n$}
    }[0.0em][6.0ex][0.7][a][b]
    \draw[thin] (anchorlt) -- +(180:0.7em);
    \node[left=0.4em of sumleft.west, anchor=center] (dots) {\rvdots};
    \node[left=0.4em of dots, anchor=center] {$m$};
    \draw[thin] (anchorlb) -- +(180:0.7em);
    \draw[thin] (anchorrt) -- +(0:0.7em);
    \node[right=0.4em of sumright.west, anchor=center] (dots) {\rvdots};
    \node[right=0.4em of dots, anchor=center] {$n$};
    \draw[thin] (anchorrb) -- +(0:0.7em);

                \end{tikzpicture}
            }
    }} \label{eqn:zxsumrules_sum} \, .
\end{equation}
\end{defn}
Each summand is written inside a \textit{bubble} and the scalar factors are written on the line combining the new summation symbol and the bubbles.
This definition naturally extends to an arbitrary number of summands. 
Summand diagrams are required have the same numbers of input $(m)$ and output $(n)$ lines as each other and
as the those of the sum object. 
Note that $m$ or $n$ are zero for diagrams representing states ($m=0$), effects ($n=0$), or constants ($m=n=0$). 
Sums of diagrams also arise in~\cite{yeung2020diagrammatic,toumi2021diagrammatic,zhao2021analyzing} in the context of differentiating diagram components (where sums arise, for example, from the product rule of calculus).
Our work is complementary to these results in that we consider the generalization to complex linear combinations of diagrams; extensions to scalars beyond the complex numbers are also possible~\cite{toumi2021diagrammatic}.

\subsubsection{Rules}
Now we state the two rules needed for the extension of ZX-calculus to linear combinations.
\begin{enumerate}
    \item {Diagram Pull Rule}
        
        The first rule 
        applies if diagrams in a linear combination are equal up to a certain subdiagram ($A$ and $B$ below).
        Then we can write a single diagram containing the linear combination of the beforementioned subdiagrams
        \combineddiagplaceholdconfig
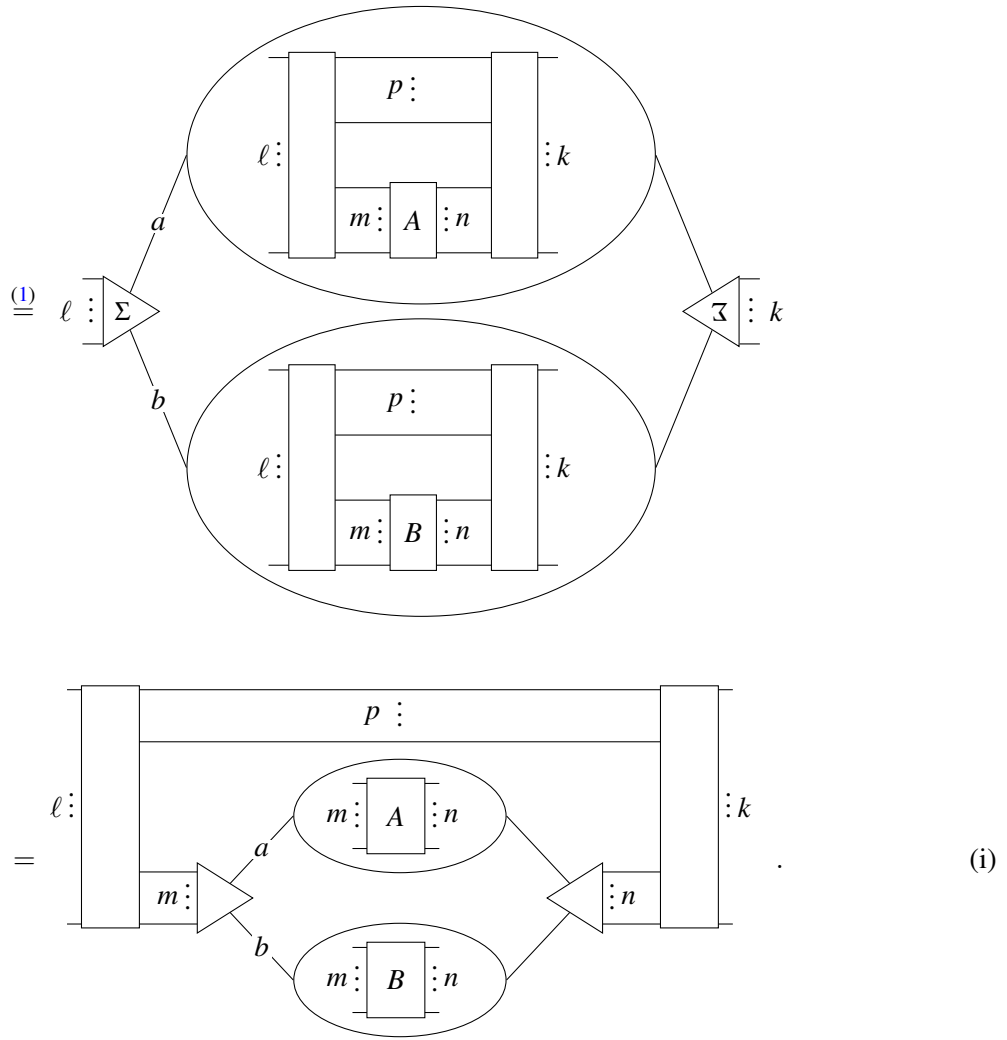
        \begin{align}
            &
            a \, \cdot \, \;
            \vcenter{\hbox{
                \scalebox{1.0}{
                    \combineddiagplacehold{$A$}
                }
            }}
            +
            b \, \cdot \, \;
            \vcenter{\hbox{
                \scalebox{1.0}{
                    \combineddiagplacehold{$B$}
                }
            }} \\[2em]
            \stackrel{\eqref{eqn:zxsumrules_sum}}{=} 
            &
            \vcenter{\hbox{
                \scalebox{1.0}{
                    \externalize{zxsumrules_partial_sum_all_combined}{%
                        \begin{tikzpicture}[node distance=5ex and 2.5em]
\def\sumwidth{22em}
\def\yscale{1.3}
\node[name=anchorlt, invisible];
\node[on grid, below=of anchorlt, name=anchorlb, invisible];
\node[right=\sumwidth of anchorlt.center, name=anchorrt, invisible];
\node[on grid, below=of anchorrt, name=anchorrb, invisible];
\node at ($(anchorlt.center)!0.5!(anchorlb.center)$) [yscale=\yscale, name=sumleft, leftsummator] ;
\node[right=\sumwidth of sumleft.west, yscale=\yscale, name=sumright, rightsummator];
\combineddiagplaceholdconfig
\zxsumtwo{%
    \combineddiagplacehold{$A$}
    }{%
    \combineddiagplacehold{$B$}
}[0.0em][12.0ex][1.0][a][b]
\draw[thin] (anchorlt) -- +(180:0.7em);
\node[left=0.4em of sumleft.west, anchor=center] (dots) {\rvdots};
\node[left=0.4em of dots, anchor=center] {$\ell$};
\draw[thin] (anchorlb) -- +(180:0.7em);
\draw[thin] (anchorrt) -- +(0:0.7em);
\node[right=0.4em of sumright.west, anchor=center] (dots) {\rvdots};
\node[right=0.4em of dots, anchor=center] {$k$};
\draw[thin] (anchorrb) -- +(0:0.7em);

                        \end{tikzpicture}
                    }
                }
            }} \\[2em]
            = &
            \vcenter{\hbox{
                \scalebox{1.0}{
                    \externalize{zxsumrules_partial_sum_combined}{%
                        \begin{tikzpicture}[node distance=4ex and 1.5em]
\tikzset{label/.style={outer sep=0pt, inner sep=0pt}}
\def\dx{10ex}
\def\sumwidth{14em}
\def\yscale{1.3}

\def\xsep{1.0em}
\def\ysep{0.3ex}
\def\ldist{0.2em}
\def\labelleft{$\ell$}
\def\labelmiddle{$p$}
\def\labelright{$k$}

\node[name=1-lb, invisible];
\node[on grid, name=2-lb, above=of 1-lb, invisible];
\node[name=3-lb, above=\dx of 2-lb, invisible];
\node[on grid, name=4-lb, above=of 3-lb, invisible];

\node[on grid, name=1-l, left=of 1-lb, invisible];
\node[on grid, name=4-l, left=of 4-lb, invisible];

\node[rectangle, 
      draw=black,thin, 
      fill=white,
      fit=(1-lb) (4-lb), 
      inner xsep=\xsep, 
      inner ysep=\ysep] (leftbox) {};
\node[left=\ldist of leftbox , label] (dots) {\rvdots};
\node[left=\ldist of dots, label] {\labelleft};

\node[on grid, right=of 1-lb, name=1-ll, invisible];
\node[on grid, right=of 1-ll, name=1-ls, invisible];
\node[on grid, right=of 2-lb, name=2-ll, invisible];
\node[on grid, right=of 2-ll, name=2-ls, invisible];
\node[right=\sumwidth of 1-ls.center, name=1-rs, invisible];
\node[right=\sumwidth of 2-ls.center, name=2-rs, invisible];
\node[on grid, right=of 1-rs, name=1-rl, invisible];
\node[on grid, right=of 2-rs, name=2-rl, invisible];
\node at ($(1-ls.center)!0.5!(2-ls.center)$) [yscale=\yscale, name=sumleft, leftsummator];
\node[right=\sumwidth of sumleft.west, yscale=\yscale, name=sumright, rightsummator];
\zxsumtwo{%
    \diagplacehold{$m$}{$A$}{$n$}
    }{%
    \diagplacehold{$m$}{$B$}{$n$}
}[0.0em][6.3ex][1.0][a][b]
\node[right=\ldist of sumright.west, label] (dots) {\rvdots};
\node[right=\ldist of dots, label] {$n$};
\node[left=\ldist of sumleft.west, label] (dots) {\rvdots};
\node[left=\ldist of dots, label] {$m$};

\node[on grid, name=1-rb, right=of 1-rl, invisible];
\node[on grid, name=2-rb, above=of 1-rb, invisible];
\node[name=3-rb, above=\dx of 2-rb, invisible];
\node[on grid, name=4-rb, above=of 3-rb, invisible];
\node[on grid, name=4-r, left=of 4-rb, invisible];

\node[rectangle, 
      draw=black,thin, 
      fill=white,
      fit=(1-rb) (4-rb), 
      inner xsep=\xsep, 
      inner ysep=\ysep] (rightbox) {};
\node[right=\ldist of rightbox , label] (dots) {\rvdots};
\node[right=\ldist of dots, label] {\labelright};

\node[anchor=center] at ($(3-lb.center)!0.5!(4-rb.center)$) (dots) {\rvdots};
\node[left=\ldist of dots, label] {\labelmiddle};

\node[on grid, name=1-r, right=of 1-rb, invisible];
\node[on grid, name=4-r, right=of 4-rb, invisible];

\begin{scope}[on background layer]
    \draw (1-l) -- (1-ls);
    \draw (1-rs) -- (1-r);
    \draw (2-lb) -- (2-ls);
    \draw (2-rs) -- (2-rb);
    \draw (3-lb) -- (3-rb);
    \draw (4-l) -- (4-r);
\end{scope}

                        \end{tikzpicture}
                    }
                }
            }} \label{eqn:zxsumrules_pull} \, . \tag{i}
        \end{align}

        The last equality we call the \textit{diagram pull rule}.
        This also holds if any of the $\ell$, $p$, $k$, $m$, $n$ vanish. 
        Thus describing how to pull scalars, effects and states in and out of the bubbles.
        If $p=0$, the rule describes how to pull in and out diagrams only from the left or only from the right.
        We will make heavy use of this in Section~\ref{sec:example_applications}.

    \item {Product (Composition) Rule}

        The second rule describes how to combine products 
        (i.e., compositions) 
        of linear combinations of diagrams. 
        We state the rule for a product of two linear combinations comprised of two summands each
        \begin{align}
            & 
            \biggl( 
                  a \, \cdot \, \;
                  \vcenter{\hbox{\diagplacehold{$m$}{$A$}{$n$}[2.5em]}} 
                + \, b \, \cdot \, \;
                  \vcenter{\hbox{\diagplacehold{$m$}{$B$}{$n$}[2.5em]}} 
            \biggr)
            \, \circ  \,
            \biggl( 
                  c \, \cdot \, \;
                  \vcenter{\hbox{\diagplacehold{$n$}{$C$}{$\ell$}[2.5em]}} 
                + \, d \, \cdot \, \;
                  \vcenter{\hbox{\diagplacehold{$n$}{$D$}{$\ell$}[2.5em]}} 
            \biggr)
            \\[2em]
            = & 
            \vcenter{\hbox{
                \externalize{zxsumrules_product_part_comb1}{%
                    \begin{tikzpicture}[node distance=5ex and 2.5em]
    \def\sumwidth{14em}
    \def\yscale{1.3}
    \node[name=anchorlt, invisible];
    \node[on grid, below=of anchorlt, name=anchorlb, invisible];
    \node[right=\sumwidth of anchorlt.center, name=anchorrt, invisible];
    \node[on grid, below=of anchorrt, name=anchorrb, invisible];
    \node at ($(anchorlt.center)!0.5!(anchorlb.center)$) [yscale=\yscale, name=sumleft, leftsummator] ;
    \node[right=\sumwidth of sumleft.west, yscale=\yscale, name=sumright, rightsummator];
    \zxsumtwo{%
        \diagplacehold{$m$}{$A$}{$n$}
        }{%
        \diagplacehold{$m$}{$B$}{$n$}
    }[0.0em][6.0ex][0.7][a][b]
    \draw[thin] (anchorlt) -- +(180:0.7em);
    \node[left=0.4em of sumleft.west, anchor=center] (dots) {\rvdots};
    \node[left=0.4em of dots, anchor=center] {$m$};
    \draw[thin] (anchorlb) -- +(180:0.7em);
    \draw[thin] (anchorrt) -- +(0:0.7em);
    \node[right=0.4em of sumright.west, anchor=center] (dots) {\rvdots};
    \node[right=0.4em of dots, anchor=center] {$n$};
    \draw[thin] (anchorrb) -- +(0:0.7em);

                    \end{tikzpicture}
                }
            }}
            \, \circ  \,
            \vcenter{\hbox{
                \externalize{zxsumrules_product_part_comb2}{%
                    \begin{tikzpicture}[node distance=5ex and 2.5em]
    \def\sumwidth{14em}
    \def\yscale{1.3}
    \node[name=anchorlt, invisible];
    \node[on grid, below=of anchorlt, name=anchorlb, invisible];
    \node[right=\sumwidth of anchorlt.center, name=anchorrt, invisible];
    \node[on grid, below=of anchorrt, name=anchorrb, invisible];
    \node at ($(anchorlt.center)!0.5!(anchorlb.center)$) [yscale=\yscale, name=sumleft, leftsummator] ;
    \node[right=\sumwidth of sumleft.west, yscale=\yscale, name=sumright, rightsummator];
    \zxsumtwo{%
        \diagplacehold{$n$}{$C$}{$\ell$}
        }{%
        \diagplacehold{$n$}{$D$}{$\ell$}
    }[0.0em][6.0ex][0.7][c][d]
    \draw[thin] (anchorlt) -- +(180:0.7em);
    \node[left=0.4em of sumleft.west, anchor=center] (dots) {\rvdots};
    \node[left=0.4em of dots, anchor=center] {$n$};
    \draw[thin] (anchorlb) -- +(180:0.7em);
    \draw[thin] (anchorrt) -- +(0:0.7em);
    \node[right=0.4em of sumright.west, anchor=center] (dots) {\rvdots};
    \node[right=0.4em of dots, anchor=center] {$l$};
    \draw[thin] (anchorrb) -- +(0:0.7em);

                    \end{tikzpicture}
                }
            }}
            \\[2em]
            = &
            \vcenter{\hbox{
                \externalize{zxsumrules_product_full_comb}{%
                    \begin{tikzpicture}[node distance=5ex and 2.5em]
    \def\sumwidth{25em}
    \def\yscale{1.3}
    \node[name=anchorlt, invisible];
    \node[on grid, below=of anchorlt, name=anchorlb, invisible];
    \node[right=\sumwidth of anchorlt.center, name=anchorrt, invisible];
    \node[on grid, below=of anchorrt, name=anchorrb, invisible];
    \node at ($(anchorlt.center)!0.5!(anchorlb.center)$) [yscale=\yscale, name=sumleft, leftsummator] ;
    \node[right=\sumwidth of sumleft.west, yscale=\yscale, name=sumright, rightsummator];
    \zxsumconfig(0, 0)[0.40em][6.0ex][0.7]
    \zxsumfour{%
        \diagplaceholdproduct{$m$}{$A$}{$C$}{$\ell$}
        }{%
        \diagplaceholdproduct{$m$}{$A$}{$D$}{$\ell$}
        }{%
        \diagplaceholdproduct{$m$}{$B$}{$C$}{$\ell$}
        }{%
        \diagplaceholdproduct{$m$}{$B$}{$D$}{$\ell$}
    }[ac][ad][bc][bd]
    \draw[thin] (anchorlt) -- +(180:0.7em);
    \node[left=0.4em of sumleft.west, anchor=center] (dots) {\rvdots};
    \node[left=0.4em of dots, anchor=center] {$m$};
    \draw[thin] (anchorlb) -- +(180:0.7em);
    \draw[thin] (anchorrt) -- +(0:0.7em);
    \node[right=0.4em of sumright.west, anchor=center] (dots) {\rvdots};
    \node[right=0.4em of dots, anchor=center] {$\ell$};
    \draw[thin] (anchorrb) -- +(0:0.7em);

                    \end{tikzpicture}
                }
            }} \label{eqn:zxsumrules_product_of_sums} \tag{ii} \,.
        \end{align}  
    The product rule 
    extends in the obvious way  
    to the case of more than two factors or summands.
\end{enumerate}
Several additional rules are given in Appendix~\ref{sec:additional_rules}.

\subsection{ZX-Calculus for Expectation Values of Quantum Circuits}
In this section, we will present various identities within the extended ZX-calculus framework, that are useful for the analysis of parameterized quantum circuits. 
While we primarily consider  Pauli operators here, similar results may derived in different 
basis or  gate sets. 
See \cite{cowtan2020generic} for some additional useful rules regarding Pauli operator exponentials.

\subsubsection{Rotations}
First, we can write rotation operators in terms of linear combinations of Clifford gates
\begin{align}
    \ee^{\ii \gamma Z}
    &\,=\,
    \ee^{\ii \gamma}
    \begin{ZX}[ampersand replacement=\&]
        \zxN{} \rar \& \zxZ{-2\gamma} \rar \& \zxN{} 
    \end{ZX}
    \,=\,
    \vcenter{\hbox{
        \externalize{z_spider}{%
            \tikz{%
    \def\sumwidth{14em}
    \def\yscale{1.3}
    \node [yscale=\yscale, name=sumleft, leftsummator];
    \node[right=\sumwidth of sumleft.west, yscale=\yscale, name=sumright, rightsummator];
    \zxsumtwo{%
        \begin{ZX}
            \\[0.3ex]
                \zxN{} \rar &[1.0em] \zxN{} \rar & \zxN{}  
            \\[0.3ex]
        \end{ZX}
        }{%
        \begin{ZX}
            \zxN{} \rar & \zxZ{\pi} \rar & \zxN{}
        \end{ZX}
    }[0.0em][5.0ex][1.0][c_\gamma][\ii s_\gamma]
    \draw[thin] (sumleft) -- +(180:2.0em);
    \draw[thin] (sumright) -- +(0:2.0em);

            }
        }
    }} \label{eqn:zspider_as_lincomb} \, ,
\end{align}
\begin{align}
    \ee^{\ii \beta X}
    &\,=\,
    \ee^{\ii \beta}
    \begin{ZX}[ampersand replacement=\&]
        \zxN{} \rar \& \zxX{-2\beta} \rar \& \zxN{} 
    \end{ZX}
   \,=\,
    \vcenter{\hbox{
        \externalize{x_spider}{%
            \tikz{%
    \def\sumwidth{14em}
    \def\yscale{1.3}
    \node [yscale=\yscale, name=sumleft, leftsummator];
    \node[right=\sumwidth of sumleft.west, yscale=\yscale, name=sumright, rightsummator];
    \zxsumtwo{%
        \begin{ZX}
            \\[0.3ex]
                \zxN{} \rar &[1.0em] \zxN{} \rar & \zxN{}  
            \\[0.3ex]
        \end{ZX}
        }{%
        \begin{ZX}
            \zxN{} \rar & \zxX{\pi} \rar & \zxN{}
        \end{ZX}
    }[0.0em][4.0ex][1.0][c_\beta][\ii s_\beta]
    \draw[thin] (sumleft) -- +(180:2.0em);
    \draw[thin] (sumright) -- +(0:2.0em);

            }
        }
    }} \label{eqn:xspider_as_lincomb} \, .
\end{align}
In both cases the proof easily follows from the identity $\ee^{\ii\alpha A}=\cos\alpha I +\ii \sin\alpha A$ for operators satisfying $A^2=I$. 
We use $c_\alpha:=\cos(\alpha)$ and $s_\alpha:=\sin(\alpha)$ throughout.

\subsubsection{Phase-Gadgets}
Important for parameterized quantum circuits are multi-qubit rotations, so-called \textit{phase-gadgets} (cf.~\cite{cowtan2020generic}),
for example
\begin{align}
    \ee^{\ii \gamma Z_u Z_v}
    & \,=\,
    \sqrt{2} \ee^{\ii \gamma}
    \begin{ZX}[ampersand replacement=\&]
        \zxN{} \rar \& \zxZ{} \rar \dar \& \zxN{}          \& \zxN{} \\
                    \& \zxX{} \rar      \& \zxZ{-2 \gamma} \&        \\
        \zxN{} \rar \& \zxZ{} \rar \uar \& \zxN{}          \& \zxN{} 
    \end{ZX}
   \,=\,
    \vcenter{\hbox{
        \externalize{phase_gadget}{%
            \tikz{%
    \def\sumwidth{14em}
    \def\yscale{1.3}
    \def\yshift{1.5ex}
    \node [yscale=\yscale, name=sumleft, leftsummator];
    \node[right=\sumwidth of sumleft.west, yscale=\yscale, name=sumright, rightsummator];
    \zxsumtwo{%
        \begin{ZX}
            \\[0.2ex]
                \zxN{} \rar &[1.0em] \zxN{} \rar & \zxN{} \\[\zxSRow,\zxSRow]
                \zxN{} \rar &[1.0em] \zxN{} \rar & \zxN{}  
            \\[0.2ex]
        \end{ZX}
        }{%
        \begin{ZX}[row sep=0.2ex]
            \zxN{} \rar & \zxZ{\pi} \rar & \zxN{} \\
            \zxN{} \rar & \zxZ{\pi} \rar & \zxN{} 
        \end{ZX}
    }[0.0em][6.0ex][1.0][c_\gamma][\ii s_\gamma]
    \draw[thin] ($(sumleft.west)  + (0,  \yshift)$) -- node[at end, xshift=-1.0em] {\small $u$} +(180:1.0em);
    \draw[thin] ($(sumleft.west)  + (0, -\yshift)$) -- node[at end, xshift=-1.0em] {\small $v$} +(180:1.0em);
    \draw[thin] ($(sumright.west) + (0,  \yshift)$) -- +(0:1.0em);
    \draw[thin] ($(sumright.west) + (0, -\yshift)$) -- +(0:1.0em);

            }
        }
    }} \label{eqn:phase_gadget_as_lincomb} \, .
\end{align}
A proof of the first equality is given in \cite[Corollary 3.4]{cowtan2020generic}. 
The second equality is derived similarly to \eqref{eqn:zspider_as_lincomb}, \eqref{eqn:xspider_as_lincomb}. 
In particular, for the analysis of QAOA expectation values, we will encounter conjugates of phase-gadgets in conjunction with $\pi$-X-spiders.
We will make heavy use of the following identity which is proven in~Appendix~\ref{sec:app_proof_conjugates_cancel_phase_gadgets}.
\begin{equation}\label{eqn:conjugates_cancel_phase_gadgets}
    \vcenter{\hbox{
        \externalizezx{conjugates_cancel_phase_gadgets_lhs}{%
            \def\dotdist{0.7em}
\begin{ZX}[ampersand replacement=\&]
                               \& \zxN{} \ar[rd, start anchor={[xshift=\dotdist]}] 
    \& \dots                   \& \zxN{} \ar[ld, start anchor={[xshift=-\dotdist]}] \ar[rd, start anchor={[xshift=\dotdist]}] 
    \& \dots                   \& \zxN{} \ar[ld, start anchor={[xshift=-\dotdist]}]          
    \& 
    \\
       \zxN{} \rar             \& \zxN{} \rar 
    \& \zxZ{} \rar             \& \zxX{t\pi} \rar                                   
    \& \zxZ{} \rar             \& \zxN{} \rar             
    \& \zxN{}
    \\
                               \& \zxZ-{\gamma} \rar  
    \& \zxX{\ell\pi} \uar \dar \&                                                   
    \& \zxX{r\pi} \uar \dar    \& \zxZ{\gamma} \lar      
    \& \zxN{}
    \\
       \zxN{} \rar             \& \zxN{} \rar   
    \& \zxZ{} \rar             \& \zxX{b\pi} \rar   
    \& \zxZ{} \rar             \& \zxN{} \rar             
    \& \zxN{}
    \\
                               \& \zxN{} \ar[ru, start anchor={[xshift=\dotdist]}] 
    \& \dots                   \& \zxN{} \ar[lu, start anchor={[xshift=-\dotdist]}] \ar[ru, start anchor={[xshift=\dotdist]}] 
    \& \dots                   \& \zxN{} \ar[lu, start anchor={[xshift=-\dotdist]}]          
    \& 
\end{ZX}

        }
    }}
    =
    \vcenter{\hbox{
        \externalizezx{conjugates_cancel_phase_gadgets_rhs}{%
            \begin{tikzpicture}
   \node (odd) {%
       $\frac{\ee^{\ii \gamma}}{\sqrt{2}}$
       \begin{ZX}[ampersand replacement=\&]
       \zxN{} \ar[rd]          \& \dots 
    \& \zxN{} \ar[ld]          \&                    
    \&                         \&                 
    \&                         \&
    \& \zxN{} \ar[rd]          \& \dots                        
    \& \zxN{} \ar[ld]      
    \\
       \zxN{}                  \& \zxZ{} \rar \lar
    \& \zxN{} \rar             \& \zxX{(t+r)\pi} \rar
    \& \zxN{} \rar             \& \zxZ{} \rar \dar
    \& \zxN{} \rar             \& \zxX{r\pi} \rar             
    \& \zxN{} \rar             \& \zxZ{} \rar             
    \& \zxN{}
    \\
                               \&
    \&                         \&
    \&                         \& \zxX{} \rar 
    \& \zxZ-{2\gamma}           \& 
    \&                         \&
    \&
    \\
       \zxN{}                  \& \zxZ{} \rar \lar
    \& \zxN{} \rar             \& \zxX{(b+1)\pi} \rar  
    \& \zxN{} \rar             \& \zxZ{} \rar \uar
    \& \zxN{} \rar             \& \zxX{\pi} \rar             
    \& \zxN{} \rar             \& \zxZ{} \rar            
    \& \zxN{}
    \\
       \zxN{} \ar[ru]          \& \dots 
    \& \zxN{} \ar[lu]          \&                    
    \&                         \&                 
    \&                         \&
    \& \zxN{} \ar[ru]          \& \dots                        
    \& \zxN{} \ar[lu]      
\end{ZX}

    };
   \node (even) [below=2ex of odd] {%
       $\frac{1}{2}$
       \begin{ZX}[ampersand replacement=\&]
       \zxN{} \ar[rd]          \& \dots                   
    \& \zxN{} \ar[ld]          \&
    \& \zxN{} \ar[rd]          \& \dots                   
    \& \zxN{} \ar[ld]          
    \\
                               \& \zxZ{} \rar \lar
    \& \zxN{} \rar             \& \zxX{t\pi} \rar
    \& \zxN{}                  \& \zxZ{} \rar \lar
    \&
    \\[\zxwRow, \zxwRow]
                               \& \zxZ{} \rar \lar
    \& \zxN{} \rar             \& \zxX{b\pi} \rar  
    \& \zxN{}                  \& \zxZ{} \rar \lar
    \&
    \\
       \zxN{} \ar[ru]          \& \dots                   
    \& \zxN{} \ar[lu]          \&
    \& \zxN{} \ar[ru]          \& \dots                   
    \& \zxN{} \ar[lu]          
\end{ZX}

    };
    \draw[decorate, decoration={brace,mirror}, thick] (odd.north west) -- ({$(even.south west)$} -| {$(odd.north west)$});

    \node(labelodd) [right=1em of odd]{%
        if $(t+l+b+r)$ odd
    };
    \node(labeleven) [anchor=west] at ({$(even.east)$} -| {$(labelodd.west)$}){%
        if $(t+l+b+r)$ even
    };
\end{tikzpicture}

        }
    }} \, .
\end{equation}

Phase-gadgets can be combined to implement so-called \textit{phase polynomials}, 
i.e., parameterized exponentials of diagonal Hamiltonians such as utilized in QAOA circuits~\cite{de2020architecture,cowtan2020generic,hadfield2019quantum,hadfield2021representation}.  

\subsubsection{Lightcones}
For quantum circuits of limited depth or connectivity, 
it is often the case when computing a particular quantity that 
a significant fraction of the gates and qubits can be ignored or discarded due to having no effect, 
in analogy with spacelike-separated events in relativity. 
Naturally, the same principle may be fruitfully applied to diagrammatic analysis.

Given an observable $C=\sum_j C_j$, typically each $C_j$ acts nontrivially on a subset of $\ell<n$ qubits. 
Hence, depending on the structure of the problem and given quantum circuit ansatz $U\ket{\psi_0}$, the $n$-qubit expectation
values $\langle C_j \rangle$ may be equivalently reduced to ones over $L$ qubits,
$\ell \leq L \leq n$, by in each case restricting the quantum circuit in the natural way. 
This phenomena is generally known 
as the \textit{lightcone} or \textit{causal cone} rule~\cite{evenbly2009algorithms,farhi2014quantum, streif2019training,hadfield2021analytical}, and is clearly exhibited with the ZX-calculus. 
For example, if $\ket{\psi_0}$ is a product state and $U$ consists of only $1$-local gates, 
then $L=\ell$ independently of the circuit depth (cf. the example of Section~\ref{sec:example_applications_single_qubit}). 
For QAOA applied to MaxCut, $\ell=2$ and it is easily shown that the lightcone after each $q$th QAOA layer consists of the restriction to the subgraph within distance $q$ of the given edge~\cite{farhi2014quantum,hadfield2021analytical}, i.e., its size $L$ depends on the vertex degrees in the graph neighborhood.
Hence, importantly, for QAOA or similar layered ansatz we may apply the lightcone rule layer-by-layer.
Applying this restriction, the inner operator for a MaxCut QAOA expectation value reads
\begin{align}
    \mathcal{O}^p_{uv} 
    &:= 
    \prod_{\ell=1}^p \,
    \ee^{\ii\gamma_\ell C} \ee^{\ii \tilde \beta_\ell B} \,
    Z_u Z_v \,
    \prod_{k=p}^1 \,
    \ee^{-\ii\tilde \beta_k B}  \ee^{-\ii\gamma_k C} 
    \\ 
    &=
     \vcenter{\hbox{
        \externalize{light_cone_rule}{%
            \begin{tikzpicture}[every node/.style={font=\small}, node distance=4ex and 2.5em]
\def\yshiftdots{0.1ex}
\def\xshiftdots{0.5em}
\def\layerboxwidth{1.0em}
\def\pslayerlabelsep{0.2em}
\def\mixlayerlabelsep{0.07em}

\node[name=u-l, invisible];
\node[on grid, name=v-l, above=of u-l, invisible];
\node[on grid, name=n1b-l, above=of v-l, invisible];
\node[on grid, name=n1t-l, above=of n1b-l, invisible];
\node[on grid, name=np1b-l, above=of n1t-l, invisible];
\node[on grid, name=np1t-l, above=of np1b-l, invisible];
\node[on grid, name=npb-l, above=of np1t-l, invisible];
\node[on grid, name=npt-l, above=of npb-l, invisible];
\node[on grid, name=mpb-l, above=of npt-l, invisible];
\node[on grid, name=mpt-l, above=of mpb-l, invisible];

\node[scale=1.0, anchor=center, xshift=\xshiftdots, thick] at ($(mpt-l.center)!0.5!(mpb-l.center)$) (dots) {\rvdots};
\node[left=0.5em of dots.west, anchor=east] {$\mathcal{M}^{p}_{uv}$};

\node[scale=1.0, anchor=center, xshift=\xshiftdots, thick] at ($(npt-l.center)!0.5!(npb-l.center)$) (dots) {\rvdots};
\node[left=0.5em of dots.west, anchor=east] {$\mathcal{N}^{p}_{uv}$};

\node[scale=1.0, anchor=center, xshift=\xshiftdots, thick] at ($(np1t-l.center)!0.5!(np1b-l.center)$) (dots) {\rvdots};
\node[left=0.5em of dots.west, anchor=east] {$\mathcal{N}^{p-1}_{uv}$};

\node[scale=1.0, anchor=center, xshift=\xshiftdots, thick] at ($(np1b-l.center)!0.5!(n1t-l.center)$) (dots) {\rvdots};
\node[left=1.0em of dots.west, anchor=east] {\rvdots};

\node[scale=1.0, anchor=center, xshift=\xshiftdots, thick] at ($(n1t-l.center)!0.5!(n1b-l.center)$) (dots) {\rvdots};
\node[left=0.5em of dots.west, anchor=east] {$\mathcal{N}^{1}_{uv}$};

\node[left=0.5em of v-l.west, anchor=east] {$v$};
\node[left=0.5em of u-l.west, anchor=east] {$u$};

\node[on grid, name=u-l1g, right=of u-l, invisible];
\node[on grid, name=np1t-l1g, right=of np1t-l, invisible];
\node[on grid, name=np1b-l1g, right=of np1b-l, invisible];
\node[on grid, name=n1t-l1g, right=of n1t-l, invisible];
\node[on grid, name=n1b-l1g, right=of n1b-l, invisible];
\node[on grid, name=npb-l1g, right=of npb-l, invisible];
\node[on grid, name=npt-l1g, right=of npt-l, invisible];
\pslayer{%
        $\zxMinus$\footnotesize$\gamma_1$
    }{%
        (u-l1g) (np1t-l1g)
    }{%
        (u-l1g) (npt-l1g)
    }{%
        inner ysep=0.5ex, inner xsep=\layerboxwidth
    }{%
        inner sep=\pslayerlabelsep
}

\node[on grid, name=u-l1b, right=of u-l1g, invisible];
\node[on grid, name=np1t-l1b, right=of np1t-l1g, invisible];
\mixlayer{%
        $\zxMinus$\footnotesize$\beta_1$
    }{%
        (u-l1b) (np1t-l1b)
    }{%
        inner ysep=0.5ex, inner xsep=\layerboxwidth
    }{%
        inner sep=\mixlayerlabelsep
}

\node[on grid, name=u-ldots, inner xsep=1pt, right=of u-l1b]{$\cdots$};
\node[on grid, name=v-ldots, inner xsep=1pt, above=of u-ldots]{$\cdots$};
\node[on grid, name=n1b-ldots, inner xsep=1pt, above=of v-ldots]{$\cdots$};
\node[on grid, name=n1t-ldots, inner xsep=1pt, above=of n1b-ldots]{$\cdots$};

\node[on grid, name=u-lpg, right=of u-ldots, invisible];
\node[on grid, name=v-lpg, right=of v-ldots, invisible];
\node[on grid, name=n1t-lpg, right=of n1t-ldots, invisible];
\pslayer{%
        $\zxMinus$\footnotesize$\gamma_p$
    }{%
        (u-lpg) (v-lpg)
    }{%
        (u-lpg) (n1t-lpg)
    }{%
        inner ysep=0.5ex, inner xsep=\layerboxwidth
    }{%
        inner sep=\pslayerlabelsep
}

\node[on grid, name=u-lpb, right=of u-lpg, invisible];
\node[on grid, name=v-lpb, right=of v-lpg, invisible];
\mixlayer{%
        $\zxMinus$\footnotesize$\beta_p$
    }{%
        (u-lpb) (v-lpb)
    }{%
        inner ysep=0.5ex, inner xsep=\layerboxwidth
    }{%
        inner sep=\mixlayerlabelsep
}

\node[on grid, name=u-m, right=of u-lpb, zpi];
\node[on grid, name=v-m, right=of v-lpb, zpi];
\node[on grid, name=n1b-m, above=of v-m, invisible];
\node[on grid, name=n1t-m, above=of n1b-m, invisible];
\node[on grid, name=np1b-m, above=of n1t-m, invisible];
\node[scale=1.0, anchor=center, thick] at ($(np1b-m.center)!0.5!(n1t-m.center)$) (dots) {\rvdots};

\node[on grid, name=u-rpb, right=of u-m, invisible];
\node[on grid, name=v-rpb, right=of v-m, invisible];
\mixlayer{%
        \footnotesize $\beta_p$
    }{%
        (u-rpb) (v-rpb)
    }{%
        inner ysep=0.5ex, inner xsep=\layerboxwidth
    }{%
        inner sep=\mixlayerlabelsep
}

\node[on grid, name=u-rpg, right=of u-rpb, invisible];
\node[on grid, name=v-rpg, right=of v-rpb, invisible];
\node[on grid, name=n1b-rpg, above=of v-rpg, invisible];
\node[on grid, name=n1t-rpg, above=of n1b-rpg, invisible];
\pslayer{%
        \footnotesize $\gamma_p$
    }{%
        (u-rpg) (v-rpg)
    }{%
        (u-rpg) (n1t-rpg)
    }{%
        inner ysep=0.5ex, inner xsep=\layerboxwidth
    }{%
        inner sep=\pslayerlabelsep
}

\node[on grid, name=u-rdots, inner xsep=1pt, right=of u-rpg]{$\cdots$};
\node[on grid, name=v-rdots, inner xsep=1pt, above=of u-rdots]{$\cdots$};
\node[on grid, name=n1b-rdots, inner xsep=1pt, above=of v-rdots]{$\cdots$};
\node[on grid, name=n1t-rdots, inner xsep=1pt, above=of n1b-rdots]{$\cdots$};

\node[on grid, name=u-r1b, right=of u-rdots, invisible];
\node[on grid, name=n1t-r1b, right=of n1t-rdots, invisible];
\node[on grid, name=np1b-r1b, above=of n1t-r1b, invisible];
\node[on grid, name=np1t-r1b, above=of np1b-r1b, invisible];
\mixlayer{%
        \footnotesize $\beta_1$
    }{%
        (u-r1b) (np1t-r1b)
    }{%
        inner ysep=0.5ex, inner xsep=\layerboxwidth
    }{%
        inner sep=\mixlayerlabelsep
}

\node[on grid, name=u-r1g, right=of u-r1b, invisible];
\node[on grid, name=v-r1g, above=of u-r1g, invisible];
\node[on grid, name=n1b-r1g, above=of v-r1g, invisible];
\node[on grid, name=n1t-r1g, above=of n1b-r1g, invisible];
\node[on grid, name=np1b-r1g, above=of n1t-r1g, invisible];
\node[on grid, name=np1t-r1g, above=of np1b-r1g, invisible];
\node[on grid, name=npb-r1g, above=of np1t-r1g, invisible];
\node[on grid, name=npt-r1g, above=of npb-r1g, invisible];
\pslayer{%
        \footnotesize $\gamma_1$
    }{%
        (u-r1g) (np1t-r1g)
    }{%
        (u-r1g) (npt-r1g)
    }{%
        inner ysep=0.5ex, inner xsep=\layerboxwidth
    }{%
        inner sep=\pslayerlabelsep
}
\node[on grid, name=u-r, right=of u-r1g, invisible];
\node[on grid, name=v-r, above=of u-r, invisible];
\node[on grid, name=n1b-r, above=of v-r, invisible];
\node[on grid, name=n1t-r, above=of n1b-r, invisible];
\node[on grid, name=np1b-r, above=of n1t-r, invisible];
\node[on grid, name=np1t-r, above=of np1b-r, invisible];
\node[on grid, name=npb-r, above=of np1t-r, invisible];
\node[on grid, name=npt-r, above=of npb-r, invisible];
\node[on grid, name=mpb-r, above=of npt-r, invisible];
\node[on grid, name=mpt-r, above=of mpb-r, invisible];

\node[scale=1.0, anchor=center, xshift=-\xshiftdots, thick] at ($(mpt-r.center)!0.5!(mpb-r.center)$) (dots) {\rvdots};

\node[scale=1.0, anchor=center, xshift=-\xshiftdots, thick] at ($(npt-r.center)!0.5!(npb-r.center)$) (dots) {\rvdots};

\node[scale=1.0, anchor=center, xshift=-\xshiftdots, thick] at ($(np1t-r.center)!0.5!(np1b-r.center)$) (dots) {\rvdots};

\node[scale=1.0, anchor=center, xshift=-\xshiftdots, thick] at ($(np1b-r.center)!0.5!(n1t-r.center)$) (dots) {\rvdots};

\node[scale=1.0, anchor=center, xshift=-\xshiftdots, thick] at ($(n1t-r.center)!0.5!(n1b-r.center)$) (dots) {\rvdots};

\begin{scope}[on background layer]
    \draw (u-l) -- (u-ldots);
    \draw (u-ldots) -- (u-rdots);
    \draw (u-rdots) -- (u-r);

    \draw (v-l) -- (v-ldots);
    \draw (v-ldots) -- (v-rdots);
    \draw (v-rdots) -- (v-r);

    \draw (n1t-l) -- (n1t-ldots);
    \draw (n1t-ldots) -- (n1t-rdots);
    \draw (n1t-rdots) -- (n1t-r);

    \draw (n1b-l) -- (n1b-ldots);
    \draw (n1b-ldots) -- (n1b-rdots);
    \draw (n1b-rdots) -- (n1b-r);

    \draw (np1t-l) -- (np1t-r);
    \draw (np1b-l) -- (np1b-r);

    \draw (npb-l) -- (npb-r);
    \draw (npt-l) -- (npt-r);

    \draw (mpb-l) -- (mpb-r);
    \draw (mpt-l) -- (mpt-r);
\end{scope}

            \end{tikzpicture}
        }
    }}\label{eqn:light_cone_rule} \, ,
\end{align}
where we used placeholder diagrams for the \textit{reduced} phase-separation layer
\begin{align}
     \vcenter{\hbox{
        \externalize{ps_layer_definition}{%

\begin{tikzpicture}[every node/.style={font=\small}, node distance=3ex and 1.5em]
    \def\xshiftdots{0.5em}
    \def\xshiftdotsdouble{1.0em}
    \def\layerboxwidth{1.0em}
    \def\rotate{30}
    \def\pslayerlabelsep{0.2em}
    \def\mixlayerlabelsep{0.07em}
    \tikzset{dots/.style={%
            scale=0.6,
            ultra thick}
    }

\node[name=1-l, invisible];
\node[on grid, name=2-l, above=of 1-l, invisible];
\node[on grid, name=3-l, above=of 2-l, invisible];
\node[on grid, name=4-l, above=of 3-l, invisible];
\node[on grid, name=5-l, above=of 4-l, invisible];
\node[on grid, name=6-l, above=of 5-l, invisible];
\node[on grid, name=7-l, above=of 6-l, invisible];
\node[on grid, name=8-l, above=of 7-l, invisible];
\node[on grid, name=9-l, above=of 8-l, invisible];
\node[on grid, name=10-l, above=of 9-l, invisible];
\node[on grid, name=11-l, above=of 10-l, invisible];
\node[on grid, name=12-l, above=of 11-l, invisible];

\node[on grid, name=1-m, right=of 1-l, invisible];
\node[on grid, name=4-m, right=of 4-l, invisible];
\node[on grid, name=12-m, right=of 12-l, invisible];

\node[on grid, name=1-r, right=of 1-m, invisible];
\node[on grid, name=2-r, above=of 1-r, invisible];
\node[on grid, name=3-r, above=of 2-r, invisible];
\node[on grid, name=4-r, above=of 3-r, invisible];
\node[on grid, name=5-r, above=of 4-r, invisible];
\node[on grid, name=6-r, above=of 5-r, invisible];
\node[on grid, name=7-r, above=of 6-r, invisible];
\node[on grid, name=8-r, above=of 7-r, invisible];
\node[on grid, name=9-r, above=of 8-r, invisible];
\node[on grid, name=10-r, above=of 9-r, invisible];
\node[on grid, name=11-r, above=of 10-r, invisible];
\node[on grid, name=12-r, above=of 11-r, invisible];

\draw[thick, decorate, decoration={brace,raise=0.5em}] (6-l) -- 
    node[midway,left=0.5em] (bracket) {$N_L$}
(12-l);
\draw[thick, decorate, decoration={brace,raise=0.5em}] (1-l) -- 
    node[midway,left=0.5em] (bracket) {$L$}
(4-l);
\node[dots, xshift=\xshiftdots] at ($(12-l.center)!0.5!(11-l.center)$) (dots) {\rvdots};
\node[dots, xshift=\xshiftdots] at ($(11-l.center)!0.5!(10-l.center)$) (dots) {\rvdots};
\node[dots, xshift=\xshiftdots] at ($(10-l.center)!0.5!(9-l.center)$) (dots) {\rvdots};
\node[dots, xshift=\xshiftdots] at ($(9-l.center)!0.5!(8-l.center)$) (dots) {\rvdots};
\node[dots, xshift=\xshiftdots] at ($(7-l.center)!0.5!(6-l.center)$) (dots) {\rvdots};
\node[dots, xshift=\xshiftdots] at ($(1-l.center)!0.5!(3-l.center)$) (dots) {\rvdots};

\node[dots, xshift=-\xshiftdots] at ($(12-r.center)!0.5!(11-r.center)$) (dots) {\rvdots};
\node[dots, xshift=-\xshiftdots] at ($(11-r.center)!0.5!(10-r.center)$) (dots) {\rvdots};
\node[dots, xshift=-\xshiftdots] at ($(10-r.center)!0.5!(9-r.center)$) (dots) {\rvdots};
\node[dots, xshift=-\xshiftdots] at ($(9-r.center)!0.5!(8-r.center)$) (dots) {\rvdots};
\node[dots, xshift=-\xshiftdots] at ($(7-r.center)!0.5!(6-r.center)$) (dots) {\rvdots};
\node[dots, xshift=-\xshiftdots] at ($(1-r.center)!0.5!(3-r.center)$) (dots) {\rvdots};

\pslayer{%
        \footnotesize $\gamma$
    }{%
        (1-m) (4-m)
    }{%
        (1-m) (12-m)
    }{%
        inner ysep=0.5ex, inner xsep=\layerboxwidth
    }{%
        inner sep=\pslayerlabelsep
}

\begin{scope}[on background layer]
    \draw (1-l) -- (1-r);
    \draw (3-l) -- (3-r);
    \draw (4-l) -- (4-r);
    \draw (6-l) -- (6-r);
    \draw (7-l) -- (7-r);
    \draw (8-l) -- (8-r);
    \draw (9-l) -- (9-r);
    \draw (10-l) -- (10-r);
    \draw (11-l) -- (11-r);
    \draw (12-l) -- (12-r);
\end{scope}

    \node[name=anchor, right=5em of 1-r, invisible];
    \node[anchor=center] at ($(12-r.center)!0.5!(anchor.center)$) (eq) {$:=\sqrt{2}^{n_L}$};


\node at (anchor.center) [name=1-l, invisible];
\node[on grid, name=2-l, above=of 1-l, invisible];
\node[on grid, name=3-l, above=of 2-l, invisible];
\node[on grid, name=4-l, above=of 3-l, invisible];
\node[on grid, name=5-l, above=of 4-l, invisible];
\node[on grid, name=6-l, above=of 5-l, invisible];
\node[on grid, name=7-l, above=of 6-l, invisible];
\node[on grid, name=8-l, above=of 7-l, invisible];
\node[on grid, name=9-l, above=of 8-l, invisible];
\node[on grid, name=10-l, above=of 9-l, invisible];
\node[on grid, name=11-l, above=of 10-l, invisible];
\node[on grid, name=12-l, above=of 11-l, invisible];
\node[dots, xshift=\xshiftdots] at ($(12-l.center)!0.5!(11-l.center)$) (dots) {\rvdots};
\node[dots, xshift=\xshiftdots] at ($(11-l.center)!0.5!(10-l.center)$) (dots) {\rvdots};
\node[dots, xshift=\xshiftdots] at ($(10-l.center)!0.5!(9-l.center)$) (dots) {\rvdots};
\node[dots, xshift=\xshiftdots] at ($(9-l.center)!0.5!(8-l.center)$) (dots) {\rvdots};
\node[dots, xshift=\xshiftdots] at ($(7-l.center)!0.5!(6-l.center)$) (dots) {\rvdots};
\node[dots, xshift=\xshiftdots] at ($(1-l.center)!0.5!(3-l.center)$) (dots) {\rvdots};

\node[on grid, name=4-Ll, right=of 4-l, zspider] {};
\node[on grid, name=3-Ll, right=of 3-l, zspider] {};
\draw (4-Ll) -- (3-Ll) node[midway, xspider, anchor=center] (x) {};
\draw (x) -- +(180:2ex) node[gamma=$\gamma$, zxstyletight];
\node[on grid, name=1-Ll, right=of 1-l, invisible];

\node[dots, xshift=\xshiftdots] at ($(3-l.center)!0.5!(1-l.center)$) (dots) {\rvdots};
\node[dots, on grid, name=4-Lm, right=of 4-Ll] {$\cdots$};
\node[dots, on grid, name=3-Lm, right=of 3-Ll] {$\cdots$};

\node[on grid, name=4-Lr, right=of 4-Lm, zspider] {};
\node[on grid, name=3-Lr, below=of 4-Lr, invisible];
\node[on grid, name=2-Lr, below=of 3-Lr, invisible];
\node[on grid, name=1-Lr, below=of 2-Lr, zspider] {};
\draw (4-Lr) -- (1-Lr) node[midway, xspider, anchor=center] (x) {};
\draw (x) -- +(180:1em) node[gamma=$\gamma$, zxstyletight];

\draw[thick, decorate, decoration={brace,mirror,raise=1.0ex}] (1-Ll) -- 
    node[midway,below=1.5ex] (label) {%
    \tikz[baseline=0.5ex]{%
        \node[zspider] (bottom) {}; 
        \node[zspider, above=2ex of bottom] (top) {}; 
        \draw (bottom) -- (top) node[midway, xspider, anchor=center] (x) {};
        \draw (x) -- +(180:1em) node[gamma=$\gamma$, zxstyletight];
    }
    $\; \in L \times L \cap E$}
(1-Lr);


\node[dots, on grid, name=4-Ldots, right=of 4-Lr] {$\cdots$};
\node[dots, on grid, name=3-Ldots, right=of 3-Lr] {$\cdots$};
\node[dots, on grid, name=1-Ldots, right=of 1-Lr] {$\cdots$};

\node[on grid, name=4-NLl, right=of 4-Ldots, zspider] {};
\node[on grid, name=3-NLl, right=of 3-Ldots, invisible];
\node[on grid, name=1-NLl, right=of 1-Ldots, invisible];
\node[on grid, name=5-NLl, above=of 4-NLl, invisible];
\node[on grid, name=6-NLl, above=of 5-NLl, zspider] {};
\draw (4-NLl) -- (6-NLl) node[midway, xspider, anchor=center] (x) {};
\draw (x) -- +(180:1em) node[gamma=$\gamma$, zxstyletight];

\node[dots, name=4-NLm, on grid, right=of 4-NLl] {$\cdots$};

\node[on grid, name=4-NLr, right=of 4-NLm, zspider] {};
\node[on grid, name=5-NLr, above=of 4-NLr, invisible];
\node[on grid, name=6-NLr, above=of 5-NLr, invisible];
\node[on grid, name=7-NLr, above=of 6-NLr, zspider] {};
\draw (4-NLr) -- (7-NLr) node[pos=0.33, xspider, anchor=center] (x) {};
\draw (x) -- +(180:1em) node[gamma=$\gamma$, zxstyletight];

\node[dots, anchor=center, rotate=\rotate] at ($(6-NLl)!0.5!(7-NLr)$) {$\cdots$};


\node[on grid, name=4-NMLl, right=of 4-NLr, invisible];
\node[on grid, name=4-NMLl-shift, left=\xshiftdots of 4-NMLl, zspider] {};
\node[on grid, name=3-NMLl, below=of 4-NMLl, invisible];
\node[on grid, name=3-NMLl-shift, right=\xshiftdots of 3-NMLl, zspider] {};
\node[on grid, name=8-NMLl, above right=of 7-NLr, zspider] {};
\draw (4-NMLl-shift) -- (8-NMLl) node[pos=0.62, xspider, anchor=center] (x) {};
\draw (x) -- +(180:0.7em) node[gamma=$\gamma$, zxstyletight];
\draw (3-NMLl-shift) -- (8-NMLl) node[pos=0.5, xspider, anchor=center] (x) {};
\draw (x) -- +(0:0.7em) node[gamma=$\gamma$, zxstyletight];
\node[dots, name=4-NMLm, on grid, right=of 4-NMLl] {$\cdots$};
\node[dots, name=3-NMLm, on grid, right=of 3-NMLl] {$\cdots$};
\node[on grid, name=4-NMLr, right=of 4-NMLm, invisible];
\node[on grid, name=4-NMLr-shift, left=\xshiftdots of 4-NMLr, zspider] {};
\node[on grid, name=3-NMLr, below=of 4-NMLr, invisible];
\node[on grid, name=3-NMLr-shift, right=\xshiftdots of 3-NMLr, zspider] {};
\node[on grid, name=8-NMLm, above right=of 8-NMLl, invisible];
\node[on grid, name=9-NMLr, right=of 8-NMLm, zspider] {};
\draw (4-NMLr-shift) -- (9-NMLr) node[pos=0.5, xspider, anchor=center] (x) {};
\draw (x) -- +(180:0.7em) node[gamma=$\gamma$, zxstyletight];
\draw (3-NMLr-shift) -- (9-NMLr) node[pos=0.41, xspider, anchor=center] (x) {};
\draw (x) -- +(0:0.7em) node[gamma=$\gamma$, zxstyletight];
\node[on grid, name=9-MLdots, right=of 9-NMLr, invisible];
\node[dots, anchor=center, rotate=\rotate] at ($(8-NMLl)!0.5!(9-NMLr)$) {$\cdots$};
\node[dots, on grid, name=4-MLdots, right=of 4-NMLr] {$\cdots$};
\node[dots, on grid, name=3-MLdots, right=of 3-NMLr] {$\cdots$};
\node[dots, on grid, name=2-MLdots, below=of 3-MLdots] {\rvdots};


\node[on grid, name=4-NMRl, right=of 4-MLdots, invisible];
\node[on grid, name=4-NMRl-shift, left=\xshiftdots of 4-NMRl, zspider] {};
\node[on grid, name=3-NMRl, below=of 4-NMRl, invisible];
\node[on grid, name=2-NMRl, below=of 3-NMRl, invisible];
\node[on grid, name=1-NMRl, below=of 2-NMRl, invisible];
\node[on grid, name=1-NMRl-shift, right=\xshiftdots of 1-NMRl, zspider] {};
\node[on grid, name=10-NMRl, above right=of 9-MLdots, zspider] {};
\draw (4-NMRl-shift) -- (10-NMRl) node[pos=0.6, xspider, anchor=center] (x) {};
\draw (x) -- +(180:0.7em) node[gamma=$\gamma$, zxstyletight];
\draw (1-NMRl-shift) -- (10-NMRl) node[pos=0.5, xspider, anchor=center] (x) {};
\draw (x) -- +(0:0.7em) node[gamma=$\gamma$, zxstyletight];
\node[dots, anchor=center, rotate=\rotate] at ($(9-NMLr)!0.5!(10-NMRl)$) {$\cdots$};
\node[dots, name=4-NMRm, on grid, right=of 4-NMRl] {$\cdots$};
\node[dots, name=3-NMRm, on grid, right=of 3-NMRl] {$\cdots$};
\node[dots, name=1-NMRm, on grid, right=of 1-NMRl] {$\cdots$};
\node[on grid, name=4-NMRr, right=of 4-NMRm, invisible];
\node[on grid, name=4-NMRr-shift, left=\xshiftdots of 4-NMRr, zspider] {};
\node[on grid, name=3-NMRr, below=of 4-NMRr, invisible];
\node[on grid, name=2-NMRr, below=of 3-NMRr, invisible];
\node[on grid, name=1-NMRr, below=of 2-NMRr, invisible];
\node[on grid, name=1-NMRr-shift, right=\xshiftdots of 1-NMRr, zspider] {};
\node[on grid, name=5-NMRr, above=of 4-NMRr, invisible];
\node[on grid, name=6-NMRr, above=of 5-NMRr, invisible];
\node[on grid, name=7-NMRr, above=of 6-NMRr, invisible];
\node[on grid, name=8-NMRr, above=of 7-NMRr, invisible];
\node[on grid, name=9-NMRr, above=of 8-NMRr, invisible];
\node[on grid, name=10-NMRr, above=of 9-NMRr, invisible];
\node[on grid, name=11-NMRr, above=of 10-NMRr, zspider] {};
\draw (4-NMRr-shift) -- (11-NMRr) node[pos=0.5, xspider, anchor=center] (x) {};
\draw (x) -- +(180:0.7em) node[gamma=$\gamma$, zxstyletight];
\draw (1-NMRr-shift) -- (11-NMRr) node[pos=0.45, xspider, anchor=center] (x) {};
\draw (x) -- +(0:0.7em) node[gamma=$\gamma$, zxstyletight];
\node[dots, anchor=center, rotate=\rotate] at ($(10-NMRl)!0.5!(11-NMRr)$) {$\cdots$};


\node[dots, on grid, name=1-MRdots, right=of 1-NMRr] {$\cdots$};
\node[on grid, name=2-MRdots, above=of 1-MRdots, invisible];
\node[dots, on grid, name=3-MRdots, above=of 2-MRdots] {$\cdots$};
\node[dots, on grid, name=4-MRdots, above=of 3-MRdots] {$\cdots$};
\node[on grid, name=5-MRdots, above=of 4-MRdots, invisible];
\node[dots, on grid, name= 6-MRdots, above=of  5-MRdots] {$\cdots$};
\node[dots, on grid, name= 7-MRdots, above=of  6-MRdots] {$\cdots$};
\node[dots, on grid, name= 8-MRdots, above=of  7-MRdots] {$\cdots$};
\node[dots, on grid, name= 9-MRdots, above=of  8-MRdots] {$\cdots$};
\node[dots, on grid, name=10-MRdots, above=of  9-MRdots] {$\cdots$};
\node[dots, on grid, name=11-MRdots, above=of 10-MRdots] {$\cdots$};



\node[on grid, name=1-NR, right=of 1-MRdots, zspider] {};
\node[on grid, name=2-NR, above=of 1-NR, invisible];
\node[on grid, name=3-NR, above=of 2-NR, invisible];
\node[on grid, name=4-NR, above=of 3-NR, invisible];
\node[on grid, name=5-NR, above=of 4-NR, invisible];
\node[on grid, name=6-NR, above=of 5-NR, invisible];
\node[on grid, name=7-NR, above=of 6-NR, invisible];
\node[on grid, name=8-NR, above=of 7-NR, invisible];
\node[on grid, name=9-NR, above=of 8-NR, invisible];
\node[on grid, name=10-NR, above=of 9-NR, invisible];
\node[on grid, name=11-NR, above=of 10-NR, invisible];
\node[on grid, name=12-NR, above=of 11-NR,  zspider] {};

\node[dots, anchor=center, rotate=\rotate] at ($(11-NMRr)!0.5!(12-NR)$) {$\cdots$};
\draw (1-NR) -- (12-NR) node[pos=0.37, xspider, anchor=center] (x) {};
\draw (x) -- +(180:0.7em) node[gamma=$\gamma$, zxstyletight];


\node[on grid, name=1-r, right=of 1-NR, invisible];
\node[on grid, name=2-r, above=of 1-r, invisible];
\node[on grid, name=3-r, above=of 2-r, invisible];
\node[on grid, name=4-r, above=of 3-r, invisible];
\node[on grid, name=5-r, above=of 4-r, invisible];
\node[on grid, name=6-r, above=of 5-r, invisible];
\node[on grid, name=7-r, above=of 6-r, invisible];
\node[on grid, name=8-r, above=of 7-r, invisible];
\node[on grid, name=9-r, above=of 8-r, invisible];
\node[on grid, name=10-r, above=of 9-r, invisible];
\node[on grid, name=11-r, above=of 10-r, invisible];
\node[on grid, name=12-r, above=of 11-r, invisible];
\node[dots, xshift=-\xshiftdots] at ($(12-r.center)!0.5!(11-r.center)$) (dots) {\rvdots};
\node[dots, xshift=-\xshiftdots] at ($(11-r.center)!0.5!(10-r.center)$) (dots) {\rvdots};
\node[dots, xshift=-\xshiftdots] at ($(10-r.center)!0.5!(9-r.center)$) (dots) {\rvdots};
\node[dots, xshift=-\xshiftdots] at ($(9-r.center)!0.5!(8-r.center)$) (dots) {\rvdots};
\node[dots, xshift=-\xshiftdots] at ($(7-r.center)!0.5!(6-r.center)$) (dots) {\rvdots};
\node[dots, xshift=-\xshiftdots] at ($(1-r.center)!0.5!(3-r.center)$) (dots) {\rvdots};
\begin{scope}[on background layer]
    \draw (4-l) -- ([xshift=\xshiftdotsdouble]4-Ll.center);
    \draw ([xshift=-\xshiftdotsdouble]4-Lr.center) -- ([xshift=\xshiftdotsdouble]4-Lr.center);
    \draw ([xshift=-\xshiftdotsdouble]4-NLl.center) -- ([xshift=\xshiftdotsdouble]4-NLl.center);
    \draw ([xshift=-\xshiftdotsdouble]4-NLr.center) -- ([xshift=\xshiftdotsdouble]4-NMLl.center);
    \draw ([xshift=-\xshiftdotsdouble]4-NMLr.center) -- ([xshift=\xshiftdotsdouble]4-NMLr.center);
    \draw ([xshift=-\xshiftdotsdouble]4-NMRl.center) -- ([xshift=\xshiftdotsdouble]4-NMRl.center);
    \draw ([xshift=-\xshiftdotsdouble]4-NMRr.center) -- ([xshift=\xshiftdotsdouble]4-NMRr.center);
    \draw ([xshift=-\xshiftdotsdouble]4-NR.center) -- (4-r);

    \draw (3-l) -- ([xshift=\xshiftdotsdouble]3-Ll.center);
    \draw ([xshift=-\xshiftdotsdouble]3-Lr.center) -- ([xshift=\xshiftdotsdouble]3-Lr.center);
    \draw ([xshift=-\xshiftdotsdouble]3-NLl.center) -- ([xshift=\xshiftdotsdouble]3-NMLl.center);
    \draw ([xshift=-\xshiftdotsdouble]3-NMLr.center) -- ([xshift=\xshiftdotsdouble]3-NMLr.center);
    \draw ([xshift=-\xshiftdotsdouble]3-NMRl.center) -- ([xshift=\xshiftdotsdouble]3-NMRl.center);
    \draw ([xshift=-\xshiftdotsdouble]3-NMRr.center) -- ([xshift=\xshiftdotsdouble]3-NMRr.center);
    \draw ([xshift=-\xshiftdotsdouble]3-NR.center) -- (3-r);

    \draw (1-l) -- (1-Ldots);
    \draw ([xshift=-\xshiftdotsdouble]1-NLl.center) -- ([xshift=\xshiftdotsdouble]1-NMRl.center);
    \draw ([xshift=-\xshiftdotsdouble]1-NMRr.center) -- ([xshift=\xshiftdotsdouble]1-NMRr.center);
    \draw ([xshift=-\xshiftdotsdouble]1-NR.center) -- (1-r);

    \draw (6-l) -- ([xshift=\xshiftdotsdouble]6-NMRr.center);
    \draw ([xshift=-\xshiftdotsdouble]6-NR.center) -- (6-r);
    \draw (7-l) -- ([xshift=\xshiftdotsdouble]7-NMRr.center);
    \draw ([xshift=-\xshiftdotsdouble]7-NR.center) -- (7-r);
    \draw (8-l) -- ([xshift=\xshiftdotsdouble]8-NMRr.center);
    \draw ([xshift=-\xshiftdotsdouble]8-NR.center) -- (8-r);
    \draw (9-l) -- ([xshift=\xshiftdotsdouble]9-NMRr.center);
    \draw ([xshift=-\xshiftdotsdouble]9-NR.center) -- (9-r);
    \draw (10-l) -- ([xshift=\xshiftdotsdouble]10-NMRr.center);
    \draw ([xshift=-\xshiftdotsdouble]10-NR.center) -- (10-r);
    \draw (11-l) -- ([xshift=\xshiftdotsdouble]11-NMRr.center);
    \draw ([xshift=-\xshiftdotsdouble]11-NR.center) -- (11-r);
    \draw (12-l) -- (12-r);
\end{scope}

\end{tikzpicture}

        }
    }}\label{eqn:ps_layer_definition} \, ,
\end{align}
and the \textit{reduced} mixing layer
\begin{align}
     \vcenter{\hbox{
        \externalize{mixing_layer_definition}{%

\begin{tikzpicture}[every node/.style={font=\small}, node distance=3ex and 1.5em]
    \def\xshiftdots{0.5em}
    \def\xshiftdotsdouble{1.0em}
    \def\layerboxwidth{1.0em}
    \def\rotate{30}
    \def\pslayerlabelsep{0.2em}
    \def\mixlayerlabelsep{0.07em}
    \tikzset{dots/.style={%
            scale=0.6,
            ultra thick}
    }

\node[name=1-l, invisible];
\node[on grid, name=2-l, above=of 1-l, invisible];
\node[on grid, name=3-l, above=of 2-l, invisible];

\node[on grid, name=1-m, right=of 1-l, invisible];
\node[on grid, name=2-m, right=of 2-l, invisible];
\node[on grid, name=3-m, right=of 3-l, invisible];

\node[on grid, name=1-r, right=of 1-m, invisible];
\node[on grid, name=2-r, right=of 2-m, invisible];
\node[on grid, name=3-r, right=of 3-m, invisible];

\node[dots, xshift=\xshiftdots] at (2-l.center) (dots) {\rvdots};
\node[dots, xshift=-\xshiftdots] at (2-r.center) (dots) {\rvdots};

\mixlayer{%
        \footnotesize $\beta$
    }{%
        (1-m) (3-m)
    }{%
        inner ysep=0.5ex, inner xsep=\layerboxwidth
    }{%
        inner sep=\mixlayerlabelsep
}

\begin{scope}[on background layer]
    \draw (1-l) -- (1-r);
    \draw (3-l) -- (3-r);
\end{scope}

    \node[name=anchor, right=2em of 1-r, invisible];
    \node[anchor=center] at ($(3-r.center)!0.5!(anchor.center)$) {$:=$};


\node at (anchor.center) [on grid, name=1-l, invisible];
\node[on grid, name=2-l, above=of 1-l, invisible];
\node[on grid, name=3-l, above=of 2-l, invisible];

\node[on grid, name=1-m, right=of 1-l, xspider, zxstyle] {$\zxMinus 2 \tilde \beta$};
\node[on grid, name=2-m, right=of 2-l, dots] {\rvdots};
\node[on grid, name=3-m, right=of 3-l, xspider, zxstyle] {$\zxMinus 2 \tilde \beta$};

\node[on grid, name=1-r, right=of 1-m, invisible];
\node[on grid, name=2-r, right=of 2-m, invisible];
\node[on grid, name=3-r, right=of 3-m, invisible];

\begin{scope}[on background layer]
    \draw (1-l) -- (1-r);
    \draw (3-l) -- (3-r);
\end{scope}

    \node[name=anchor, right=2em of 1-r, invisible];
    \node[anchor=center] at ($(3-r.center)!0.5!(anchor.center)$) {$=$};


\node at (anchor.center) [on grid, name=1-l, invisible];
\node[on grid, name=2-l, above=of 1-l, invisible];
\node[on grid, name=3-l, above=of 2-l, invisible];

\node[on grid, name=1-m, right=of 1-l, xspider, zxstyle] {$\beta$};
\node[on grid, name=2-m, right=of 2-l, dots] {\rvdots};
\node[on grid, name=3-m, right=of 3-l, xspider, zxstyle] {$\beta$};

\node[on grid, name=1-r, right=of 1-m, invisible];
\node[on grid, name=2-r, right=of 2-m, invisible];
\node[on grid, name=3-r, right=of 3-m, invisible];

\begin{scope}[on background layer]
    \draw (1-l) -- (1-r);
    \draw (3-l) -- (3-r);
\end{scope}

\end{tikzpicture}

        }
    }}\label{eqn:mixing_layer_definition}
\end{align}
where $\beta:=-2\tilde \beta$ for convenience.
For the reduced phase-separation layer we have used the MaxCut cost function Hamiltonian
$
C = \frac{1}{2} \sum_{uv \in E} \left(1 - Z_u Z_v\right)  
$
and
\begin{align}
    \ee^{\ii \gamma C} 
    &\,=\,
    \prod_{(u,v) \in E} \ee^{\frac{\ii \gamma}{2}} \ee^{\frac{-\ii \gamma}{2} Z_u Z_v}     
    \,\stackrel{\eqref{eqn:phase_gadget_as_lincomb}}{=}\,
    \sqrt{2}^{|E|}
    \prod_{(u,v) \in E}
    \begin{ZX}[ampersand replacement=\&]
           \zxN{} \rar \&[\zxwCol] \zxZ{} \rar 
        \& \zxN{} \rar \& \zxN{}
        \& u
        \\
                       \& \zxX{} \uar \dar \rar 
        \& \zxZ{\gamma}            \&
        \&
        \\
           \zxN{} \rar \& \zxZ{} \rar 
        \& \zxN{} \rar \& \zxN{}
        \& v
    \end{ZX}
    \, .
\end{align}
This leads to the factor $\sqrt{2}^{n_L}$ in~\eqref{eqn:ps_layer_definition}, where $n_L$ is the number of phase-gadgets in the right diagram of~\eqref{eqn:ps_layer_definition}.
Also, we implicitly used the neighborhood of a set of nodes $L$,
    $N_{L} := \bigcup_{\ell \in L} \text{nbhd}(\ell)$, 
the exclusive $p$-th neighborhood of $\{u, v\}$, recursively defined by
\begin{align}
    \mathcal{N}_{uv}^p &:= \bigcup_{i, j \in \mathcal{N}_{uv}^{p-1} \times \mathcal{N}_{uv}^{p-1} \cap E}
    N_{\{i, j\}} \setminus \cup_{k=0}^{p-1} \mathcal{N}_{uv}^k  \, ,
\end{align}
where $\mathcal{N}_{uv}^0 := \{u, v\}$, 
as well as the complement 
    $\mathcal{M}_{uv}^p = \mathcal{N}_{uv}^p \setminus E$. 

While here we have considered QAOA circuits as a demonstrative example, the same principle may be applied to or formalized for more general ans\"atze and observables.

\section{
Application to Combinatorial Optimization}%
\label{sec:example_applications}

Expectation values of quantum circuit observables -- i.e., constants -- may be represented with ZX-diagrams,
as has been previously observed in \cite[Eq. 7]{zhao2021analyzing}. 
In doing so, in some cases the structure of the original problem 
may be directly reflected in the structure of the corresponding ZX-diagrams. 
This is demonstrated by two examples in this section, in which 
apply our ZX-calculus extension to 
calculate cost expectation values for a particular ansatz for combinatorial optimization. 
The purpose of this section is twofold.
First, we want to demonstrate that calculations with parameterized quantum circuits, like the finding an analytical expression for expectation values, can sometimes become 
more intuitive  and simplified by using ZX-calculus in conjunction with our extension to linear combinations.
Second, we show that our extension is indeed necessary to achieve the aforementioned task diagrammatically by, for instance, providing means to \say{commute} X- and Z-spiders (cf.~\eqref{eqn:single_qubit_rotation_exp_value}), while explicitly keeping track of all resulting terms.

We show how the cost function expectation value $\langle C\rangle$ may be computed and analyzed using our extended ZX-calculus. Recall that given a decomposition of the cost Hamiltonian $C=\sum C_\ell$ it suffices to compute the $\langle C_\ell \rangle$ values independently, which typically correspond to similar diagrams. In particular (sub)graph symmetry can be 
exploited to reduced the number of unique diagrams required \cite{farhi2014quantum,shaydulin2021classical,shaydulin2021exploiting}. 
Generally the quantity $\langle C\rangle$ is important in parameter setting, as well as bounding algorithm performance 
such as the approximation ratio achieved~\cite{hadfield2018quantum}.

\subsection{Independent Single-Qubit Rotations Ansatz}%
\label{sec:example_applications_single_qubit}
We begin with 
a simple but important example.
Consider an arbitrary cost function and corresponding cost (diagonal) Hamiltonian $C$ on $n$ qubits we seek to extremize, together with the simple depth-1 ansatz consisting of a free single-qubit Pauli-$Y$ rotation on each qubit, applied to the initial state $\ket{00\dots 0}=\ket{0}^{\otimes n}$, 
\begin{equation}
    \vcenter{\hbox{
        \begin{quantikz}[thin lines, row sep={1.0ex}]
            \lstick{\ket{0}}        & \qw  & \gate{R_Y(\alpha_1)} & \qw  \\ 
            \lstick{\ket{0}}        & \qw  & \gate{R_Y(\alpha_2)} & \qw  \\ 
            \vdots                  &      & \vdots             &      \\
            \lstick{\ket{0}}        & \qw  & \gate{R_Y(\alpha_n)} & \qw  
        \end{quantikz}
    }} 
    \,\,=\,\,
    \frac{1}{\sqrt{2^n}}
    \vcenter{\hbox{
        \begin{ZX}[row sep=1.9ex, column sep=1em]
            \zxX{} \rar & \zxFracZ-{\pi}{2} \rar & \zxX{\alpha} \rar & \zxFracZ{\pi}{2} \rar & \zxN{} \\
            \zxX{} \rar & \zxFracZ-{\pi}{2} \rar & \zxX{\alpha} \rar & \zxFracZ{\pi}{2} \rar & \zxN{} \\
            \vdots      &                       &  \vdots           &                        & \\
            \zxX{} \rar & \zxFracZ-{\pi}{2} \rar & \zxX{\alpha} \rar & \zxFracZ{\pi}{2} \rar & \zxN{} 
        \end{ZX}
    }}\label{eqn:single_qubit_rotation_ansatz} \,.
\end{equation}
For example, consider an arbitrary instance of MaxCut, 
a prototypical NP-hard optimization problem,
though the same argument we show here applies similarly to many other problems. 
For a graph with edge set $E$ the cost Hamiltonian is $C=\tfrac{|E|}2-\tfrac12\sum_{(uv)\in E}Z_uZ_v$.
As demonstrated in Equation~\eqref{eqn:single_qubit_rotation_exp_value} the derivation of each
$\langle Z_u Z_v \rangle$
becomes very simple with ZX-calculus.
\def\xshiftdots{0.5em}
\tikzset{dots/.style={%
        scale=0.6,
        ultra thick}
}
\begin{equation}
        \begin{tikzpicture}[node distance=3.2ex and 2.0em]


\node[xspider, name=1-l] {};
\node[on grid, right=of 1-l,  name=1-lml, zspider, zxstyle] {$\zxMinus \frac{\pi}{2}$};
\node[on grid, right=of 1-lml, name=1-lmr, xspider, zxstyle] {$\alpha_u$};
\node[on grid, right=of 1-lmr, name=1-m, zspider, zxstyle] {$\pi$};
\node[on grid, right=of 1-m, name=1-rml, xspider, zxstyle] {$\zxMinus \alpha_u$};
\node[on grid, right=of 1-rml,  name=1-rmr, zspider, zxstyle] {$\frac{\pi}{2}$};
\node[on grid, right=of 1-rmr,  name=1-r, xspider] {};

\node[on grid, above=of 1-l, name=2-l, xspider] {};
\node[on grid, right=of 2-l,  name=2-lml, zspider, zxstyle] {$\zxMinus \frac{\pi}{2}$};
\node[on grid, right=of 2-lml, name=2-lmr, xspider, zxstyle] {$\alpha_v$};
\node[on grid, right=of 2-lmr, name=2-m, zspider, zxstyle] {$\pi$};
\node[on grid, right=of 2-m, name=2-rml, xspider, zxstyle] {$\zxMinus \alpha_v$};
\node[on grid, right=of 2-rml,  name=2-rmr, zspider, zxstyle] {$\frac{\pi}{2}$};
\node[on grid, right=of 2-rmr,  name=2-r, xspider] {};

\node[on grid, above=of 2-l, name=3-l, xspider] {};
\node[on grid, right=of 3-l, name=3-r, xspider] {};

\node[on grid, above=of 3-l, name=4-l, invisible];

\node[on grid, above=of 4-l, name=5-l, xspider] {};
\node[on grid, right=of 5-l, name=5-r, xspider] {};
\node[on grid, above=of 2-r, name=anchorr, xspider] {};

\node at ($(3-l)!0.5!(5-r)$) {\rvdots};
\begin{scope}[on background layer]
    \draw (1-l) -- (1-r);
    \draw (2-l) -- (2-r);
    \draw (3-l) -- (3-r);
    \draw (5-l) -- (5-r);
\end{scope}

\node[left=2ex of 3-l]  {%
     $\langle Z_u Z_v \rangle \;= \frac{1}{2^n}$
};
\node[right=1.5em of anchorr] {%
    $
    =  
    \;
    c_{\alpha_u} 
    c_{\alpha_v} 
    $
};
\draw[thick, decorate, decoration={brace,mirror,raise=2.5ex}] (1-l) -- 
    node[midway] (label) {}
(1-r);
\node[baseline=default,below=3ex of label.north, anchor=north west, xshift=-2em] {
    \begin{tikzpicture}
        \node (step01) {%
            \begin{tikzpicture}[node distance=1ex and 2.5em]%
\node[xspider, name=l] {};
\node[on grid, right=of l,  name=lm, zspider, zxstyle] {$\frac{\pi}{2}$};
\node[on grid, right=of lm, name=m, xspider, zxstyle] {$\zxMinus 2\alpha_u$};
\node[on grid, right=of m,  name=rm, zspider, zxstyle] {$\frac{\pi}{2}$};
\node[on grid, right=of rm,  name=r, xspider] {};
\node[on grid, left=of l] {$\ee^{\ii \alpha_u}$};
\begin{scope}[on background layer]
    \draw (l) -- (r);
\end{scope}
            \end{tikzpicture}
        };
        \node (eq1) [left=1ex of step01] {$=$};
        \node [above=-0.5ex of eq1.north, anchor=south] {\spiderrule\pirule};
        \node (step02) [below=0.1ex of step01.south west, anchor=north west] {%
            \begin{tikzpicture}[node distance=3ex and 2.0em]

\node[xspider, name=l] {};
\node[on grid, right=of l,  name=lm, zspider, zxstyle] {$\frac{\pi}{2}$};
\node[on grid, right=of lm, name=sumleft, leftsummator];
\node[on grid, right=10em of sumleft, name=sumright, rightsummator];
\node[on grid, right=of sumright.west,  name=rm, zspider, zxstyle] {$\frac{\pi}{2}$};
\node[on grid, right=of rm,  name=r, xspider] {};

\zxsumtwo{%
    \begin{ZX}
        \\[0.3ex]
            \zxN{} \rar &[1.0em] \zxN{} \rar & \zxN{}  
        \\[0.3ex]
    \end{ZX}
    }{%
    \begin{ZX}
        \zxN{} \rar & \zxX{\pi} \rar & \zxN{}
    \end{ZX}
}[0.0em][4.0ex][1.0][c_{\alpha_u}][\ii s_{\alpha_u}]

\begin{scope}[on background layer]
    \draw (l) -- (sumleft.west);
    \draw (sumright.west) -- (r);
\end{scope}

            \end{tikzpicture}
        };
        \node (eq2) [left=1ex of step02] {$=$};
        \node [above=0.01ex of eq2] {\eqref{eqn:xspider_as_lincomb}};
        \node (step03) [below=0.1ex of step02.south west, anchor=north west] {%
            \begin{tikzpicture}[node distance=3ex and 2.0em]

\node[name=sumleft, leftsummator];
\node[on grid, right=18em of sumleft, name=sumright, rightsummator];

\zxsumtwo{%
    \begin{ZX}
        \\[0.3ex]
            \zxX{} \rar &[5em] \zxN{} \rar & \zxX{}
        \\[0.3ex]
    \end{ZX}
    }{%
    \begin{ZX}
        \zxX{} \rar & \zxFracZ{\pi}{2} \rar & \zxX{\pi} \rar & \zxFracZ{\pi}{2} \rar & \zxX{}
    \end{ZX}
}[0.0em][4.0ex][1.0][c_{\alpha_u}][\ii s_{\alpha_u}]

\node[name=anchorl, below=2ex of bubblebottom.south west, invisible];
\node[name=anchorr, below=2ex of bubblebottom.south east, invisible];
\draw[thick, decorate, decoration={brace,mirror,raise=0.3ex}] (anchorl) -- 
    node[midway, below=1.0ex] (label) {$=0$}
(anchorr);

            \end{tikzpicture}
        };
        \node (eq3) [left=1ex of step03, yshift=2ex] {$=$};
        \node (eq4) [right=1ex of step03, yshift=2ex] {$=$};
        \node (step04) [right=0.5ex of eq4] {%
            $2 c_{\alpha_u}$
        };
    \end{tikzpicture}
};

        \end{tikzpicture}
    \label{eqn:single_qubit_rotation_exp_value}\,.
\end{equation}
Here, again, $s_\alpha=\sin(\alpha)$ and $c_\alpha=\cos(\alpha)$, 
and each underbrace used refers only to the subdiagram directly above.
Note that after the second step the usage of linear combinations to handle the X-spider with phase $(-2\alpha_u)$ provides a way to continue the calculation, which would not be possible within the conventional ZX-framework. From the permutation symmetry of the ansatz, the expectation value $\langle Z_i Z_j\rangle$ of each edge is of the same form~\cite{shaydulin2021classical}. 
Hence we have 
\begin{equation} \label{eq:npansatz}
   \langle C\rangle= \tfrac{|E|}2- \tfrac12 \sum_{(u, v)\in E} \cos(\alpha_u) \cos(\alpha_v), 
\end{equation}
which implies 
\begin{equation}
    \max_{\alpha} \langle C\rangle = \max_{\alpha\in\{0,\pi\}^n} \langle C\rangle =  \max_x c(x)= c(y^*),
\end{equation}
where 
we have used the observation that
angles $\alpha^* \in \{0,\pi\}^n$ encode a 
bit string $y^*$ via $y^*_i=\tfrac12-\tfrac12 \cos(\alpha^*_i)$. 
Hence, as any globally optimal angles must directly encode an optimal solution to the MaxCut instance,
the expectation value $\langle C\rangle$ is NP-hard to optimize. 
Indeed, for MaxCut, \eqref{eq:npansatz} reproduces the quantity of Equation~1 of~\cite{bittel2021training} (up to an affine shift).  
This result is used throughout~\cite{bittel2021training} via further reductions to show that optimizing a number of other classes of PQCs is 
NP-hard in general.  
We have similarly demonstrated that the single-qubit rotations ansatz is NP-hard to optimize for problems such as MaxCut, but via a compact derivation using ZX-diagrams.

\subsection{QAOA$_1$ for MaxCut on a Simple Graph}%
\begin{figure}[htpb]
    \centering
    \externalize{qaoa_max_cut_example}{%
        \begin{tikzpicture}
           \node (lhs) {$G$ \; \; = }; 
           \node[right=2pt of lhs] {%
                \begin{tikzpicture}[scale=0.4]
                    \node[draw, circle, label={left:\small 1}, inner sep=2pt] at (0, 0) (1) {};
                    \node[draw, circle, label={right:\small 2}, inner sep=2pt] at (1, -1) (2) {};
                    \node[draw, circle, label={left:\small 3}, inner sep=2pt] at (-1, -2) (3) {};
                    \node[draw, circle, label={right:\small 4}, inner sep=2pt] at (0, -3) (4) {};
                    \draw (1) -- (2);
                    \draw (2) -- (3);
                    \draw (3) -- (4);
                    \draw (2) -- (4);
                \end{tikzpicture}
            };
        \end{tikzpicture}
    }
    \caption{Simple example graph to consider for MaxCut with QAOA}%
    \label{fig:qaoa_exp_value_simple_example}
\end{figure}
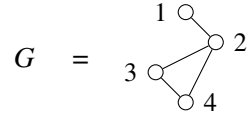

\label{sec:example_applications_qaoa_example_graph}
Next we turn to QAOA~\cite{farhi2014quantum,hadfield2019quantum}, 
for which we continue our use of MaxCut as a running example. 
For simplicity we consider QAOA$_1$, the lowest depth realization, 
which is indicative of the $p>1$ case due to the alternating structure of the ansatz. 
Recall that for a QAOA state the MaxCut expectation value reads
$\langle C \rangle
= \tfrac{|E|}{2} - \tfrac12\sum_{i, j\in E} \langle Z_i Z_j \rangle \, .$
We begin with the specific graph $G$ 
of Figure~\ref{fig:qaoa_exp_value_simple_example}, before we consider ring graphs in Section~\ref{sec:example_applications_qaoa_ring}, and arbitrary graphs in Appendix~\ref{sec:app_qaoa}.
Observe how the structure of the graph directly reappears in the diagrams below, 
which reflects the fact that the QAOA phase operator is derived from the cost Hamiltonian. 
For deeper QAOA circuits, the graph structure will again appear at each layer in the diagrammatic representation. 
Hence ZX-calculus provides a toolkit toward directly incorporating or better understanding the relationship between the cost function and a given parameterized quantum algorithm. 

Here we demonstrate the edge expectation value calculation for QAOA$_{1}$,  
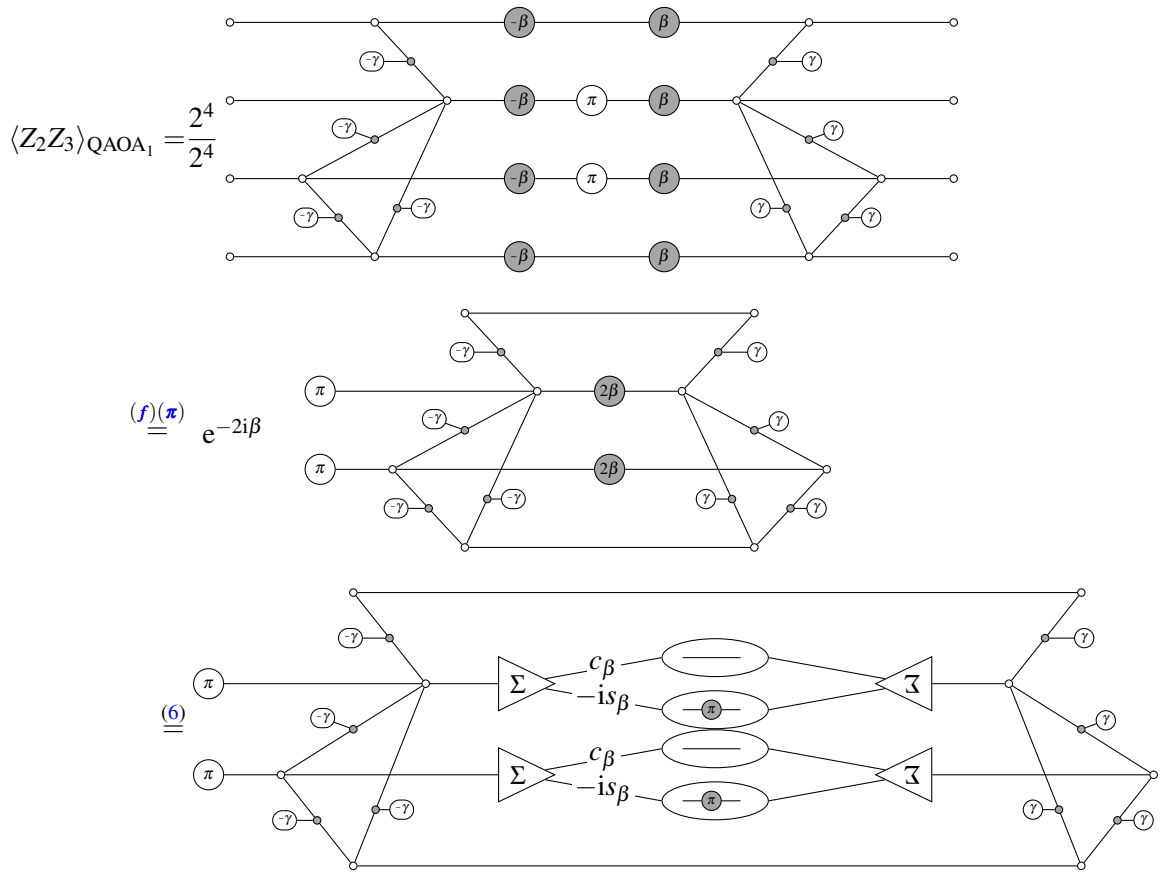
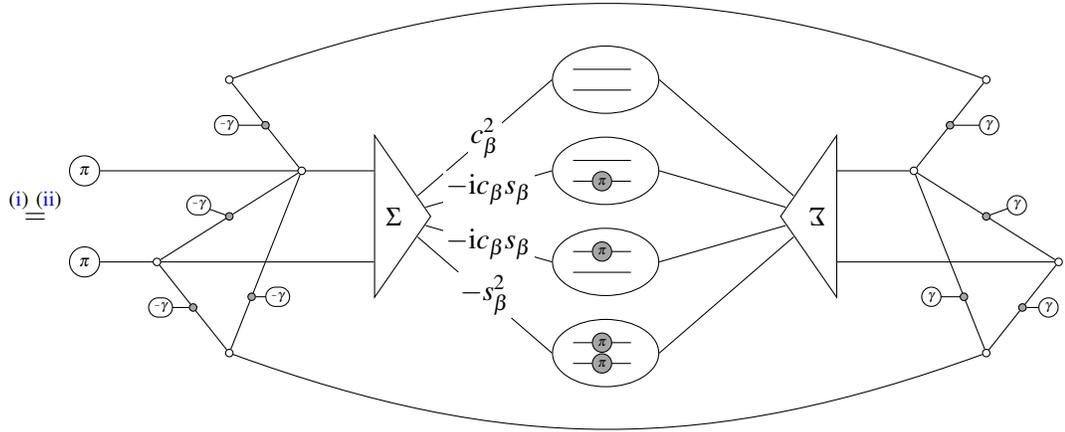
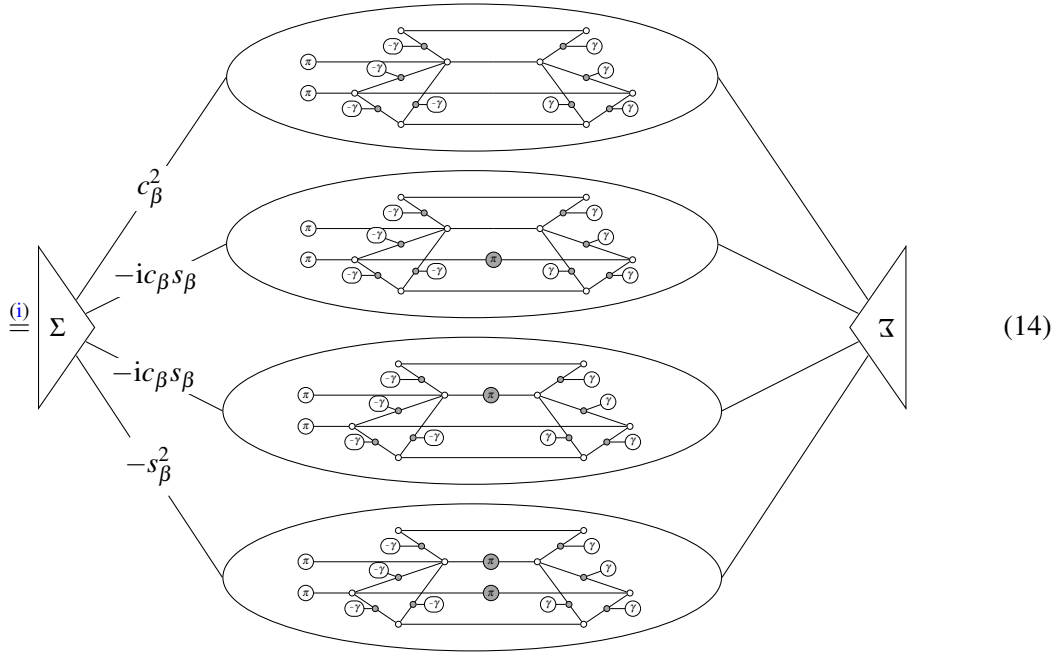
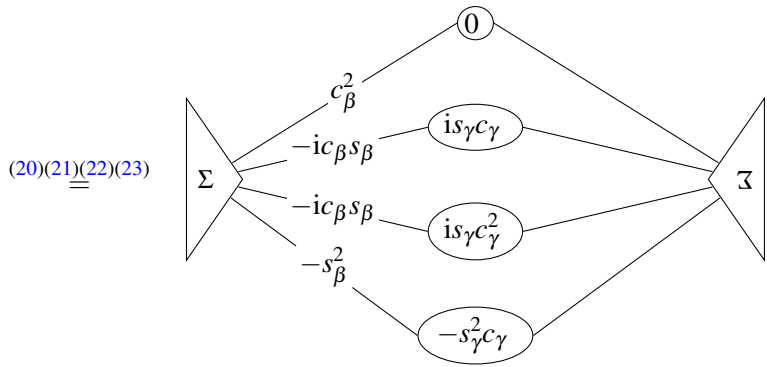
\begin{align}
    \langle Z_2 Z_3 \rangle_{\text{QAOA}_1}
    = & 
    \frac{2^4}{2^4}
    \vcenter{\hbox{
        \externalize{qaoa_max_cut_example_01}{%

        }
    }} \, .
\end{align}
The remaining expectation values can be similarly computed for each of the edges in $E$ to give $\langle C\rangle$.  
In the third step above, we could not have easily continued within the conventional ZX-calculus framework.
Whenever one needs to pull parameterized X-spiders through parameterized Z-spiders or vice versa, our extension is utilized.
The detailed calculation of the four contributions used in the last step 
is given in Appendix~\ref{sec:app_qaoa_example_graph}.
Note that calculation of the general $n$-qubit case (cf.~Appendix~\ref{sec:app_qaoa}) is surprisingly concise compared to the special case of 4-qubits considered here.



We consider a hardware-efficient ansatz and two 
more general QAOA examples in Appendix~\ref{sec:additional_examples}.

\section{Outlook}
\label{sec:summary_and_outlook}

We introduced an extension of the ZX-calculus to 
conveniently incorporate linear combinations of ZX-diagrams. 
Moreover we demonstrated how this generalized diagrammatic framework can be applied to the analysis of parameterized quantum circuits,
in particular to the calculation of observable expectation values. 
Further quantities of interest such as gradients may be similarly derived, 
as well as more complicated PQC phenomenon such as barren plateaus studied,
by combining our framework with several distinct but complementary recent ZX-calculus advances~\cite{zhao2021analyzing,jeandel2022addition,wang2022differentiating}.
Software implementation of these 
results may facilitate novel 
approaches 
for automatic contraction of 
diagrams related to PQCs, including but not limited to expectation values. 
A concrete next step is to rigorously derive such algorithms 
and carefully analyze problems and PQC classes where they may yield advantages.

Future research could further formalize our approach as well as 
integrate it with other variants of ZX-calculus, 
like ZH-calculus~\cite{backens2018zh} or the ZX-framework for qudits~\cite{ranchin2014depicting,wang2014qutrit}.
In particular the latter could facilitate novel insights into performance analysis of quantum alternating operator ans\"atze~\cite{hadfield2019quantum} for problems like graph-coloring~\cite{wang2020xymixers} and beyond~\cite{stollenwerk2020toward}.
Similarly, our approach could be likewise applied to applications beyond combinatorial optimization, like variational quantum eigensolvers for quantum 
chemistry applications~\cite{cowtan2020generic}.
Generally, it is of interest to explore to what extent diagrammatic approaches may ultimately aid in the design and analysis of better performing parameterized quantum circuit ans\"atze, 
as well as help with important related challenges such as alleviating the cost of parameter setting, avoiding undesirable features such as barren plateaus, or tailoring ansatz design to a given set of hardware constraints.

\section*{Acknowledgments}
\label{sec:acks}
SH is grateful for support from the NASA Ames Research Center, from NASA Academic Mission Services (NAMS) under Contract No. NNA16BD14C, and from the DARPA ONISQ program under interagency agreement IAA 8839, Annex 114.

\bibliographystyle{eptcs}
\bibliography{references}


\appendix

\appendix
\section{Additional Rules for Linear Combinations of ZX-Diagrams}%
\label{sec:additional_rules}
We introduce  several additional rules which 
are 
useful for the calculation of expectation values for PQC 
which we 
utilize in the derivations to follow.
\paragraph{Scalar-pull rule}
First, scalars can be pulled through the bubble. 
I.e.\ it does not matter if we write them to the left or right of the bubbles.
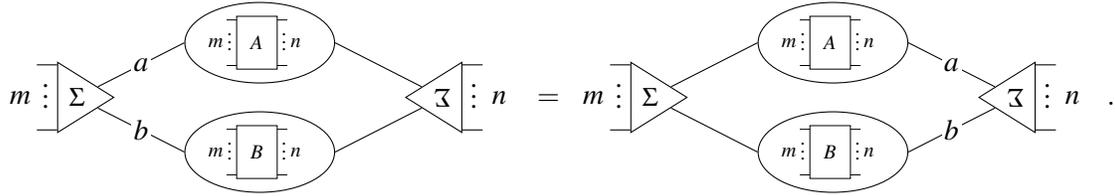
\begin{equation}
    \vcenter{\hbox{
        \externalize{zxsumrules_scalar_pull_lhs}{%
            \begin{tikzpicture}[node distance=5ex and 2.5em]
\def\sumwidth{14em}
\def\yscale{1.3}
\node[name=anchorlt, invisible];
\node[on grid, below=of anchorlt, name=anchorlb, invisible];
\node[right=\sumwidth of anchorlt.center, name=anchorrt, invisible];
\node[on grid, below=of anchorrt, name=anchorrb, invisible];
\node at ($(anchorlt.center)!0.5!(anchorlb.center)$) [yscale=\yscale, name=sumleft, leftsummator] ;
\node[right=\sumwidth of sumleft.west, yscale=\yscale, name=sumright, rightsummator];
\zxsumtwo{%
\diagplacehold{$m$}{$A$}{$n$}
}{%
\diagplacehold{$m$}{$B$}{$n$}
}[0.0em][6.0ex][0.7][a][b]
\draw[thin] (anchorlt) -- +(180:0.7em);
\node[left=0.4em of sumleft.west, anchor=center] (dots) {\rvdots};
\node[left=0.4em of dots, anchor=center] {$m$};
\draw[thin] (anchorlb) -- +(180:0.7em);
\draw[thin] (anchorrt) -- +(0:0.7em);
\node[right=0.4em of sumright.west, anchor=center] (dots) {\rvdots};
\node[right=0.4em of dots, anchor=center] {$n$};
\draw[thin] (anchorrb) -- +(0:0.7em);

            \end{tikzpicture}
        }
    }} 
    =
    \vcenter{\hbox{
        \externalize{zxsumrules_scalar_pull_rhs}{%
            \begin{tikzpicture}[node distance=5ex and 2.5em]
    \def\sumwidth{14em}
    \def\yscale{1.3}
    \node[name=anchorlt, invisible];
    \node[on grid, below=of anchorlt, name=anchorlb, invisible];
    \node[right=\sumwidth of anchorlt.center, name=anchorrt, invisible];
    \node[on grid, below=of anchorrt, name=anchorrb, invisible];
    \node at ($(anchorlt.center)!0.5!(anchorlb.center)$) [yscale=\yscale, name=sumleft, leftsummator] ;
    \node[right=\sumwidth of sumleft.west, yscale=\yscale, name=sumright, rightsummator];
    \zxsumtwo{%
        \diagplacehold{$m$}{$A$}{$n$}
        }{%
        \diagplacehold{$m$}{$B$}{$n$}
    }[0.0em][6.0ex][0.7][a][b]r
    \draw[thin] (anchorlt) -- +(180:0.7em);
    \node[left=0.4em of sumleft.west, anchor=center] (dots) {\rvdots};
    \node[left=0.4em of dots, anchor=center] {$m$};
    \draw[thin] (anchorlb) -- +(180:0.7em);
    \draw[thin] (anchorrt) -- +(0:0.7em);
    \node[right=0.4em of sumright.west, anchor=center] (dots) {\rvdots};
    \node[right=0.4em of dots, anchor=center] {$n$};
    \draw[thin] (anchorrb) -- +(0:0.7em);

            \end{tikzpicture}
        }
    }} \label{eqn:zxsumrules_scalarpull} \,.
\end{equation}
\paragraph{Linear combinations for states and effects}
Since we can put the scalar factor left or right of the bubbles, we can simplify linear combinations in the case of states or effects. 
For states (no inputs), we can cut the left half of the diagram
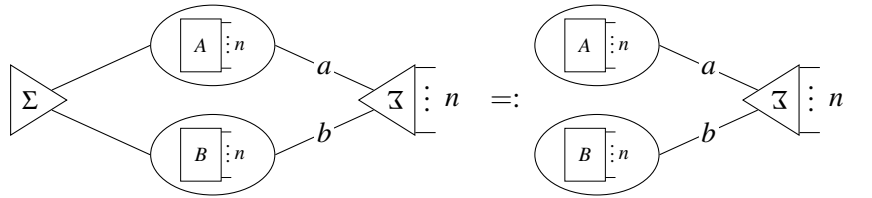
\begin{equation}
    \vcenter{\hbox{
        \externalize{zxsumrules_cut_left_summator_lhs}{%
            \begin{tikzpicture}[node distance=5ex and 2.5em]
\def\sumwidth{14em}
\def\yscale{1.3}
\node[name=anchorlt, invisible];
\node[on grid, below=of anchorlt, name=anchorlb, invisible];
\node[right=\sumwidth of anchorlt.center, name=anchorrt, invisible];
\node[on grid, below=of anchorrt, name=anchorrb, invisible];
\node at ($(anchorlt.center)!0.5!(anchorlb.center)$) [yscale=\yscale, name=sumleft, leftsummator] ;
\node[right=\sumwidth of sumleft.west, yscale=\yscale, name=sumright, rightsummator];
\zxsumtwo{%
    \diagplaceholdnoinput{$A$}{$n$}
    }{%
    \diagplaceholdnoinput{$B$}{$n$}
}[0.0em][6.0ex][0.7][a][b]r
\draw[thin] (anchorrt) -- +(0:0.7em);
\node[right=0.4em of sumright.west, anchor=center] (dots) {\rvdots};
\node[right=0.4em of dots, anchor=center] {$n$};
\draw[thin] (anchorrb) -- +(0:0.7em);

            \end{tikzpicture}
        }
    }} 
    =:
    \vcenter{\hbox{
        \externalize{zxsumrules_cut_left_summator_rhs}{%
            \begin{tikzpicture}[node distance=5ex and 2.5em]
\def\sumwidth{7em}
\def\yscale{1.3}
\node[name=anchorlt, invisible];
\node[on grid, below=of anchorlt, name=anchorlb, invisible];
\node[right=\sumwidth of anchorlt.center, name=anchorrt, invisible];
\node[on grid, below=of anchorrt, name=anchorrb, invisible];
\node at ($(anchorlt.center)!0.5!(anchorlb.center)$) [name=sumleft, invisible] ;
\node[right=\sumwidth of sumleft.west, yscale=\yscale, name=sumright, rightsummator];
\zxsumtwo{%
    \diagplaceholdnoinput{$A$}{$n$}
    }{%
    \diagplaceholdnoinput{$B$}{$n$}
}[0.0em][6.0ex][0.7][a][b]rc
\draw[thin] (anchorrt) -- +(0:0.7em);
\node[right=0.4em of sumright.west, anchor=center] (dots) {\rvdots};
\node[right=0.4em of dots, anchor=center] {$n$};
\draw[thin] (anchorrb) -- +(0:0.7em);

            \end{tikzpicture}
        }
    }} \label{eqn:zxsumrules_cutleftsummator}\, .
\end{equation}
For effects (no outputs), we can cut the right half of the diagram
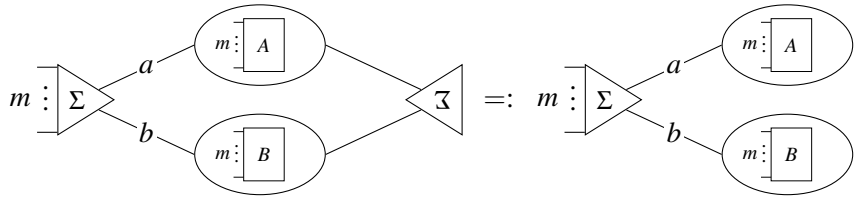
\begin{equation}
    \vcenter{\hbox{
        \externalize{zxsumrules_cut_right_summator_lhs}{%
            \begin{tikzpicture}[node distance=5ex and 2.5em]
\def\sumwidth{14em}
\def\yscale{1.3}
\node[name=anchorlt, invisible];
\node[on grid, below=of anchorlt, name=anchorlb, invisible];
\node[right=\sumwidth of anchorlt.center, name=anchorrt, invisible];
\node[on grid, below=of anchorrt, name=anchorrb, invisible];
\node at ($(anchorlt.center)!0.5!(anchorlb.center)$) [yscale=\yscale, name=sumleft, leftsummator] ;
\node[right=\sumwidth of sumleft.west, yscale=\yscale, name=sumright, rightsummator];
\zxsumtwo{%
    \diagplaceholdnooutput{$m$}{$A$}
    }{%
    \diagplaceholdnooutput{$m$}{$B$}
}[0.0em][6.0ex][0.7][a][b]
\draw[thin] (anchorlt) -- +(180:0.7em);
\node[left=0.4em of sumleft.west, anchor=center] (dots) {\rvdots};
\node[left=0.4em of dots, anchor=center] {$m$};
\draw[thin] (anchorlb) -- +(180:0.7em);

            \end{tikzpicture}
        }
    }} 
    =:
    \vcenter{\hbox{
        \externalize{zxsumrules_cut_right_summator_rhs}{%
            \begin{tikzpicture}[node distance=5ex and 2.5em]
\def\sumwidth{7em}
\def\yscale{1.3}
\node[name=anchorlt, invisible];
\node[on grid, below=of anchorlt, name=anchorlb, invisible];
\node[right=\sumwidth of anchorlt.center, name=anchorrt, invisible];
\node[on grid, below=of anchorrt, name=anchorrb, invisible];
\node at ($(anchorlt.center)!0.5!(anchorlb.center)$) [yscale=\yscale, name=sumleft, leftsummator] ;
\node[right=\sumwidth of sumleft.west, name=sumright, invisible];
\zxsumtwo{%
    \diagplaceholdnooutput{$m$}{$A$}
    }{%
    \diagplaceholdnooutput{$m$}{$B$}
}[0.0em][6.0ex][0.7][a][b]c
\draw[thin] (anchorlt) -- +(180:0.7em);
\node[left=0.4em of sumleft.west, anchor=center] (dots) {\rvdots};
\node[left=0.4em of dots, anchor=center] {$m$};
\draw[thin] (anchorlb) -- +(180:0.7em);

            \end{tikzpicture}
        }
    }} \label{eqn:zxsumrules_cutrightsummator} \, .
\end{equation}
\paragraph{Direct connection of diagrams (no bubbles)}
We can also completely drop the bubbles and continue the input and output wires through the sum symbols
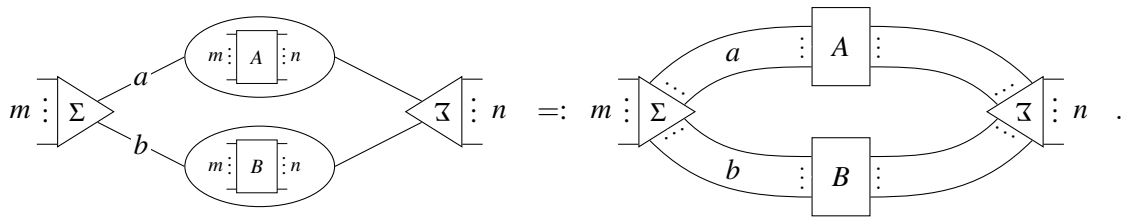
\begin{equation}
    \vcenter{\hbox{
        \externalize{zxsumrules_direct_notation_lhs}{%
            \begin{tikzpicture}[node distance=5ex and 2.5em]
\def\sumwidth{14em}
\def\yscale{1.3}
\node[name=anchorlt, invisible];
\node[on grid, below=of anchorlt, name=anchorlb, invisible];
\node[right=\sumwidth of anchorlt.center, name=anchorrt, invisible];
\node[on grid, below=of anchorrt, name=anchorrb, invisible];
\node at ($(anchorlt.center)!0.5!(anchorlb.center)$) [yscale=\yscale, name=sumleft, leftsummator] ;
\node[right=\sumwidth of sumleft.west, yscale=\yscale, name=sumright, rightsummator];
\zxsumtwo{%
    \diagplacehold{$m$}{$A$}{$n$}
    }{%
    \diagplacehold{$m$}{$B$}{$n$}
}[0.0em][6.0ex][0.7][a][b]
\draw[thin] (anchorlt) -- +(180:0.7em);
\node[left=0.4em of sumleft.west, anchor=center] (dots) {\rvdots};
\node[left=0.4em of dots, anchor=center] {$m$};
\draw[thin] (anchorlb) -- +(180:0.7em);
\draw[thin] (anchorrt) -- +(0:0.7em);
\node[right=0.4em of sumright.west, anchor=center] (dots) {\rvdots};
\node[right=0.4em of dots, anchor=center] {$n$};
\draw[thin] (anchorrb) -- +(0:0.7em);

            \end{tikzpicture}
        }
    }} 
    =:
    \vcenter{\hbox{
        \externalize{zxsumrules_direct_notation_rhs}{%
            \begin{tikzpicture}[node distance=5ex and 2.5em]
\def\sumwidth{14em}
\def\yscale{1.3}
\def\yshift{0.7em}
\def\xdotpad{0.3em}
\def\ydotpad{0.3ex}
\node[name=anchorlt, invisible];
\node[on grid, below=of anchorlt, name=anchorlb, invisible];
\node[right=\sumwidth of anchorlt.center, name=anchorrt, invisible];
\node[on grid, below=of anchorrt, name=anchorrb, invisible];
\node at ($(anchorlt.center)!0.5!(anchorlb.center)$) [yscale=\yscale, name=sumleft, leftsummator] ;
\node[right=\sumwidth of sumleft.west, yscale=\yscale, name=sumright, rightsummator];

\node at ($(anchorlt.center)!0.5!(anchorrt.center)$) [yshift=5ex, name=tt, invisible] ;
\node at ($(anchorlb.center)!0.5!(anchorrb.center)$) [yshift=5ex, name=tb, invisible] ;
\node[rectangle, 
      draw=black,thin, 
      fill=white,
      fit=(tt) (tb), 
      inner xsep=1.0em, 
      inner ysep=0.5ex] (topbox) {};
\node[fill=white, inner sep=0pt, anchor=center] at ($(topbox.north west)!0.5!(topbox.south east)$) {$A$};

\draw[thin] ($(sumleft.north west)!0.2!(sumleft.east)$) to [out=45, in=180] ([yshift=\yshift]topbox.west) ;
\draw[thin, draw=white] ($(sumleft.north west)!0.5!(sumleft.east)$) to [out=45, in=180] node[fill=white, midway, inner sep=1pt, anchor=center] {$a$} (topbox.west) ;
\draw[thin] ($(sumleft.north west)!0.8!(sumleft.east)$) to [out=45, in=180] ([yshift=-\yshift]topbox.west) ;
\node at ($(sumleft.north west)!0.5!(sumleft.east)$) [xshift=\xdotpad, yshift=\ydotpad, rotate=-30, anchor=center] {\footnotesize $\cdots$};
\node at ($(topbox.north west)!0.5!(topbox.south west)$) [xshift=-\xdotpad, anchor=center, rotate=90]  {\footnotesize $\cdots$};

\draw[thin] ($(sumright.south west)!0.2!(sumright.east)$) to [out=135, in=0] ([yshift=\yshift]topbox.east) ;
\draw[thin] ($(sumright.south west)!0.8!(sumright.east)$) to [out=135, in=0] ([yshift=-\yshift]topbox.east) ;
\node at ($(sumright.south west)!0.5!(sumright.east)$) [xshift=-\xdotpad, yshift=\ydotpad, rotate=30, anchor=center] {\footnotesize $\cdots$};
\node at ($(topbox.north east)!0.5!(topbox.south east)$) [xshift=\xdotpad, anchor=center, rotate=90]  {\footnotesize $\cdots$};

\node at ($(anchorlt.center)!0.5!(anchorrt.center)$) [yshift=-5ex, name=bt, invisible] ;
\node at ($(anchorlb.center)!0.5!(anchorrb.center)$) [yshift=-5ex, name=bb, invisible] ;
\node[rectangle, 
      draw=black,thin, 
      fill=white,
      fit=(bt) (bb), 
      inner xsep=1.0em, 
      inner ysep=0.5ex] (bottombox) {};
\node[fill=white, inner sep=0pt, anchor=center] at ($(bottombox.north west)!0.5!(bottombox.south east)$) {$B$};

\draw[thin] ($(sumleft.south west)!0.2!(sumleft.east)$) to [out=-45, in=180] ([yshift=-\yshift]bottombox.west) ;
\draw[thin, draw=white] ($(sumleft.south west)!0.5!(sumleft.east)$) to [out=-45, in=180] node[fill=white, midway, inner sep=1pt, anchor=center] {$b$} (bottombox.west) ;
\draw[thin] ($(sumleft.south west)!0.8!(sumleft.east)$) to [out=-45, in=180] ([yshift=\yshift]bottombox.west) ;
\node at ($(sumleft.south west)!0.5!(sumleft.east)$) [xshift=\xdotpad, yshift=-\ydotpad, rotate=30, anchor=center] {\footnotesize $\cdots$};
\node at ($(bottombox.north west)!0.5!(bottombox.south west)$) [xshift=-\xdotpad, anchor=center, rotate=90]  {\footnotesize $\cdots$};

\draw[thin] ($(sumright.north west)!0.2!(sumright.east)$) to [out=-135, in=0] ([yshift=-\yshift]bottombox.east) ;
\draw[thin] ($(sumright.north west)!0.8!(sumright.east)$) to [out=-135, in=0] ([yshift=\yshift]bottombox.east) ;
\node at ($(sumright.north west)!0.5!(sumright.east)$) [xshift=-\xdotpad, yshift=-\ydotpad, rotate=-30, anchor=center] {\footnotesize $\cdots$};
\node at ($(bottombox.north east)!0.5!(bottombox.south east)$) [xshift=\xdotpad, anchor=center, rotate=90]  {\footnotesize $\cdots$};

\draw[thin] (anchorlt) -- +(180:0.7em);
\node[left=0.4em of sumleft.west, anchor=center] (dots) {\rvdots};
\node[left=0.4em of dots, anchor=center] {$m$};
\draw[thin] (anchorlb) -- +(180:0.7em);
\draw[thin] (anchorrt) -- +(0:0.7em);
\node[right=0.4em of sumright.west, anchor=center] (dots) {\rvdots};
\node[right=0.4em of dots, anchor=center] {$n$};
\draw[thin] (anchorrb) -- +(0:0.7em);

            \end{tikzpicture}
        }
    }} \label{eqn:zxsumrules_direct_notation}\,.
\end{equation}
However, we will not 
require this notation 
in the examples considered 
in this paper.

\section{Additional Examples}
\label{sec:additional_examples}
Here we continue our examples from Section~\ref{sec:example_applications} and diagrammatically 
derive MaxCut expectation values for a hardware efficient ansatz as well as for QAOA$_1$ on rings and general graphs.

\subsection{Hardware Efficient Ansatz}%
\label{sec:example_applications_hweff}

We consider  a 
variant of a hardware efficient SU-2 2-local ansatz from Qiskit~\cite{qiskit2019}.
This ansatz was also studied in~\cite{funcke2021dimensional}. 
For simplicity here we consider a 3-qubit realization, 
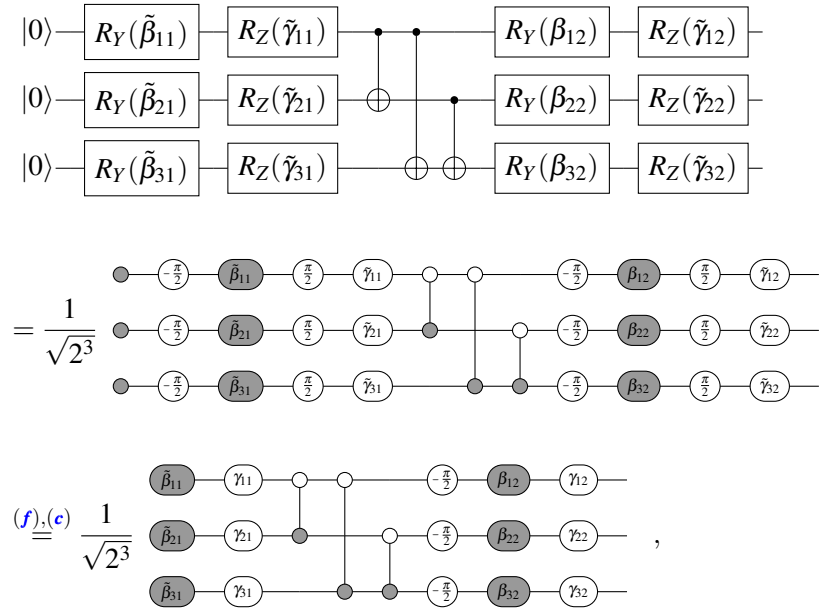
\begin{align}
    &
    \vcenter{\hbox{
        \externalizezx{hw_eff_ansatz_example_ansatz_00}{%
            \begin{tikzcd}[thin lines, row sep={1.0ex}, 
                 column sep=0.5em,
                 ampersand replacement=\&]
    \lstick{\ket{0}} 
    \& \qw  \& \gate{R_Y(\tilde\beta_{11})} 
    \& \qw \& \gate{R_Z(\tilde \gamma_{11})} \& \qw  
    \& \ctrl{1} \& \ctrl{2} \& \qw 
    \& \qw  \& \gate{R_Y(\beta_{12})} 
    \& \qw \& \gate{R_Z(\tilde \gamma_{12})} \& \qw  
    \\ 
    \lstick{\ket{0}} 
    \& \qw  \& \gate{R_Y(\tilde\beta_{21})} 
    \& \qw \& \gate{R_Z(\tilde \gamma_{21})} \& \qw  
    \& \targ{} \& \qw \& \ctrl{1} 
    \& \qw  \& \gate{R_Y(\beta_{22})} 
    \& \qw \& \gate{R_Z(\tilde \gamma_{22})} \& \qw  
    \\ 
    \lstick{\ket{0}} 
    \& \qw  \& \gate{R_Y(\tilde\beta_{31})} 
    \& \qw \& \gate{R_Z(\tilde \gamma_{31})} \& \qw  
    \& \qw \& \targ{} \& \targ{} 
    \& \qw  \& \gate{R_Y(\beta_{32})} 
    \& \qw \& \gate{R_Z(\tilde \gamma_{32})} \& \qw  
\end{tikzcd}

        }
    }} 
    \\[4ex]
    &=
    \frac{1}{\sqrt{2^3}}
    \vcenter{\hbox{
        \externalizezx{hw_eff_ansatz_example_ansatz_01}{%
            \begin{ZX}[row sep=1.9ex, column sep=1em, ampersand replacement=\&]
    \zxX{} \rar 
    \& \zxFracZ-{\pi}{2} \rar \& \zxX{\tilde\beta_{11}} \rar \& \zxFracZ{\pi}{2} \rar
    \& \zxZ{\tilde \gamma_{11}} \rar
    \& \zxZ{} \dar \rar \& \zxZ{} \dar \rar \& \zxN{} \rar 
    \& \zxFracZ-{\pi}{2} \rar \& \zxX{\beta_{12}} \rar \& \zxFracZ{\pi}{2} \rar
    \& \zxZ{\tilde \gamma_{12}} \rar
    \& \zxN{} 
    \\
    \zxX{} \rar 
    \& \zxFracZ-{\pi}{2} \rar \& \zxX{\tilde\beta_{21}} \rar \& \zxFracZ{\pi}{2} \rar
    \& \zxZ{\tilde \gamma_{21}} \rar
    \& \zxX{} \rar \& \zxN{} \dar \rar \& \zxZ{} \dar \rar 
    \& \zxFracZ-{\pi}{2} \rar \& \zxX{\beta_{22}} \rar \& \zxFracZ{\pi}{2} \rar
    \& \zxZ{\tilde \gamma_{22}} \rar
    \& \zxN{} 
    \\
    \zxX{} \rar 
    \& \zxFracZ-{\pi}{2} \rar \& \zxX{\tilde\beta_{31}} \rar \& \zxFracZ{\pi}{2} \rar
    \& \zxZ{\tilde \gamma_{31}} \rar 
    \& \zxN{} \rar \& \zxX{} \rar \& \zxX{} \rar 
    \& \zxFracZ-{\pi}{2} \rar \& \zxX{\beta_{32}} \rar \& \zxFracZ{\pi}{2} \rar
    \& \zxZ{\tilde \gamma_{32}} \rar 
    \& \zxN{}
\end{ZX}

        }
    }}
    \\[4ex]
    &
    \stackrel{\spiderrule,\copyrule}{=}
    \frac{1}{\sqrt{2^3}}
    \vcenter{\hbox{
        \externalizezx{hw_eff_ansatz_example_ansatz_02}{%
            \begin{ZX}[row sep=1.9ex, column sep=1em, ampersand replacement=\&]
    \zxX{\tilde\beta_{11}} \rar
    \& \zxZ{\gamma_{11}} \rar
    \& \zxZ{} \dar \rar \& \zxZ{} \dar \rar \& \zxN{} \rar 
    \& \zxFracZ-{\pi}{2} \rar
    \& \zxX{\beta_{12}} \rar
    \& \zxZ{\gamma_{12}} \rar
    \& \zxN{} 
    \\
    \zxX{\tilde\beta_{21}} \rar 
    \& \zxZ{\gamma_{21}} \rar
    \& \zxX{} \rar \& \zxN{} \dar \rar \& \zxZ{} \dar \rar 
    \& \zxFracZ-{\pi}{2} \rar
    \& \zxX{\beta_{22}} \rar 
    \& \zxZ{\gamma_{22}} \rar
    \& \zxN{} 
    \\
    \zxX{\tilde\beta_{31}} \rar
    \& \zxZ{\gamma_{31}} \rar 
    \& \zxN{} \rar \& \zxX{} \rar \& \zxX{} \rar 
    \& \zxFracZ-{\pi}{2} \rar
    \& \zxX{\beta_{32}} \rar
    \& \zxZ{\gamma_{32}} \rar 
    \& \zxN{}
\end{ZX}

        }
    }}\label{eqn:hw_eff_ansatz} \, ,
\end{align}
where we conveniently set 
$\gamma_{ij} := \tilde \gamma_{ij} + \frac{\pi}{2}$ 
and 
$\beta_{ij} := \frac{-\tilde \beta{ij}}{2}$.
To compute the expectation value of 
a given cost Hamiltonian,
we again requite expectation values of products of 
Pauli-Z operators. 
We demonstrate how such calculations can be performed diagrammatically by considering again MaxCut as an example. For the expectation value corresponding to a given edge $(2, 3)$ we have  
\begin{align}
    &\langle Z_2 Z_3 \rangle 
    =  
    \\[8ex]
    &
    \frac{1}{{2^3}}
    \vcenter{\hbox{
        \externalize{hw_eff_ansatz_example_01}
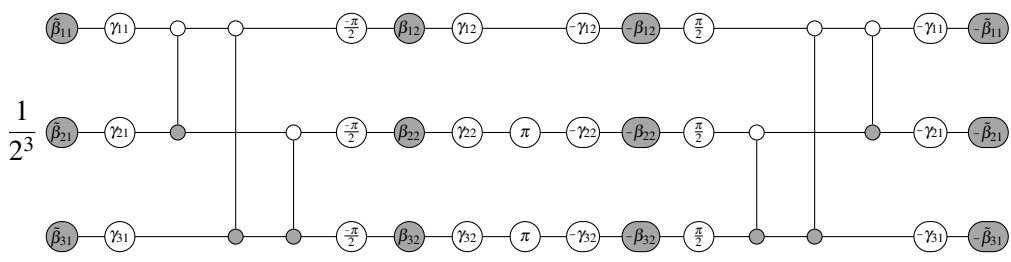
{%
            \begin{tikzpicture}[node distance=8ex and 2.0em]
\node[beta=$\tilde\beta_{11}$, name=b11l];
\node[on grid, below=of b11l, beta=$\tilde\beta_{21}$, name=b21l];
\node[on grid, below=of b21l, beta=$\tilde\beta_{31}$, name=b31l];

\node[on grid, right=of b11l, gamma=$\gamma_{11}$, name=g11l];
\node[on grid, right=of b21l, gamma=$\gamma_{21}$, name=g21l];
\node[on grid, right=of b31l, gamma=$\gamma_{31}$, name=g31l];

\node[on grid, right=of g11l, name=e1ll, zspider=2pt] {};
\node[on grid, right=of g21l, name=e2ll, xspider=2pt] {};
\node[on grid, right=of g31l, name=e3ll, invisible];

\node[on grid, right=of e1ll, name=e1lm, zspider=2pt] {};
\node[on grid, right=of e2ll, name=e2lm, invisible];
\node[on grid, right=of e3ll, name=e3lm, xspider=2pt] {};

\node[on grid, right=of e1lm, name=e1lr, invisible];
\node[on grid, right=of e2lm, name=e2lr, zspider=2pt] {};
\node[on grid, right=of e3lm, name=e3lr, xspider=2pt] {};

\node[on grid, right=of e1lr, name=p1l, gamma=$\frac{\zxMinus\pi}{2}$];
\node[on grid, right=of e2lr, name=p2l, gamma=$\frac{\zxMinus\pi}{2}$];
\node[on grid, right=of e3lr, name=p3l, gamma=$\frac{\zxMinus\pi}{2}$];

\node[on grid, right=of p1l, beta=$\beta_{12}$, name=b12l];
\node[on grid, right=of p2l, beta=$\beta_{22}$, name=b22l];
\node[on grid, right=of p3l, beta=$\beta_{32}$, name=b32l];

\node[on grid, right=of b12l, gamma=$\gamma_{12}$, name=g12l];
\node[on grid, right=of b22l, gamma=$\gamma_{22}$, name=g22l];
\node[on grid, right=of b32l, gamma=$\gamma_{32}$, name=g32l];

\node[on grid, right=of g12l, name=o1, invisible];
\node[on grid, right=of g22l, name=o2, zpi];
\node[on grid, right=of g32l, name=o3, zpi];

\node[on grid, right=of o1, gamma=$\zxMinus\gamma_{12}$, name=g12r];
\node[on grid, right=of o2, gamma=$\zxMinus\gamma_{22}$, name=g22r];
\node[on grid, right=of o3, gamma=$\zxMinus\gamma_{32}$, name=g32r];

\node[on grid, right=of g12r, beta=$\zxMinus\beta_{12}$, name=b12r];
\node[on grid, right=of g22r, beta=$\zxMinus\beta_{22}$, name=b22r];
\node[on grid, right=of g32r, beta=$\zxMinus\beta_{32}$, name=b32r];

\node[on grid, right=of b12r, name=p1r, gamma=$\frac{\pi}{2}$];
\node[on grid, right=of b22r, name=p2r, gamma=$\frac{\pi}{2}$];
\node[on grid, right=of b32r, name=p3r, gamma=$\frac{\pi}{2}$];

\node[on grid, right=of p1r, name=e1rl, invisible];
\node[on grid, right=of p2r, name=e2rl, zspider=2pt] {};
\node[on grid, right=of p3r, name=e3rl, xspider=2pt] {};

\node[on grid, right=of e1rl, name=e1rm, zspider=2pt] {};
\node[on grid, right=of e2rl, name=e2rm, invisible];
\node[on grid, right=of e3rl, name=e3rm, xspider=2pt] {};

\node[on grid, right=of e1rm, name=e1rr, zspider=2pt] {};
\node[on grid, right=of e2rm, name=e2rr, xspider=2pt] {};
\node[on grid, right=of e3rm, name=e3rr, invisible];

\node[on grid, right=of e1rr, gamma=$\zxMinus\gamma_{11}$, name=g11r];
\node[on grid, right=of e2rr, gamma=$\zxMinus\gamma_{21}$, name=g21r];
\node[on grid, right=of e3rr, gamma=$\zxMinus\gamma_{31}$, name=g31r];

\node[on grid, right=of g11r, beta=$\zxMinus\tilde\beta_{11}$, name=b11r];
\node[on grid, right=of g21r, beta=$\zxMinus\tilde\beta_{21}$, name=b21r];
\node[on grid, right=of g31r, beta=$\zxMinus\tilde\beta_{31}$, name=b31r];

\begin{scope}[on background layer]
    \draw (b11l) -- (b11r);
    \draw (b21l) -- (b21r);
    \draw (b31l) -- (b31r);

    \draw (e1ll) -- (e2ll);
    \draw (e1lm) -- (e3lm);
    \draw (e2lr) -- (e3lr);

    \draw (e2rl) -- (e3rl);
    \draw (e1rm) -- (e3rm);
    \draw (e1rr) -- (e2rr);
\end{scope}

            \end{tikzpicture}
        }
    }}
    \\[8ex]
    \stackrel{\spiderrule, \pirule}{=}
    & 
    \frac{1}{{2^3}}
    \vcenter{\hbox{
        \externalize{hw_eff_ansatz_example_02}
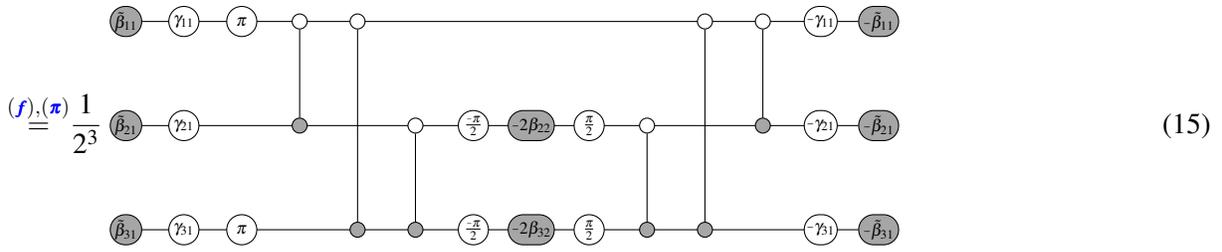
{%
            \begin{tikzpicture}[node distance=8ex and 2.0em]

\node[beta=$\tilde\beta_{11}$, name=b11l];
\node[on grid, below=of b11l, beta=$\tilde\beta_{21}$, name=b21l];
\node[on grid, below=of b21l, beta=$\tilde\beta_{31}$, name=b31l];

\node[on grid, right=of b11l, gamma=$\gamma_{11}$, name=g11l];
\node[on grid, right=of b21l, gamma=$\gamma_{21}$, name=g21l];
\node[on grid, right=of b31l, gamma=$\gamma_{31}$, name=g31l];

\node[on grid, right=of g11l, name=pi1, zpi] {};
\node[on grid, right=of g21l, name=pi2, invisible] {};
\node[on grid, right=of g31l, name=pi3, zpi];

\node[on grid, right=of pi1, name=e1ll, zspider=2pt] {};
\node[on grid, right=of pi2, name=e2ll, xspider=2pt] {};
\node[on grid, right=of pi3, name=e3ll, invisible];

\node[on grid, right=of e1ll, name=e1lm, zspider=2pt] {};
\node[on grid, right=of e2ll, name=e2lm, invisible];
\node[on grid, right=of e3ll, name=e3lm, xspider=2pt] {};

\node[on grid, right=of e1lm, name=e1lr, invisible];
\node[on grid, right=of e2lm, name=e2lr, zspider=2pt] {};
\node[on grid, right=of e3lm, name=e3lr, xspider=2pt] {};

\node[on grid, right=of e1lr, name=p1l, invisible];
\node[on grid, right=of e2lr, name=p2l, gamma=$\frac{\zxMinus\pi}{2}$];
\node[on grid, right=of e3lr, name=p3l, gamma=$\frac{\zxMinus\pi}{2}$];

\node[on grid, right=of p1l, name=b12r, invisible];
\node[on grid, right=of p2l, beta=$\zxMinus2\beta_{22}$, name=b22r];
\node[on grid, right=of p3l, beta=$\zxMinus2\beta_{32}$, name=b32r];

\node[on grid, right=of b12r, name=p1r, invisible];
\node[on grid, right=of b22r, name=p2r, gamma=$\frac{\pi}{2}$];
\node[on grid, right=of b32r, name=p3r, gamma=$\frac{\pi}{2}$];

\node[on grid, right=of p1r, name=e1rl, invisible];
\node[on grid, right=of p2r, name=e2rl, zspider=2pt] {};
\node[on grid, right=of p3r, name=e3rl, xspider=2pt] {};

\node[on grid, right=of e1rl, name=e1rm, zspider=2pt] {};
\node[on grid, right=of e2rl, name=e2rm, invisible];
\node[on grid, right=of e3rl, name=e3rm, xspider=2pt] {};

\node[on grid, right=of e1rm, name=e1rr, zspider=2pt] {};
\node[on grid, right=of e2rm, name=e2rr, xspider=2pt] {};
\node[on grid, right=of e3rm, name=e3rr, invisible];

\node[on grid, right=of e1rr, gamma=$\zxMinus\gamma_{11}$, name=g11r];
\node[on grid, right=of e2rr, gamma=$\zxMinus\gamma_{21}$, name=g21r];
\node[on grid, right=of e3rr, gamma=$\zxMinus\gamma_{31}$, name=g31r];

\node[on grid, right=of g11r, beta=$\zxMinus\tilde\beta_{11}$, name=b11r];
\node[on grid, right=of g21r, beta=$\zxMinus\tilde\beta_{21}$, name=b21r];
\node[on grid, right=of g31r, beta=$\zxMinus\tilde\beta_{31}$, name=b31r];

\begin{scope}[on background layer]
    \draw (b11l) -- (b11r);
    \draw (b21l) -- (b21r);
    \draw (b31l) -- (b31r);

    \draw (e1ll) -- (e2ll);
    \draw (e1lm) -- (e3lm);
    \draw (e2lr) -- (e3lr);

    \draw (e2rl) -- (e3rl);
    \draw (e1rm) -- (e3rm);
    \draw (e1rr) -- (e2rr);
\end{scope}

            \end{tikzpicture}\label{eq:hwexpeczz}
        }
    }}
    \\[8ex]
    \stackrel{\eqref{eqn:xspider_as_lincomb}}{=}
    & 
    \frac{1}{{2^3}}
    \vcenter{\hbox{
        \externalize{hw_eff_ansatz_example_03}
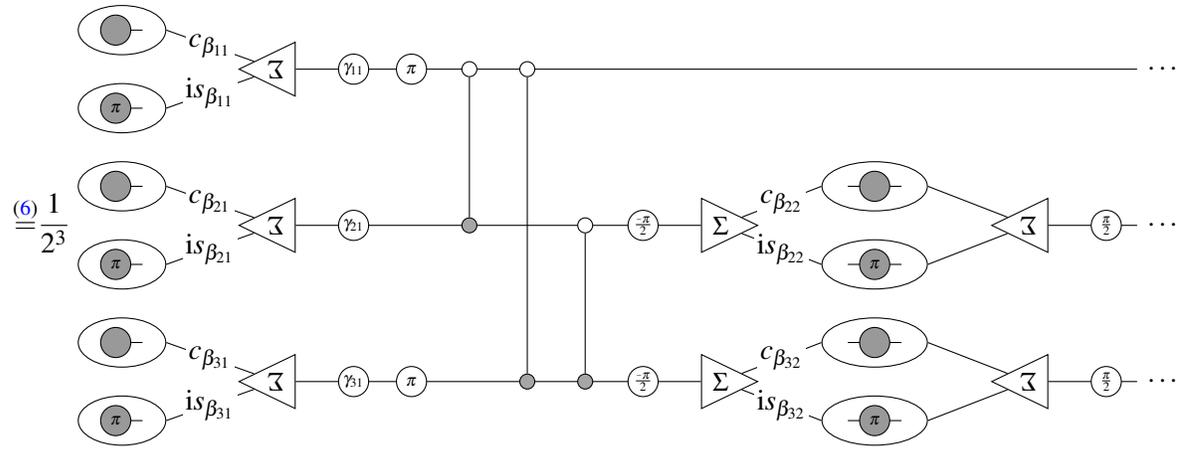
{%
            \begin{tikzpicture}[node distance=12ex and 2.0em]
                

\def\sumwidth{6em}
\def\sumwidthfull{12em}
\def\sumheight{3ex}

\node[name=b11ll, invisible];
\node[on grid, below=of b11ll, name=b21ll, invisible];
\node[on grid, below=of b21ll, name=b31ll, invisible];

\node[right=\sumwidth of b11ll, name=b11lr, invisible];
\node[right=\sumwidth of b21ll, name=b21lr, invisible];
\node[right=\sumwidth of b31ll, name=b31lr, invisible];

\node at (b11ll.center) [name=sumleft, invisible];
\node at (b11lr.center) [name=sumright, rightsummator];
\zxsumtwo{%
    \begin{ZX}
        \zxX{\phantom{\pi}} \rar & \zxN{}
    \end{ZX}
    }{%
    \begin{ZX}
        \zxX{\pi} \rar & \zxN{}
    \end{ZX}
}[0.0em][\sumheight][1.0][c_{\beta_{11}}][\ii s_{\beta_{11}}]rc

\node at (b21ll.center) [name=sumleft, invisible];
\node at (b21lr.center) [name=sumright, rightsummator];
\zxsumtwo{%
    \begin{ZX}
        \zxX{\phantom{\pi}} \rar & \zxN{}
    \end{ZX}
    }{%
    \begin{ZX}
        \zxX{\pi} \rar & \zxN{}
    \end{ZX}
}[0.0em][\sumheight][1.0][c_{\beta_{21}}][\ii s_{\beta_{21}}]rc

\node at (b31ll.center) [name=sumleft, invisible];
\node at (b31lr.center) [name=sumright, rightsummator];
\zxsumtwo{%
    \begin{ZX}
        \zxX{\phantom{\pi}} \rar & \zxN{}
    \end{ZX}
    }{%
    \begin{ZX}
        \zxX{\pi} \rar & \zxN{}
    \end{ZX}
}[0.0em][\sumheight][1.0][c_{\beta_{31}}][\ii s_{\beta_{31}}]rc

\node[on grid, right=of b11lr, gamma=$\gamma_{11}$, name=g11l];
\node[on grid, right=of b21lr, gamma=$\gamma_{21}$, name=g21l];
\node[on grid, right=of b31lr, gamma=$\gamma_{31}$, name=g31l];

\node[on grid, right=of g11l, name=pi1, zpi] {};
\node[on grid, right=of g21l, name=pi2, invisible] {};
\node[on grid, right=of g31l, name=pi3, zpi];

\node[on grid, right=of pi1, name=e1ll, zspider=2pt] {};
\node[on grid, right=of pi2, name=e2ll, xspider=2pt] {};
\node[on grid, right=of pi3, name=e3ll, invisible];

\node[on grid, right=of e1ll, name=e1lm, zspider=2pt] {};
\node[on grid, right=of e2ll, name=e2lm, invisible];
\node[on grid, right=of e3ll, name=e3lm, xspider=2pt] {};

\node[on grid, right=of e1lm, name=e1lr, invisible];
\node[on grid, right=of e2lm, name=e2lr, zspider=2pt] {};
\node[on grid, right=of e3lm, name=e3lr, xspider=2pt] {};

\node[on grid, right=of e1lr, name=p1l, invisible];
\node[on grid, right=of e2lr, name=p2l, gamma=$\frac{\zxMinus\pi}{2}$];
\node[on grid, right=of e3lr, name=p3l, gamma=$\frac{\zxMinus\pi}{2}$];

\node[on grid, right=of p1l, name=b1ml, invisible];
\node[on grid, right=of p2l, name=b2ml, invisible];
\node[on grid, right=of p3l, name=b3ml, invisible];

\node[right=\sumwidthfull of b1ml, name=b1mr, invisible];
\node[right=\sumwidthfull of b2ml, name=b2mr, invisible];
\node[right=\sumwidthfull of b3ml, name=b3mr, invisible];

\node at (b2ml.center) [name=sumleft, leftsummator];
\node at (b2mr.center) [name=sumright, rightsummator];
\zxsumtwo{%
    \begin{ZX}
        \zxN{} \rar & \zxX{\phantom{\pi}} \rar & \zxN{}
    \end{ZX}
    }{%
    \begin{ZX}
        \zxN{} \rar & \zxX{\pi} \rar & \zxN{}
    \end{ZX}
}[0.0em][\sumheight][1.0][c_{\beta_{22}}][\ii s_{\beta_{22}}]

\node at (b3ml.center) [name=sumleft, leftsummator];
\node at (b3mr.center) [name=sumright, rightsummator];
\zxsumtwo{%
    \begin{ZX}
        \zxN{} \rar & \zxX{\phantom{\pi}} \rar & \zxN{}
    \end{ZX}
    }{%
    \begin{ZX}
        \zxN{} \rar & \zxX{\pi} \rar & \zxN{}
    \end{ZX}
}[0.0em][\sumheight][1.0][c_{\beta_{32}}][\ii s_{\beta_{32}}]

\node[on grid, right=of b1mr, name=p1r, invisible];
\node[on grid, right=of b2mr, name=p2r, gamma=$\frac{\pi}{2}$];
\node[on grid, right=of b3mr, name=p3r, gamma=$\frac{\pi}{2}$];

\node[on grid, right=of p1r, name=dots1] {$\cdots$};
\node[on grid, right=of p2r, name=dots2] {$\cdots$};
\node[on grid, right=of p3r, name=dots3] {$\cdots$};

\begin{scope}[on background layer]
    \draw (b11lr) -- (dots1);
    \draw (b21lr) -- (b2ml);
    \draw (b2mr) -- (dots2);
    \draw (b31lr) -- (b3ml);
    \draw (b3mr) -- (dots3);

    \draw (e1ll) -- (e2ll);
    \draw (e1lm) -- (e3lm);
    \draw (e2lr) -- (e3lr);
\end{scope}

            \end{tikzpicture}
        }
    }}
    \\[4ex]
    \stackrel{\spiderrule, \eqref{eqn:zxsumrules_pull}}{=}
    & 
    \frac{1}{{2^3}}
    \vcenter{\hbox{
        \externalize{hw_eff_ansatz_example_04}
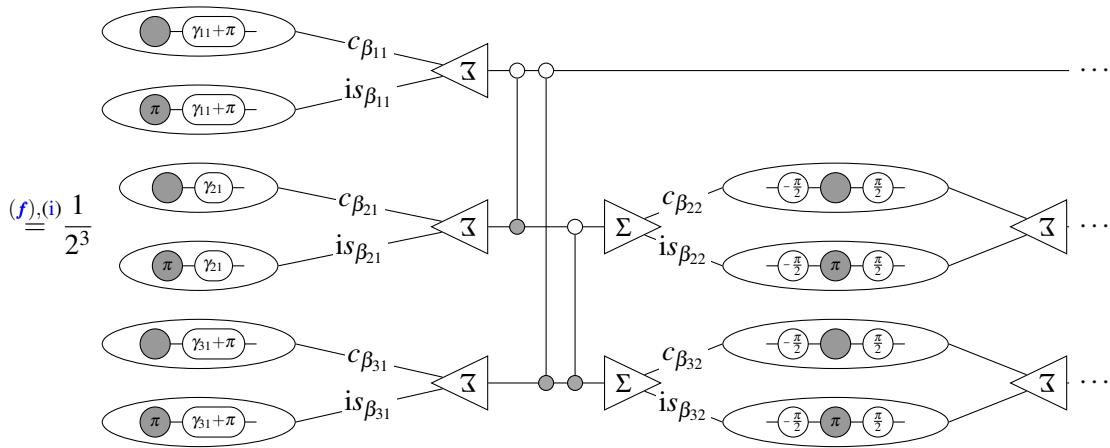
{%
            \begin{tikzpicture}[node distance=12ex and 1.0em]

\def\sumwidth{10em}
\def\sumwidthfull{16em}
\def\sumheight{3ex}

\node[name=b11ll, invisible];
\node[on grid, below=of b11ll, name=b21ll, invisible];
\node[on grid, below=of b21ll, name=b31ll, invisible];

\node[right=\sumwidth of b11ll, name=b11lr, invisible];
\node[right=\sumwidth of b21ll, name=b21lr, invisible];
\node[right=\sumwidth of b31ll, name=b31lr, invisible];

\node at (b11ll.center) [name=sumleft, invisible];
\node at (b11lr.center) [name=sumright, rightsummator];
\zxsumtwo{%
    \begin{ZX}
        \zxX{\phantom{\pi}} \rar & \zxZ{\gamma_{11} + \pi} \rar & \zxN{}
    \end{ZX}
    }{%
    \begin{ZX}
        \zxX{\pi} \rar & \zxZ{\gamma_{11} + \pi} \rar & \zxN{}
    \end{ZX}
}[0.0em][\sumheight][1.0][c_{\beta_{11}}][\ii s_{\beta_{11}}]rc

\node at (b21ll.center) [name=sumleft, invisible];
\node at (b21lr.center) [name=sumright, rightsummator];
\zxsumtwo{%
    \begin{ZX}
        \zxX{\phantom{\pi}} \rar & \zxZ{\gamma_{21}} \rar & \zxN{}
    \end{ZX}
    }{%
    \begin{ZX}
        \zxX{\pi} \rar & \zxZ{\gamma_{21}} \rar & \zxN{}
    \end{ZX}
}[0.0em][\sumheight][1.0][c_{\beta_{21}}][\ii s_{\beta_{21}}]rc

\node at (b31ll.center) [name=sumleft, invisible];
\node at (b31lr.center) [name=sumright, rightsummator];
\zxsumtwo{%
    \begin{ZX}
        \zxX{\phantom{\pi}} \rar & \zxZ{\gamma_{31} + \pi} \rar & \zxN{}
    \end{ZX}
    }{%
    \begin{ZX}
        \zxX{\pi} \rar & \zxZ{\gamma_{31} + \pi} \rar & \zxN{}
    \end{ZX}
}[0.0em][\sumheight][1.0][c_{\beta_{31}}][\ii s_{\beta_{31}}]rc

\node[on grid, right=of b11lr, name=e1ll, zspider=2pt] {};
\node[on grid, right=of b21lr, name=e2ll, xspider=2pt] {};
\node[on grid, right=of b31lr, name=e3ll, invisible];

\node[on grid, right=of e1ll, name=e1lm, zspider=2pt] {};
\node[on grid, right=of e2ll, name=e2lm, invisible];
\node[on grid, right=of e3ll, name=e3lm, xspider=2pt] {};

\node[on grid, right=of e1lm, name=e1lr, invisible];
\node[on grid, right=of e2lm, name=e2lr, zspider=2pt] {};
\node[on grid, right=of e3lm, name=e3lr, xspider=2pt] {};

\node[on grid, right=of e1lr, name=b1ml, invisible];
\node[on grid, right=of e2lr, name=b2ml, invisible];
\node[on grid, right=of e3lr, name=b3ml, invisible];

\node[right=\sumwidthfull of b1ml, name=b1mr, invisible];
\node[right=\sumwidthfull of b2ml, name=b2mr, invisible];
\node[right=\sumwidthfull of b3ml, name=b3mr, invisible];

\node at (b2ml.center) [name=sumleft, leftsummator];
\node at (b2mr.center) [name=sumright, rightsummator];
\zxsumtwo{%
    \begin{ZX}
        \zxN{} \rar & \zxFracZ-{\pi}{2} \rar & \zxX{\phantom{\pi}} \rar & \zxFracZ{\pi}{2} \rar & \zxN{}
    \end{ZX}
    }{%
    \begin{ZX}
        \zxN{} \rar & \zxFracZ-{\pi}{2} \rar & \zxX{\pi} \rar & \zxFracZ{\pi}{2} \rar & \zxN{}
    \end{ZX}
}[0.0em][\sumheight][1.0][c_{\beta_{22}}][\ii s_{\beta_{22}}]

\node at (b3ml.center) [name=sumleft, leftsummator];
\node at (b3mr.center) [name=sumright, rightsummator];
\zxsumtwo{%
    \begin{ZX}
        \zxN{} \rar & \zxFracZ-{\pi}{2} \rar & \zxX{\phantom{\pi}} \rar & \zxFracZ{\pi}{2} \rar & \zxN{}
    \end{ZX}
    }{%
    \begin{ZX}
        \zxN{} \rar & \zxFracZ-{\pi}{2} \rar & \zxX{\pi} \rar & \zxFracZ{\pi}{2} \rar & \zxN{}
    \end{ZX}
}[0.0em][\sumheight][1.0][c_{\beta_{32}}][\ii s_{\beta_{32}}]

\node[on grid, right=of b1mr, name=dots1] {$\cdots$};
\node[on grid, right=of b2mr, name=dots2] {$\cdots$};
\node[on grid, right=of b3mr, name=dots3] {$\cdots$};

\begin{scope}[on background layer]
    \draw (b11lr) -- (dots1);
    \draw (b21lr) -- (b2ml);
    \draw (b2mr) -- (dots2);
    \draw (b31lr) -- (b3ml);
    \draw (b3mr) -- (dots3);

    \draw (e1ll) -- (e2ll);
    \draw (e1lm) -- (e3lm);
    \draw (e2lr) -- (e3lr);
\end{scope}

            \end{tikzpicture}
        }
    }}
    \\[4ex]
    \stackrel{\substack{\copyrule\\ \pirule \\ \spiderrule}}{=}
    & 
    \frac{1}{{2^3}}
    \vcenter{\hbox{
        \externalize{hw_eff_ansatz_example_05}
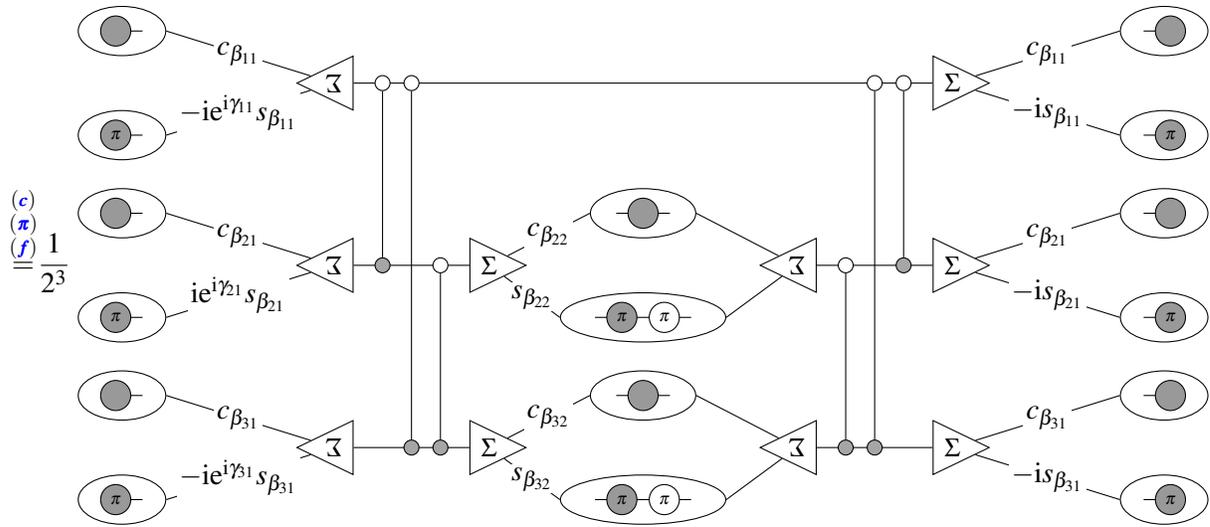
{%
            \begin{tikzpicture}[node distance=14ex and 1.0em]

\def\sumwidth{8em}
\def\sumwidthfull{12em}
\def\sumheight{4ex}

\node[name=b11ll, invisible];
\node[on grid, below=of b11ll, name=b21ll, invisible];
\node[on grid, below=of b21ll, name=b31ll, invisible];

\node[right=\sumwidth of b11ll, name=b11lr, invisible];
\node[right=\sumwidth of b21ll, name=b21lr, invisible];
\node[right=\sumwidth of b31ll, name=b31lr, invisible];

\node at (b11ll.center) [name=sumleft, invisible];
\node at (b11lr.center) [name=sumright, rightsummator];
\zxsumtwo{%
    \begin{ZX}
        \zxX{\phantom{\pi}} \rar & \zxN{}
    \end{ZX}
    }{%
    \begin{ZX}
        \zxX{\pi} \rar & \zxN{}
    \end{ZX}
}[0.0em][\sumheight][1.0][c_{\beta_{11}}][-\ii \ee^{\ii \gamma_{11}}s_{\beta_{11}}]rc

\node at (b21ll.center) [name=sumleft, invisible];
\node at (b21lr.center) [name=sumright, rightsummator];
\zxsumtwo{%
    \begin{ZX}
        \zxX{\phantom{\pi}} \rar & \zxN{}
    \end{ZX}
    }{%
    \begin{ZX}
        \zxX{\pi} \rar & \zxN{}
    \end{ZX}
}[0.0em][\sumheight][1.0][c_{\beta_{21}}][\ii \ee^{\ii \gamma_{21}} s_{\beta_{21}}]rc

\node at (b31ll.center) [name=sumleft, invisible];
\node at (b31lr.center) [name=sumright, rightsummator];
\zxsumtwo{%
    \begin{ZX}
        \zxX{\phantom{\pi}} \rar & \zxN{}
    \end{ZX}
    }{%
    \begin{ZX}
        \zxX{\pi} \rar & \zxN{}
    \end{ZX}
}[0.0em][\sumheight][1.0][c_{\beta_{31}}][-\ii \ee^{\ii \gamma_{31}} s_{\beta_{31}}]rc

\node[on grid, right=of b11lr, name=e1ll, zspider=2pt] {};
\node[on grid, right=of b21lr, name=e2ll, xspider=2pt] {};
\node[on grid, right=of b31lr, name=e3ll, invisible];

\node[on grid, right=of e1ll, name=e1lm, zspider=2pt] {};
\node[on grid, right=of e2ll, name=e2lm, invisible];
\node[on grid, right=of e3ll, name=e3lm, xspider=2pt] {};

\node[on grid, right=of e1lm, name=e1lr, invisible];
\node[on grid, right=of e2lm, name=e2lr, zspider=2pt] {};
\node[on grid, right=of e3lm, name=e3lr, xspider=2pt] {};

\node[on grid, right=of e1lr, name=b1ml, invisible];
\node[on grid, right=of e2lr, name=b2ml, invisible];
\node[on grid, right=of e3lr, name=b3ml, invisible];

\node[right=\sumwidthfull of b1ml, name=b1mr, invisible];
\node[right=\sumwidthfull of b2ml, name=b2mr, invisible];
\node[right=\sumwidthfull of b3ml, name=b3mr, invisible];

\node at (b2ml.center) [name=sumleft, leftsummator];
\node at (b2mr.center) [name=sumright, rightsummator];
\zxsumtwo{%
    \begin{ZX}
        \zxN{} \rar & \zxX{\phantom{\pi}} \rar & \zxN{}
    \end{ZX}
    }{%
    \begin{ZX}
        \zxN{} \rar & \zxX{\pi} \rar & \zxZ{\pi} \rar & \zxN{}
    \end{ZX}
}[0.0em][\sumheight][1.0][c_{\beta_{22}}][s_{\beta_{22}}]

\node at (b3ml.center) [name=sumleft, leftsummator];
\node at (b3mr.center) [name=sumright, rightsummator];
\zxsumtwo{%
    \begin{ZX}
        \zxN{} \rar & \zxX{\phantom{\pi}} \rar & \zxN{}
    \end{ZX}
    }{%
    \begin{ZX}
        \zxN{} \rar & \zxX{\pi} \rar & \zxZ{\pi} \rar & \zxN{}
    \end{ZX}
}[0.0em][\sumheight][1.0][c_{\beta_{32}}][s_{\beta_{32}}]

\node[on grid, right=of b1mr, name=e1rl, invisible];
\node[on grid, right=of b2mr, name=e2rl, zspider=2pt] {};
\node[on grid, right=of b3mr, name=e3rl, xspider=2pt] {};

\node[on grid, right=of e1rl, name=e1rm, zspider=2pt] {};
\node[on grid, right=of e2rl, name=e2rm, invisible];
\node[on grid, right=of e3rl, name=e3rm, xspider=2pt] {};

\node[on grid, right=of e1rm, name=e1rr, zspider=2pt] {};
\node[on grid, right=of e2rm, name=e2rr, xspider=2pt] {};
\node[on grid, right=of e3rm, name=e3rr, invisible];

\node[on grid, right=of e1rr, name=b11rl, invisible];
\node[on grid, right=of e2rr, name=b21rl, invisible];
\node[on grid, right=of e3rr, name=b31rl, invisible];

\node[right=\sumwidth of b11rl, name=b11rr, invisible];
\node[right=\sumwidth of b21rl, name=b21rr, invisible];
\node[right=\sumwidth of b31rl, name=b31rr, invisible];

\node at (b11rl.center) [name=sumleft, leftsummator];
\node at (b11rr.center) [name=sumright, invisible];
\zxsumtwo{%
    \begin{ZX}
        \zxN{} \rar & \zxX{\phantom{\pi}}
    \end{ZX}
    }{%
    \begin{ZX}
        \zxN{} \rar & \zxX{\pi} 
    \end{ZX}
}[0.0em][\sumheight][1.0][c_{\beta_{11}}][-\ii s_{\beta_{11}}]c

\node at (b21rl.center) [name=sumleft, leftsummator];
\node at (b21rr.center) [name=sumright, invisible];
\zxsumtwo{%
    \begin{ZX}
        \zxN{} \rar & \zxX{\phantom{\pi}}
    \end{ZX}
    }{%
    \begin{ZX}
        \zxN{} \rar & \zxX{\pi} 
    \end{ZX}
}[0.0em][\sumheight][1.0][c_{\beta_{21}}][-\ii s_{\beta_{21}}]c

\node at (b31rl.center) [name=sumleft, leftsummator];
\node at (b31rr.center) [name=sumright, invisible];
\zxsumtwo{%
    \begin{ZX}
        \zxN{} \rar & \zxX{\phantom{\pi}}
    \end{ZX}
    }{%
    \begin{ZX}
        \zxN{} \rar & \zxX{\pi} 
    \end{ZX}
}[0.0em][\sumheight][1.0][c_{\beta_{31}}][-\ii s_{\beta_{31}}]c

\begin{scope}[on background layer]
    \draw (b11lr) -- (b11rl);
    \draw (b21lr) -- (b2ml);
    \draw (b2mr) -- (b21rl);
    \draw (b31lr) -- (b3ml);
    \draw (b3mr) -- (b31rl);

    \draw (e1ll) -- (e2ll);
    \draw (e1lm) -- (e3lm);
    \draw (e2lr) -- (e3lr);

    \draw (e2rl) -- (e3rl);
    \draw (e1rm) -- (e3rm);
    \draw (e1rr) -- (e2rr);
\end{scope}

            \end{tikzpicture}
        }
    }}
    \\[4ex]
    \stackrel{\eqref{eqn:zxidentity_hweffansatz}}{=}
    &
c_{\beta_{11}}^{2} c_{\beta_{21}}^{2} c_{\beta_{22}} c_{\beta_{31}}^{2} c_{\beta_{32}} 
- \ii c_{\beta_{11}}^{2} c_{\beta_{21}}^{2} c_{\beta_{22}} c_{\beta_{31}} s_{\beta_{31}} s_{\beta_{32}} \ee^{\ii \gamma_{31}} 
\\
&
+ \ii c_{\beta_{11}}^{2} c_{\beta_{21}}^{2} c_{\beta_{22}} c_{\beta_{31}} s_{\beta_{31}} s_{\beta_{32}} 
- c_{\beta_{11}}^{2} c_{\beta_{21}}^{2} c_{\beta_{22}} c_{\beta_{32}} s_{\beta_{31}}^{2} \ee^{\ii \gamma_{31}} 
\\
&
+ \ii c_{\beta_{11}}^{2} c_{\beta_{21}} c_{\beta_{31}}^{2} s_{\beta_{21}} s_{\beta_{22}} s_{\beta_{32}} \ee^{\ii \gamma_{21}} 
- \ii c_{\beta_{11}}^{2} c_{\beta_{21}} c_{\beta_{31}}^{2} s_{\beta_{21}} s_{\beta_{22}} s_{\beta_{32}} 
\\
&
+ c_{\beta_{11}}^{2} c_{\beta_{21}} c_{\beta_{31}} c_{\beta_{32}} s_{\beta_{21}} s_{\beta_{22}} s_{\beta_{31}} \ee^{\ii \gamma_{21}} \ee^{\ii \gamma_{31}} 
+ c_{\beta_{11}}^{2} c_{\beta_{21}} c_{\beta_{31}} c_{\beta_{32}} s_{\beta_{21}} s_{\beta_{22}} s_{\beta_{31}} \ee^{\ii \gamma_{21}} 
\\
&
+ c_{\beta_{11}}^{2} c_{\beta_{21}} c_{\beta_{31}} c_{\beta_{32}} s_{\beta_{21}} s_{\beta_{22}} s_{\beta_{31}} \ee^{\ii \gamma_{31}} 
+ c_{\beta_{11}}^{2} c_{\beta_{21}} c_{\beta_{31}} c_{\beta_{32}} s_{\beta_{21}} s_{\beta_{22}} s_{\beta_{31}} 
\\
&
+ \ii c_{\beta_{11}}^{2} c_{\beta_{21}} s_{\beta_{21}} s_{\beta_{22}} s_{\beta_{31}}^{2} s_{\beta_{32}} \ee^{\ii \gamma_{21}} \ee^{\ii \gamma_{31}} 
- \ii c_{\beta_{11}}^{2} c_{\beta_{21}} s_{\beta_{21}} s_{\beta_{22}} s_{\beta_{31}}^{2} s_{\beta_{32}} \ee^{\ii \gamma_{31}} 
\\
&
+ c_{\beta_{11}}^{2} c_{\beta_{22}} c_{\beta_{31}}^{2} c_{\beta_{32}} s_{\beta_{21}}^{2} \ee^{\ii \gamma_{21}} 
+ \ii c_{\beta_{11}}^{2} c_{\beta_{22}} c_{\beta_{31}} s_{\beta_{21}}^{2} s_{\beta_{31}} s_{\beta_{32}} \ee^{\ii \gamma_{21}} \ee^{\ii \gamma_{31}} 
\\
&
- \ii c_{\beta_{11}}^{2} c_{\beta_{22}} c_{\beta_{31}} s_{\beta_{21}}^{2} s_{\beta_{31}} s_{\beta_{32}} \ee^{\ii \gamma_{21}} 
- c_{\beta_{11}}^{2} c_{\beta_{22}} c_{\beta_{32}} s_{\beta_{21}}^{2} s_{\beta_{31}}^{2} \ee^{\ii \gamma_{21}} \ee^{\ii \gamma_{31}} 
\\
&
- c_{\beta_{21}}^{2} c_{\beta_{22}} c_{\beta_{31}}^{2} s_{\beta_{11}}^{2} s_{\beta_{32}} \ee^{\ii \gamma_{11}} 
+ \ii c_{\beta_{21}}^{2} c_{\beta_{22}} c_{\beta_{31}} c_{\beta_{32}} s_{\beta_{11}}^{2} s_{\beta_{31}} \ee^{\ii \gamma_{11}} \ee^{\ii \gamma_{31}} 
\\
&
+ \ii c_{\beta_{21}}^{2} c_{\beta_{22}} c_{\beta_{31}} c_{\beta_{32}} s_{\beta_{11}}^{2} s_{\beta_{31}} \ee^{\ii \gamma_{11}} 
- c_{\beta_{21}}^{2} c_{\beta_{22}} s_{\beta_{11}}^{2} s_{\beta_{31}}^{2} s_{\beta_{32}} \ee^{\ii \gamma_{11}} \ee^{\ii \gamma_{31}} 
\\
&
+ \ii c_{\beta_{21}} c_{\beta_{31}}^{2} c_{\beta_{32}} s_{\beta_{11}}^{2} s_{\beta_{21}} s_{\beta_{22}} \ee^{\ii \gamma_{11}} \ee^{\ii \gamma_{21}} 
+ \ii c_{\beta_{21}} c_{\beta_{31}}^{2} c_{\beta_{32}} s_{\beta_{11}}^{2} s_{\beta_{21}} s_{\beta_{22}} \ee^{\ii \gamma_{11}} 
\\
&
+ c_{\beta_{21}} c_{\beta_{31}} s_{\beta_{11}}^{2} s_{\beta_{21}} s_{\beta_{22}} s_{\beta_{31}} s_{\beta_{32}} \ee^{\ii \gamma_{11}} \ee^{\ii \gamma_{21}} \ee^{\ii \gamma_{31}} 
- c_{\beta_{21}} c_{\beta_{31}} s_{\beta_{11}}^{2} s_{\beta_{21}} s_{\beta_{22}} s_{\beta_{31}} s_{\beta_{32}} \ee^{\ii \gamma_{11}} \ee^{\ii \gamma_{21}} 
\\
&
- c_{\beta_{21}} c_{\beta_{31}} s_{\beta_{11}}^{2} s_{\beta_{21}} s_{\beta_{22}} s_{\beta_{31}} s_{\beta_{32}} \ee^{\ii \gamma_{11}} \ee^{\ii \gamma_{31}} 
+ c_{\beta_{21}} c_{\beta_{31}} s_{\beta_{11}}^{2} s_{\beta_{21}} s_{\beta_{22}} s_{\beta_{31}} s_{\beta_{32}} \ee^{\ii \gamma_{11}} 
\\
&
- \ii c_{\beta_{21}} c_{\beta_{32}} s_{\beta_{11}}^{2} s_{\beta_{21}} s_{\beta_{22}} s_{\beta_{31}}^{2} \ee^{\ii \gamma_{11}} \ee^{\ii \gamma_{21}} \ee^{\ii \gamma_{31}} 
- \ii c_{\beta_{21}} c_{\beta_{32}} s_{\beta_{11}}^{2} s_{\beta_{21}} s_{\beta_{22}} s_{\beta_{31}}^{2} \ee^{\ii \gamma_{11}} \ee^{\ii \gamma_{31}} 
\\
&
+ c_{\beta_{22}} c_{\beta_{31}}^{2} s_{\beta_{11}}^{2} s_{\beta_{21}}^{2} s_{\beta_{32}} \ee^{\ii \gamma_{11}} \ee^{\ii \gamma_{21}} 
+ \ii c_{\beta_{22}} c_{\beta_{31}} c_{\beta_{32}} s_{\beta_{11}}^{2} s_{\beta_{21}}^{2} s_{\beta_{31}} \ee^{\ii \gamma_{11}} \ee^{\ii \gamma_{21}} \ee^{\ii \gamma_{31}} 
\\
&
+ \ii c_{\beta_{22}} c_{\beta_{31}} c_{\beta_{32}} s_{\beta_{11}}^{2} s_{\beta_{21}}^{2} s_{\beta_{31}} \ee^{\ii \gamma_{11}} \ee^{\ii \gamma_{21}} 
+ c_{\beta_{22}} s_{\beta_{11}}^{2} s_{\beta_{21}}^{2} s_{\beta_{31}}^{2} s_{\beta_{32}} \ee^{\ii \gamma_{11}} \ee^{\ii \gamma_{21}} \ee^{\ii \gamma_{31}} \label{eq:hwexpeczzformula}
 \, ,
\end{align}
where in the last step we have used the identity 
\begin{align}
        \vcenter{\hbox{
            \externalizezx{hw_eff_ansatz_side_calculation_lhs}{%
                \begin{ZX}[ampersand replacement=\&]
    \zxX{\ell_1 \pi} \rar 
    \& \zxZ{} \dar \rar \& \zxZ{} \dar \rar \& \zxN{} \rar 
    \& \zxN{} \rar
    \& \zxN{} \rar
    \& \zxN{} \rar \& \zxZ{} \dar \rar \& \zxZ{} \dar \rar 
    \& \zxX{r_1 \pi} 
    \\
    \zxX{\ell_2 \pi} \rar 
    \& \zxX{} \rar \& \zxN{} \dar \rar \& \zxZ{} \dar \rar 
    \& \zxX{m_2 \pi} \rar
    \& \zxZ{m_2 \pi} \rar
    \& \zxZ{} \dar \rar \& \zxN{} \dar \rar \& \zxX{}  \rar 
    \& \zxX{r_2 \pi} 
    \\
    \zxX{\ell_3 \pi} \rar 
    \& \zxN{} \rar \& \zxX{} \rar \& \zxX{} \rar 
    \& \zxX{m_3 \pi} \rar
    \& \zxZ{m_3 \pi} \rar
    \& \zxX{} \rar \& \zxX{} \rar \& \zxN{} \rar 
    \& \zxX{r_3 \pi} 
\end{ZX}

            }
        }}
        & =
        \frac{f_{m_2 m_3}^{r_1 r_2 r_3}}{2^3}
        \vcenter{\hbox{
            \externalizezx{hw_eff_ansatz_side_calculation_rhs}{
                \begin{ZX}[ampersand replacement=\&]
    \zxX{(\ell_1 + m_2 + \ell_3 + m_3 + r_3)\pi}
    \& 
    \& \zxX{(r_1 + m_2 + \ell_3 + m_3 + r_3)\pi}
    \\[\zxwRow, \zxwRow]
    \zxX[a=l3]{(\ell_2 + m_2 + r_2) \pi}
    \&
    \& 
\end{ZX}

            }
        }}
        \\
        &=
        \begin{cases}
            0   & \text{if }    \quad   
            \begin{aligned}
                        & \ell_1 + m_2 + \ell_3 + m_3 + r_3  \text{ odd} \\
                \vee \, & r_1 + m_2 + \ell_3 + m_3 + r_3  \text{ odd} \\
                \vee \, & \ell_2 + m_2 + r_2  \text{ odd}
            \end{aligned} \\
            f_{m_2 m_3}^{r_1 r_2 r_3}
            & \text{else}     
        \end{cases}%
        \label{eqn:zxidentity_hweffansatz} \, ,
\end{align}
with 
$
\ell_1, 
\ell_2, 
\ell_3,
m_2,
m_3,
r_1,
r_2,
r_3
\in \{0, 1\}^{\times 8}
$
and
$
f_{m_2 m_3}^{r_1 r_2 r_3}
:=
(-1)^{m_2 r_1 + (m_2 \oplus m_3) r_2 + m_3 r_3}
$,
which is the proven in Appendix~\ref{sec:app_proof_zxidentity_hweffansatz}.
Observe that in the second step above any dependency on the parameters $\gamma_{12},\gamma_{22}$, and $\gamma_{32}$ was immediately 
shown to cancel out (due to commuting with the diagonal cost Hamiltonian),
and likewise for $\beta_{12}$ (due to the locality of $Z_2Z_3$). 
Similar simplifications are often easily obtained from the diagrammatic perspective.

The formula \eqref{eq:hwexpeczzformula} exemplifies the significant difficulty faced in obtaining analytical results for 
PQCs, even for relatively small ans\"atze. 
Nevertheless, in our analysis the complexity remained manageable with the diagrammatic approach up until the very last step, were we applied a simple numerical procedure to collect all the surviving terms (according to~\eqref{eqn:zxidentity_hweffansatz}) of the contraction. 
Different hardware-efficient ans\"atze may be similarly considered, including ones tailored to specific hardware topology.
As mentioned, for deeper or more complicated ans\"atze, analysis may 
be aided or automated through implementation in software. 
Here \eqref{eq:hwexpeczz} demonstrates how diagrammatic approaches can yield more compact representations of expectation values (as compared to \eqref{eq:hwexpeczzformula}).


\subsection{QAOA for MaxCut on Ring graphs}%
\label{sec:example_applications_qaoa_ring}
We consider the simple example of the one-dimensional \say{ring-of-disagrees}, i.e.,  2-regular connected graphs, and rederive the 
QAOA$_1$ expectation value as previously shown in~\cite{farhi2014quantum,wang2018quantum}. 
First consider the case of QAOA with arbitrary number of layers~$p$, with $n\gg p$. 
From the problem symmetry, it suffices to consider the expectation value of a single edge term $\langle Z_iZ_{i+1} \rangle_{QAOA_p}$. 
Applying the lightcone rule~\eqref{eqn:light_cone_rule}, 
the outermost 
reduced phase-separation layer%
~\eqref{eqn:ps_layer_definition}
reads
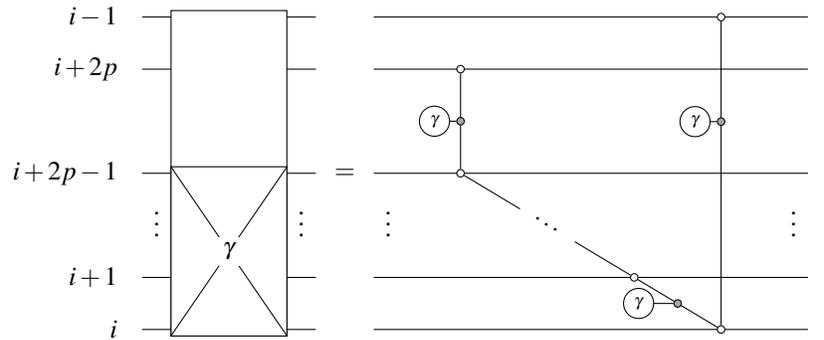
\begin{equation}
    \vcenter{\hbox{
        \externalize{qaoa_ring_ps_layer}{%

        }
    }} \label{eqn:qaoa_ring_ps_layer} \, .
\end{equation}
Hence for 
the QAOA$_p$ expectation value we obtain 
\begin{align}
    &\langle Z_i Z_{i+1} \rangle_{\text{QAOA}_p} 
    \\[4ex]
    &=  
    \frac{1}{2^{2(p+1)}}
    \vcenter{\hbox{
        \externalize{qaoa_ring_exp_val_general_01}{%
            %
        }
    }} \,.
\end{align}
Observe how the problem and 
structure 
again appears in the above diagram (i.e., $p$-neighborhoods of the edge $(i, i+1)$ are line graphs). 
Furthermore, the utility of the lightcone rule is clearly demonstrated here.
Continuing for the $p=1$ case, we get
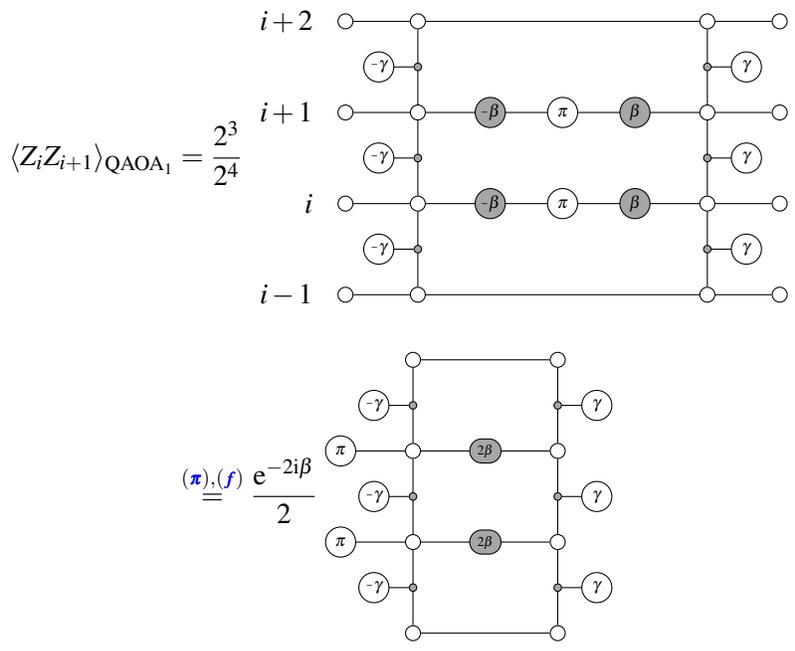
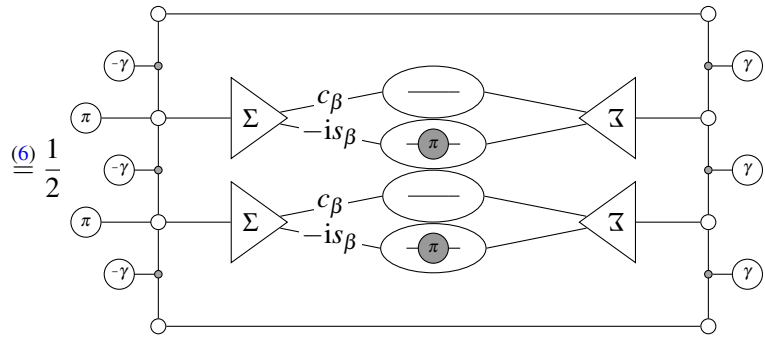
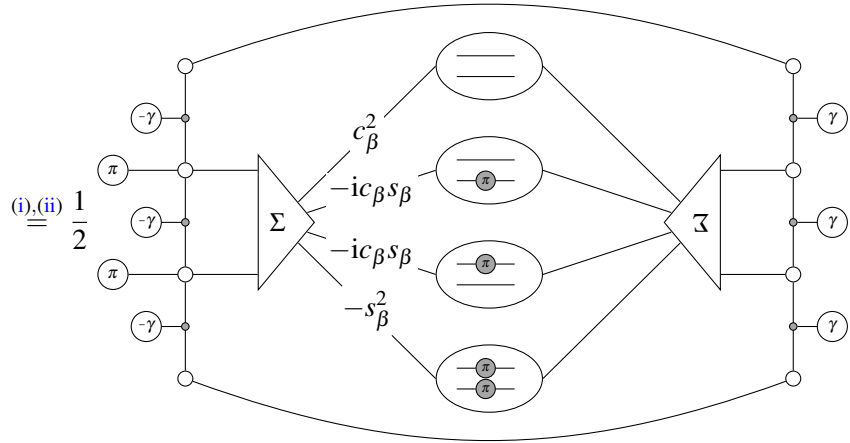
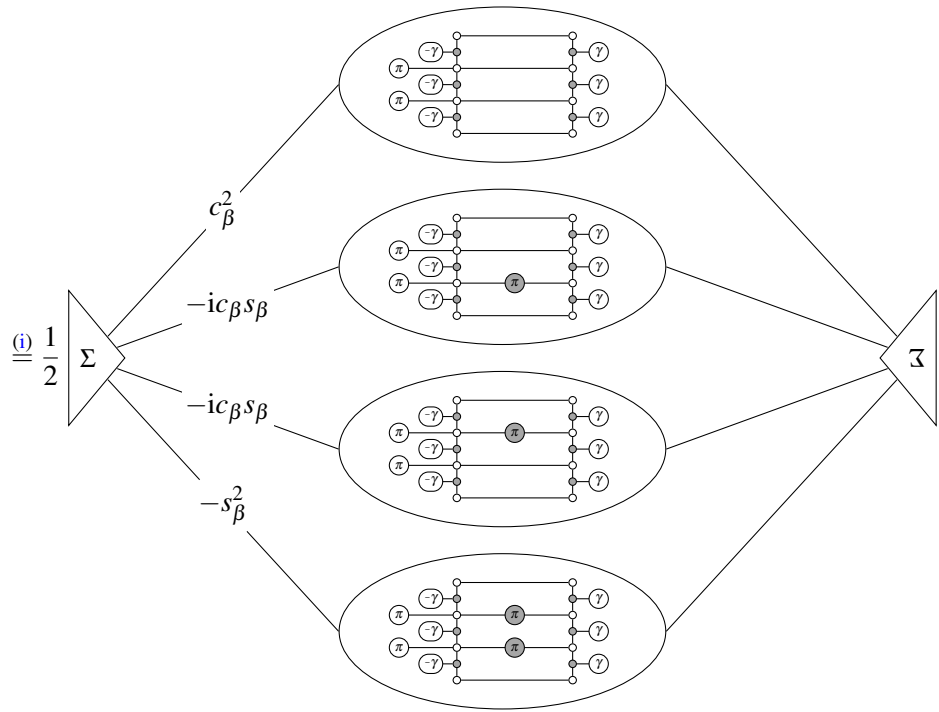
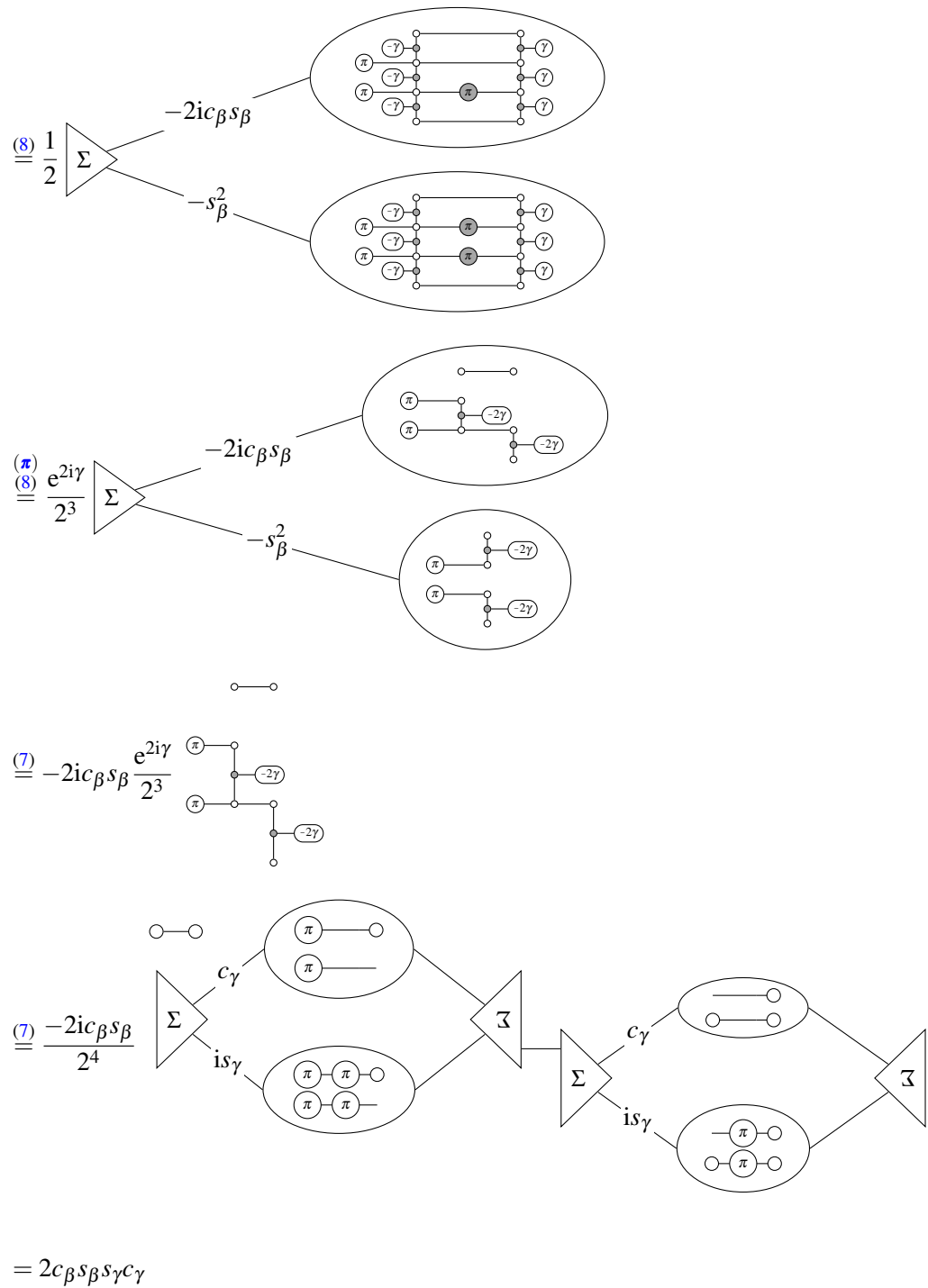
\begin{align}
    \langle Z_i Z_{i+1} \rangle_{\text{QAOA}_1} 
    &=  
    \frac{2^3}{2^4}
    \vcenter{\hbox{
        \externalize{qaoa_ring_exp_val_p1_01}{%
            %
        }
    }}
    \\[2ex]
    &  
    =
    2  c_{\beta} s_{\beta} 
    s_{\gamma} c_{\gamma} 
\end{align}
The result is consistent with that of~\cite[Thm. 1]{wang2018quantum}. 
The expression obtained for $\langle C\rangle$ is easily optimized to reproduce the performance result obtained numerically for the ring of disagrees in~\cite{farhi2014quantum}. 
Similar to the previous examples, here we saw the necessity of our extension for handling X-Z commutations in the third and seventh steps of the derivation above. 

In Appendix~\ref{sec:app_qaoa} we show the same calculation for MaxCut on general graphs, 
as obtained for QAOA$_1$ in \cite[Thm. 1]{wang2018quantum}. Similar techniques may be applied and results 
obtained for a wide variety of important problems, for instance quadratic binary optimization problems of which MaxCut is a special case.
\subsection{QAOA$_1$ for MaxCut on General Graphs}%
\label{sec:app_qaoa}
For the  QAOA expectation value for MaxCut $
\langle C \rangle
= \tfrac{|E|}{2} - \tfrac12\sum_{u, v\in E} \langle Z_u Z_v \rangle
$ 
on general graphs
we need to calculate the contributions $\langle Z_u Z_v \rangle$.
In this section, we perform the calculation for general graphs in the QAOA, $p=1$ case, reproducing results obtained in \cite{wang2018quantum}. 

Following the lightcone rule from Equation~\eqref{eqn:light_cone_rule} we obtain for $Z$-$Z$ terms in the MaxCut QAOA$_1$ expectation value on a general graph $G=(V, E)$ 
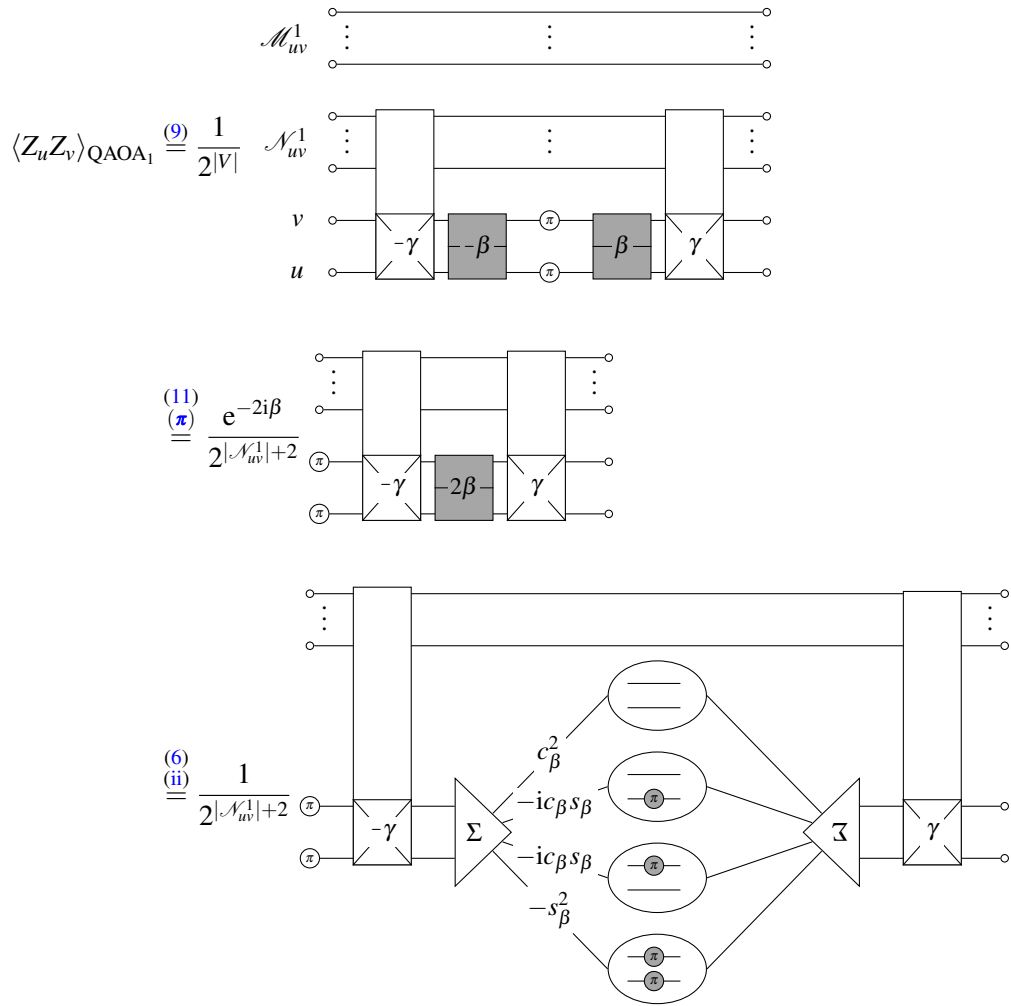
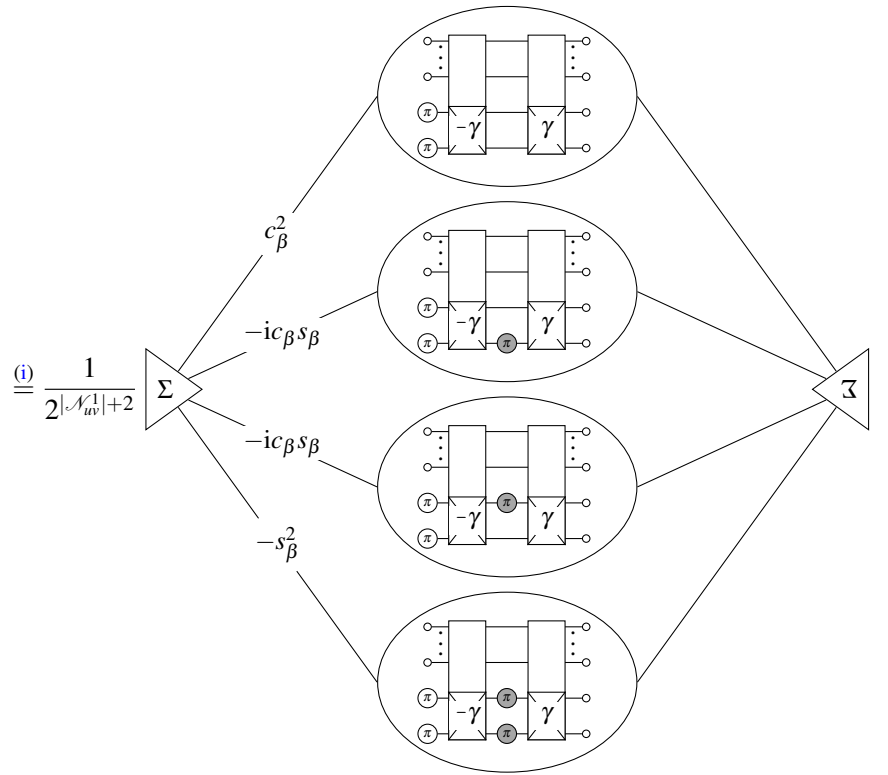
\begin{align}
    \langle Z_u Z_{v} \rangle_{\text{QAOA}_1} 
    &
    \stackrel{\eqref{eqn:light_cone_rule}}{=}
    \frac{1}{2^{|V|}}
    \vcenter{\hbox{
        \externalize{qaoa_max_cut_00}{%

        }
    }} \, .
\end{align}
The first summand vanishes and the second and third are linked by symmetry.
We continue with the second summand (the \textit{I-X}-term)
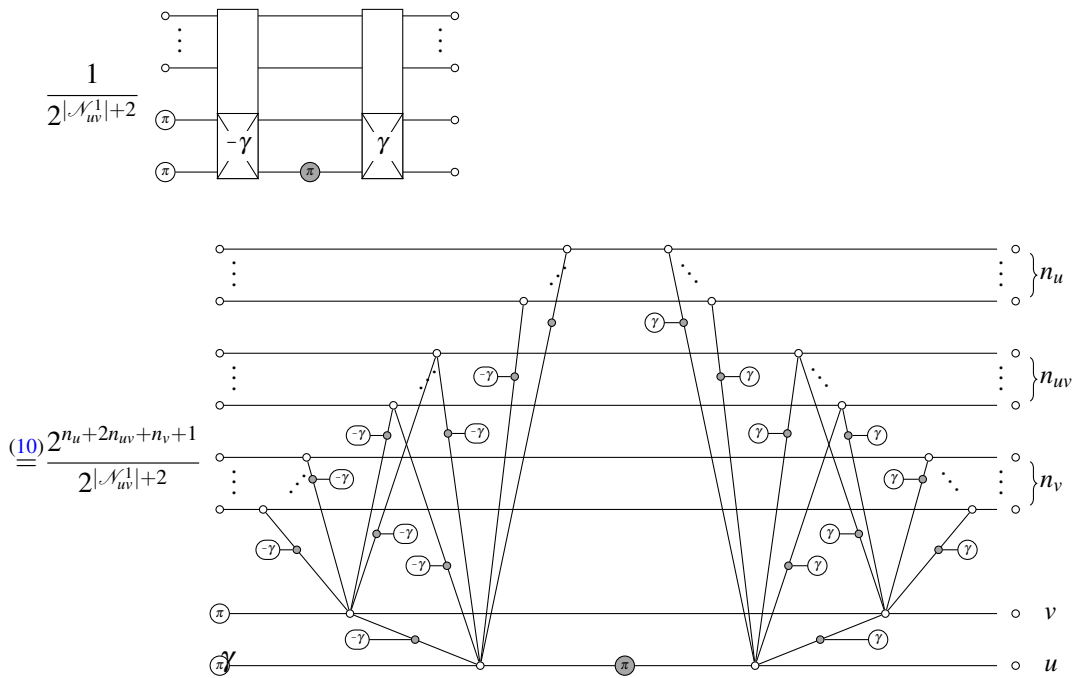
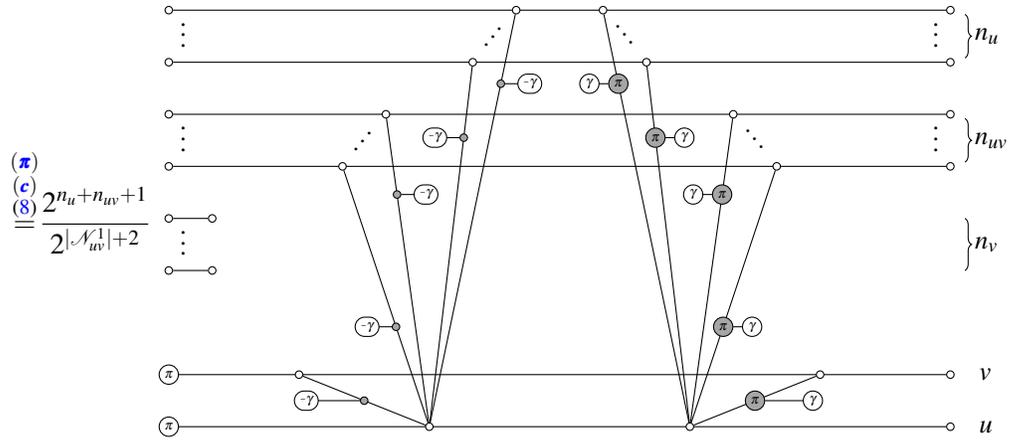
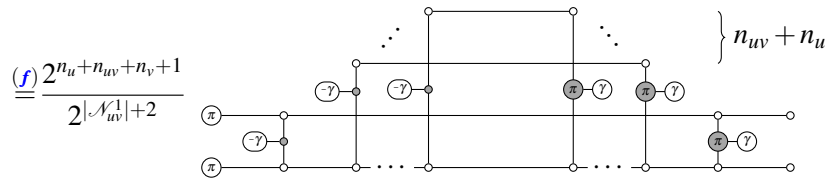
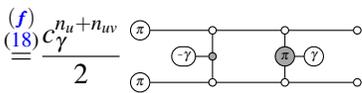
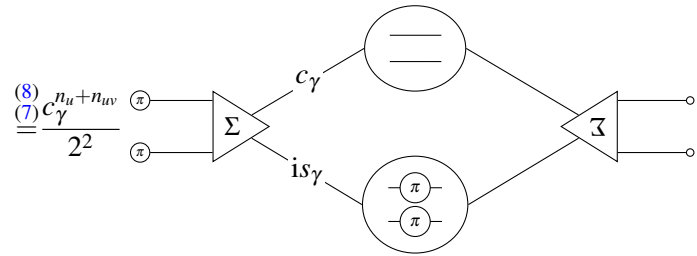
\begin{align}
    &
    \frac{1}{2^{|\mathcal{N}^1_{uv}| + 2}}
    \vcenter{\hbox{
        \externalize{qaoa_max_cut_summand_ix_00}{
            %
        }
    }}
    \\[4ex]
    =
    &
    \ii s_\gamma c_{\gamma}^{n_u + n_{uv}} \, ,
\end{align}
where we have used the size of the exclusive neighborhoods 
$n_u:=|N_{u} \setminus \{N_v \cup u\}|$,
$n_v:=|N_{v} \setminus \{N_u \cup v\}|$,
and the joined neighborhood $n_{uv}:=|N_u \cap N_v|$,
the relation
$|\mathcal{N}^1_{uv}| = n_u + n_{uv} + n_v$, 
as well as
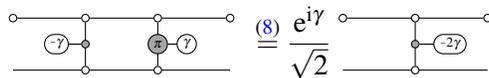
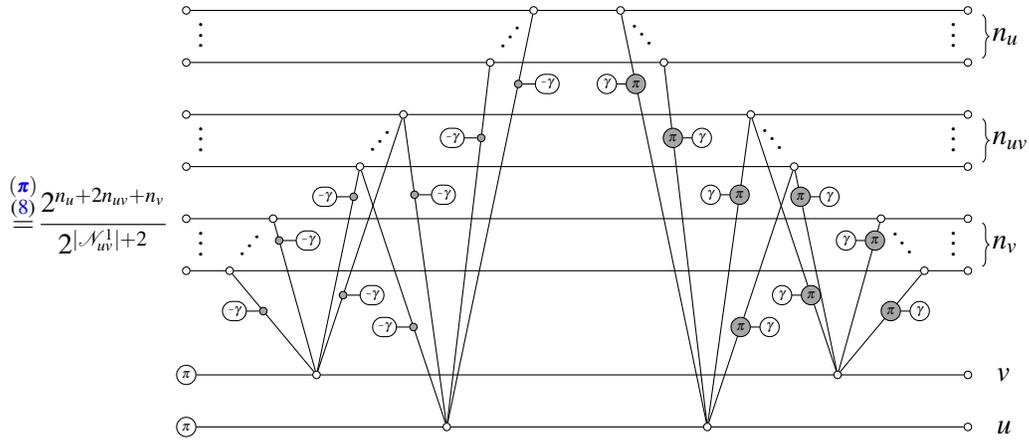
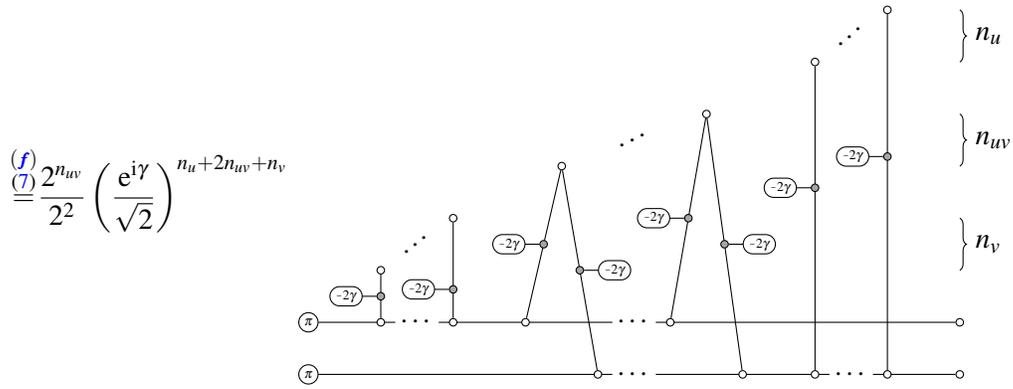
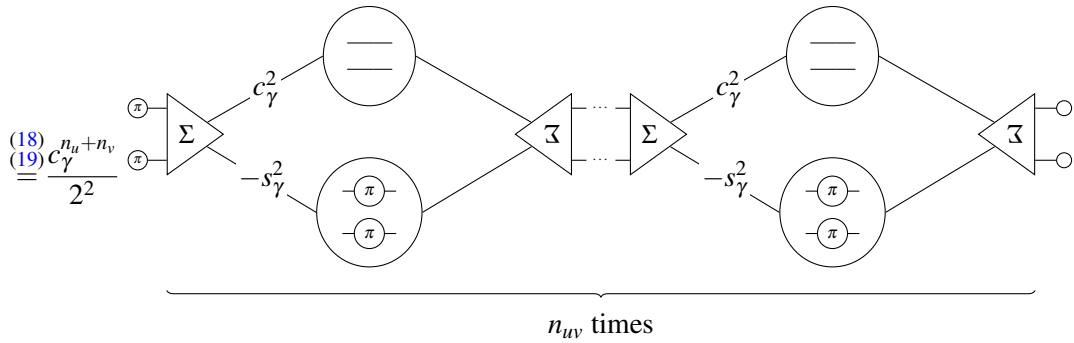
\begin{align}
    \vcenter{\hbox{
        \externalize{qaoa_max_cut_summand_ix_side_calculation_00}{%
            %
        }
    }}\label{eqn:qaoa_max_cut_summand_ix_side_calculation} \,.
\end{align}
Analogously the third summand (the \textit{X-I}-term) can be obtained as
\begin{align}
    \vcenter{\hbox{
        \externalize{qaoa_max_cut_summand_xi_00}{%
            %
        }
    }}
    =
    \ii s_\gamma c_{\gamma}^{n_v + n_{uv}} \, .
\end{align}
The fourth summand (the \textit{X-X}-term) reads
\begin{align}
    &
    \frac{1}{2^{|\mathcal{N}^1_{uv}| + 2}}
    \vcenter{\hbox{
        \externalize{qaoa_max_cut_summand_xx_00}{
            %
        }
    }}
    \\[4ex]
    \stackrel{\eqref{eqn:zxsumrules_product_of_sums}}{=}
    &
    \frac{-c_\gamma^{n_u + n_v}}{2^2}
    \biggl\{
    \biggr.
        \binom{n_{uv}}{1} 
        s_\gamma^2
        c_\gamma^{n_{uv}-2}
        \;\;\;
        \begin{ZX}[ampersand replacement=\&]
               \zxZ{\pi} \rar 
            \& \zxZ{\pi} \rar 
            \&[1.2em] \zxN{} \rar 
            \&[1.2em] \zxN{} \rar 
            \&[1.2em] \zxN{} \rar 
            \&[1.2em] \zxN{} \rar 
            \& \zxZ{}
            \\
               \zxZ{\pi} \rar 
            \& \zxZ{\pi} \rar 
            \&[1.2em] \zxN{} \rar 
            \&[1.2em] \zxN{} \rar 
            \&[1.2em] \zxN{} \rar 
            \&[1.2em] \zxN{} \rar 
            \& \zxZ{}
        \end{ZX}
    \\[3ex]
        &
        \qquad
        \qquad
        +
        \binom{n_{uv}}{3} 
        s_\gamma^6
        c_\gamma^{n_{uv}-6}
        \;\;\;
        \begin{ZX}[ampersand replacement=\&]
               \zxZ{\pi} \rar 
            \& \zxZ{\pi} \rar 
            \& \zxZ{\pi} \rar 
            \& \zxZ{\pi} \rar 
            \&[1.2em] \zxN{} \rar 
            \&[1.2em] \zxN{} \rar 
            \& \zxZ{}
            \\
               \zxZ{\pi} \rar 
            \& \zxZ{\pi} \rar 
            \& \zxZ{\pi} \rar 
            \& \zxZ{\pi} \rar 
            \&[1.2em] \zxN{} \rar 
            \&[1.2em] \zxN{} \rar 
            \& \zxZ{}
        \end{ZX}
    \\[3ex]
        &
        \qquad
        \qquad
        +
        \binom{n_{uv}}{5} 
        s_\gamma^{10}
        c_\gamma^{n_{uv}-10}
        \begin{ZX}[ampersand replacement=\&]
               \zxZ{\pi} \rar 
            \& \zxZ{\pi} \rar 
            \& \zxZ{\pi} \rar 
            \& \zxZ{\pi} \rar 
            \& \zxZ{\pi} \rar 
            \& \zxZ{\pi} \rar 
            \& \zxZ{}
            \\
               \zxZ{\pi} \rar 
            \& \zxZ{\pi} \rar 
            \& \zxZ{\pi} \rar 
            \& \zxZ{\pi} \rar 
            \& \zxZ{\pi} \rar 
            \& \zxZ{\pi} \rar 
            \& \zxZ{}
        \end{ZX}
    \\[3ex]
        &
        \qquad
        \qquad
        +
        \;
        \dots
        \qquad \qquad \qquad \qquad
        \qquad \qquad \qquad \qquad
        \qquad \qquad 
    \biggl.
    \biggr\}
    \\[3ex]
    & = 
    -c_\gamma^{n_u + n_v}
    \sum_{i=1, 3, \dots} \binom{n_{uv}}{i} (s_\gamma^2)^i (c_\gamma^2)^{n_{uv} - i } \, ,
\end{align}
where we have used
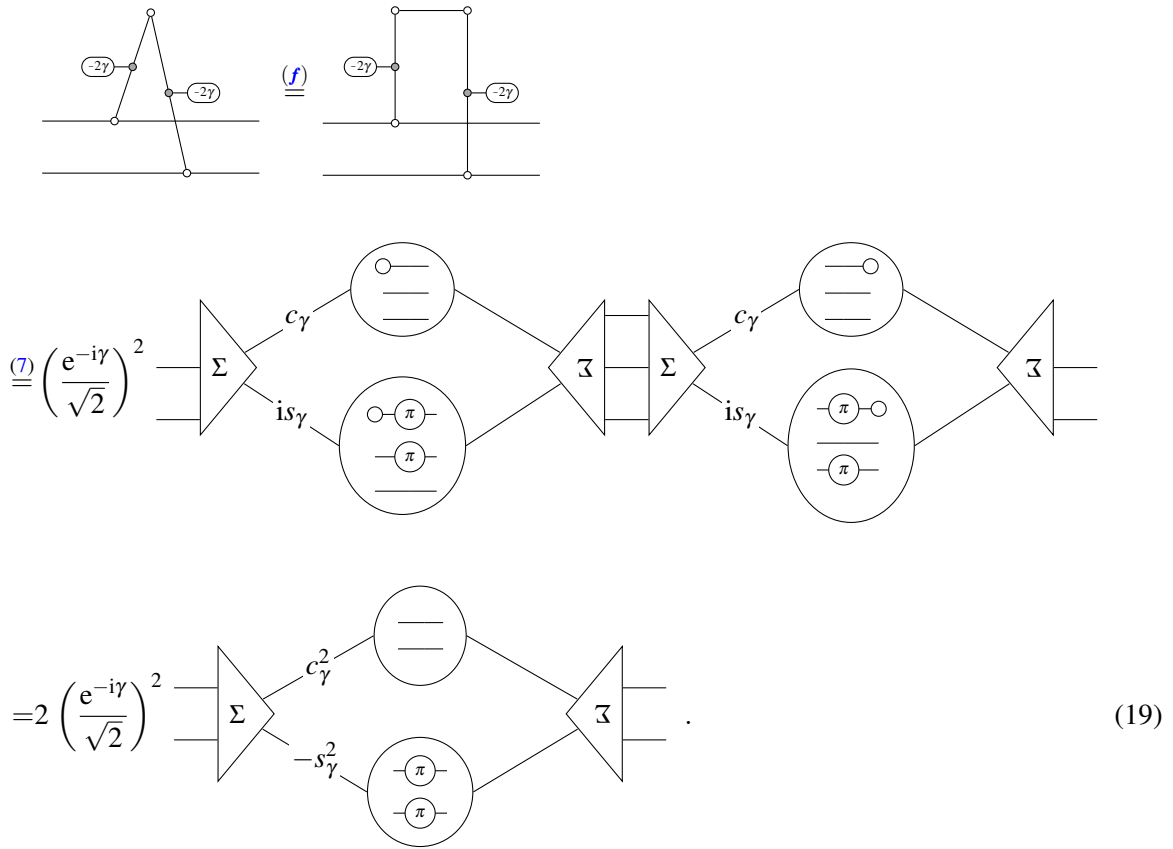
\begin{align}
    &
    \vcenter{\hbox{
        \externalize{qaoa_max_cut_summand_xx_side_calculation_00}{%
            \begin{tikzpicture}[node distance=4ex and 2.5em]
\def\yspace{8ex}

\node[name=u-ll, invisible];
\node[on grid, name=v-ll, above=of u-ll, invisible];

\node[on grid, name=v-l, right=of v-ll, zspider=1pt] {};
\node[name=w-l, above=\yspace of v-l, invisible];
\node[on grid, name=w-r, right=of w-l, invisible];
\node at ($(w-l)!0.5!(w-r)$) [on grid, name=w-m, zspider=1pt] {};
\node[on grid, name=u-r, below right=of v-l, zspider=1pt] {};
\draw (v-l) -- (w-m) node[pos=0.5, xspider, anchor=center] (x) {};
\draw (x) -- +(180:1.2em) node[gammainv=$2\gamma$, zxstyletight];
\draw (u-r) -- (w-m) node[pos=0.5, xspider, anchor=center] (x) {};
\draw (x) -- +(0:1.2em) node[gammainv=$2\gamma$, zxstyletight];

\node[on grid, name=u-rr, right=of u-r, invisible];
\node[on grid, name=v-rr, above=of u-rr, invisible];

\begin{scope}[on background layer]
    \draw (v-ll) -- (v-rr);
    \draw (u-ll) -- (u-rr);
\end{scope}

            \end{tikzpicture}
        }
    }}
    \stackrel{\spiderrule}{=}
    \vcenter{\hbox{
        \externalize{qaoa_max_cut_summand_xx_side_calculation_01}{%
            \begin{tikzpicture}[node distance=4ex and 2.5em]
\def\yspace{8ex}

\node[name=u-ll, invisible];
\node[on grid, name=v-ll, above=of u-ll, invisible];

\node[on grid, name=v-l, right=of v-ll, zspider=1pt] {};
\node[name=w-l, above=\yspace of v-l, zspider=1pt] {};
\node[on grid, name=w-r, right=of w-l, zspider=1pt] {};
\node[on grid, name=u-r, below right=of v-l, zspider=1pt] {};
\draw (v-l) -- (w-l) node[pos=0.5, xspider, anchor=center] (x) {};
\draw (x) -- +(180:1.2em) node[gammainv=$2\gamma$, zxstyletight];
\draw (u-r) -- (w-r) node[pos=0.5, xspider, anchor=center] (x) {};
\draw (x) -- +(0:1.2em) node[gammainv=$2\gamma$, zxstyletight];

\node[on grid, name=u-rr, right=of u-r, invisible];
\node[on grid, name=v-rr, above=of u-rr, invisible];

\begin{scope}[on background layer]
    \draw (w-l) -- (w-r);
    \draw (v-ll) -- (v-rr);
    \draw (u-ll) -- (u-rr);
\end{scope}

            \end{tikzpicture}
        }
    }}
    \\[4ex]
    \stackrel{\eqref{eqn:phase_gadget_as_lincomb}}{=}
    &
    \left(\frac{\ee^{-\ii \gamma}}{\sqrt{2}}\right)^2
    \vcenter{\hbox{
        \externalize{qaoa_max_cut_summand_xx_side_calculation_02}{%
            \begin{tikzpicture}[node distance=4ex and 1.5em]

\def\yspace{8ex}
\def\yscale{2.5}
\def\sumwidth{14em}

\node[name=u-l, invisible];
\node[on grid, name=v-l, above=of u-l, invisible];
\node[on grid, name=w-l, above=of v-l, invisible];

\node[on grid, name=anchorllw, right=of w-l, invisible];
\node[on grid, name=anchorllv, right=of v-l, invisible];
\node[on grid, name=anchorllu, right=of u-l, invisible];
\node[name=anchorlrw, right=\sumwidth of anchorllw.center, invisible];
\node[name=anchorlrv, right=\sumwidth of anchorllv.center, invisible];
\node[name=anchorlru, right=\sumwidth of anchorllu.center, invisible];

\node at (anchorllv.center) [yscale=\yscale, name=sumleft, leftsummator];
\node[right=\sumwidth of sumleft.west, yscale=\yscale, name=sumright, rightsummator];
\zxsumtwo{%
    \begin{ZX}[ampersand replacement=\&]
        \zxZ{} \rar \&[\zxSCol] \zxN{} \rar \&[\zxSCol] \zxN{} \\[\zxSRow]
        \zxN{} \rar \&[\zxSCol] \zxN{} \rar \&[\zxSCol] \zxN{} \\[\zxSRow,\zxSRow]
        \zxN{} \rar \&[\zxSCol] \zxN{} \rar \&[\zxSCol] \zxN{}
    \end{ZX}
    }{%
    \begin{ZX}[ampersand replacement=\&]
        \zxZ{} \rar \& \zxZ{\pi} \rar \& \zxN{} \\
        \zxN{} \rar \& \zxZ{\pi} \rar \& \zxN{} \\[\zxSRow]
        \zxN{} \rar \& \zxN{} \rar \& \zxN{}
    \end{ZX}
}[0.0em][6.0ex][1.0][c_\gamma][\ii s_\gamma]

\node[on grid, name=anchorrlw, right=of anchorlrw, invisible];
\node[on grid, name=anchorrlv, right=of anchorlrv, invisible];
\node[on grid, name=anchorrlu, right=of anchorlru, invisible];
\node[name=anchorrrw, right=\sumwidth of anchorrlw.center, invisible];
\node[name=anchorrrv, right=\sumwidth of anchorrlv.center, invisible];
\node[name=anchorrru, right=\sumwidth of anchorrlu.center, invisible];

\node at (anchorrlv.center) [yscale=\yscale, name=sumleft, leftsummator];
\node[right=\sumwidth of sumleft.west, yscale=\yscale, name=sumright, rightsummator];
\zxsumtwo{%
    \begin{ZX}[ampersand replacement=\&]
        \zxN{} \rar \&[\zxSCol] \zxN{} \rar \&[\zxSCol] \zxZ{} \\[\zxSRow]
        \zxN{} \rar \&[\zxSCol] \zxN{} \rar \&[\zxSCol] \zxN{} \\[\zxSRow,\zxSRow]
        \zxN{} \rar \&[\zxSCol] \zxN{} \rar \&[\zxSCol] \zxN{}
    \end{ZX}
    }{%
    \begin{ZX}[ampersand replacement=\&]
        \zxN{} \rar \& \zxZ{\pi} \rar \& \zxZ{} \\[\zxSRow]
        \zxN{} \rar \& \zxN{} \rar \& \zxN{} \\
        \zxN{} \rar \& \zxZ{\pi} \rar \& \zxN{} \\
    \end{ZX}
}[0.0em][6.0ex][1.0][c_\gamma][\ii s_\gamma]

\node[on grid, name=w-r, right=of anchorrrw, invisible];
\node[on grid, name=v-r, right=of anchorrrv, invisible];
\node[on grid, name=u-r, right=of anchorrru, invisible];

\begin{scope}[on background layer]
    \draw (v-l) -- (anchorllv);
    \draw (u-l) -- (anchorllu);
    \draw (anchorlrw) -- (anchorrlw);
    \draw (anchorlrv) -- (anchorrlv);
    \draw (anchorlru) -- (anchorrlu);
    \draw (anchorrrv) -- (v-r);
    \draw (anchorrru) -- (u-r);
\end{scope}

            \end{tikzpicture}
        }
    }}
    \\[4ex]
    =
    &
    2\, \left(\frac{\ee^{-\ii \gamma}}{\sqrt{2}}\right)^2
    \vcenter{\hbox{
        \externalize{qaoa_max_cut_summand_xx_side_calculation_03}{%
            \begin{tikzpicture}[node distance=2ex and 1.5em]
\def\yspace{8ex}
\def\yscale{2.5}
\def\sumwidth{14em}

\node[name=u-l, invisible];
\node[on grid, name=v-l, above=of u-l, invisible];
\node[on grid, name=w-l, above=of v-l, invisible];

\node[on grid, name=anchorlw, right=of w-l, invisible];
\node[on grid, name=anchorlv, right=of v-l, invisible];
\node[on grid, name=anchorlu, right=of u-l, invisible];
\node[name=anchorrw, right=\sumwidth of anchorlw.center, invisible];
\node[name=anchorrv, right=\sumwidth of anchorlv.center, invisible];
\node[name=anchorru, right=\sumwidth of anchorlu.center, invisible];

\node at (anchorlv.center) [yscale=\yscale, name=sumleft, leftsummator];
\node[right=\sumwidth of sumleft.west, yscale=\yscale, name=sumright, rightsummator];
\zxsumtwo{%
    \begin{ZX}[ampersand replacement=\&]
        \\[\zxSRow]
        \zxN{} \rar \&[\zxSCol,\zxSCol] \zxN{} \rar \&[\zxSCol] \zxN{} \\[\zxSRow,\zxSRow]
        \zxN{} \rar \&[\zxSCol,\zxSCol] \zxN{} \rar \&[\zxSCol] \zxN{}
        \\[\zxSRow]
    \end{ZX}
    }{%
    \begin{ZX}[ampersand replacement=\&]
        \zxN{} \rar \& \zxZ{\pi} \rar \& \zxN{} \\
        \zxN{} \rar \& \zxZ{\pi} \rar \& \zxN{}
    \end{ZX}
}[0.0em][6.0ex][1.0][c^2_\gamma][-s^2_\gamma]

\node[on grid, name=w-r, right=of anchorrw, invisible];
\node[on grid, name=v-r, right=of anchorrv, invisible];
\node[on grid, name=u-r, right=of anchorru, invisible];

\begin{scope}[on background layer]
    \draw (w-l) -- (anchorlw);
    \draw (u-l) -- (anchorlu);
    \draw (anchorrw) -- (w-r);
    \draw (anchorru) -- (u-r);
\end{scope}

            \end{tikzpicture}
        }
    }}\label{eqn:qaoa_max_cut_summand_xx_side_calculation} \, .
\end{align}
Hence, the total $Z$-$Z$-expectation value reads
\begin{align}
    \langle Z_u Z_{v} \rangle_{\text{QAOA}_1} 
    & =
    c_\beta s_\beta
    s_\gamma 
    \left(%
      c_{\gamma}^{n_u + n_{uv}}
    + c_{\gamma}^{n_v + n_{uv}}
    \right)
    +
    c_\gamma^{n_u + n_v}
    s_\beta^2
    \sum_{i=1, 3, \dots} \binom{n_{uv}}{i} (s_\gamma^2)^i (c_\gamma^2)^{n_{uv} - i }. 
\end{align}
This result is consistent with the corresponding QAOA$_1$ performance analysis of \cite{wang2018quantum}; applying the binomial theorem to write the sum above in closed form then leads directly to the result of~\cite[Thm. 1]{wang2018quantum}.

\section{Details on QAOA$_1$ for MaxCut on Simple Graph}%
\label{sec:app_qaoa_example_graph}
We calculate each of the four summands in~\eqref{eqn:qaoa_max_cut_example_terms}.
The first summand (the \textit{I-I}-term) reads
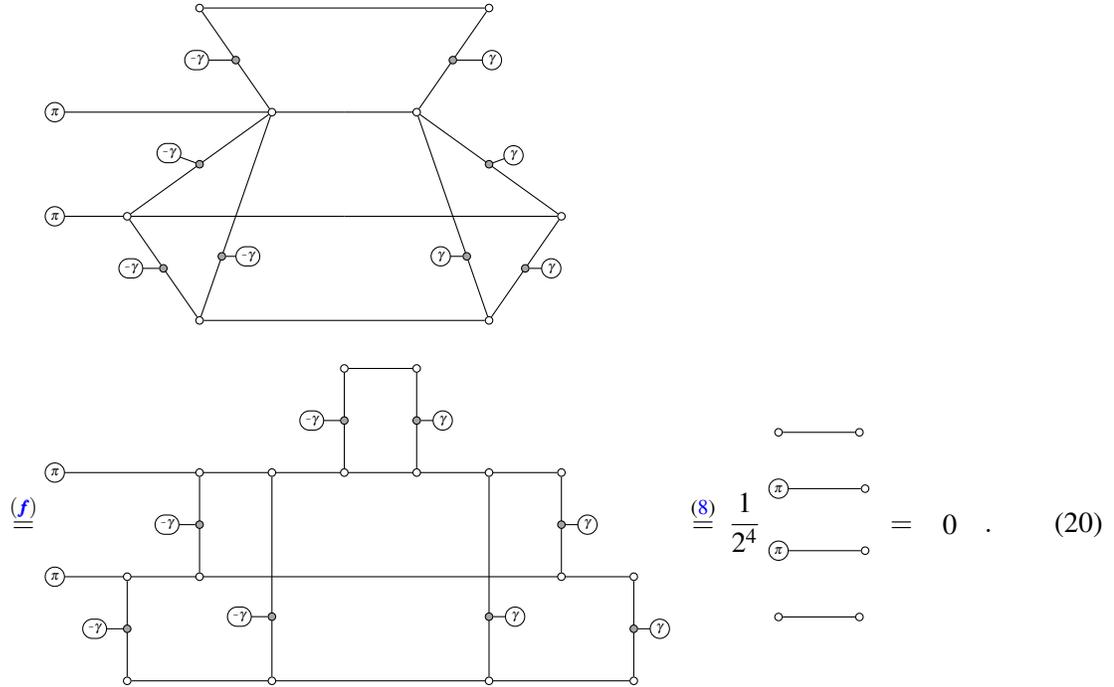
\begin{align}
     & \vcenter{\hbox{
        \externalize{qaoa_max_cut_example_summand_ii_00}{%

        }
    }} \label{eqn:qaoa_max_cut_example_summand_ii} \, .
\end{align}

The second summand (the \textit{I-X}-term) reads
\begin{align}
     & \vcenter{\hbox{
        \externalize{qaoa_max_cut_example_summand_ix_00}{%
            %
        }
    }}
    \label{eqn:qaoa_max_cut_example_summand_ix}
\end{align} \,.

The third summand (the \textit{X-I}-term) reads
\begin{align}
     & \vcenter{\hbox{
        \externalize{qaoa_max_cut_example_summand_xi_00}{%
            %
        }
    }}
    \label{eqn:qaoa_max_cut_example_summand_xi}
\end{align} \, .

The fourth summand (the \textit{X-X}-term) reads
\begin{align}
     & \vcenter{\hbox{
        \externalize{qaoa_max_cut_example_summand_xx_00}{%
            %
        }
    }}\label{eqn:qaoa_max_cut_example_summand_xx} \, .
\end{align}

\section{Proofs of useful ZX-diagram Identities}
\subsection{Phase-gadget identity}%
\label{sec:app_proof_conjugates_cancel_phase_gadgets}
\begin{proof}[Proof of~\eqref{eqn:conjugates_cancel_phase_gadgets}]

First, we can us the spider fusion rule to write
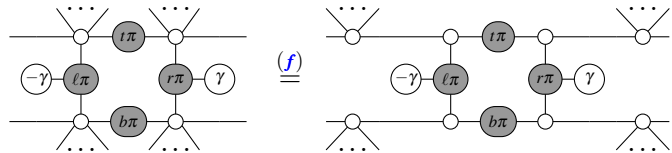
\begin{align}
    \vcenter{\hbox{
        \externalizezx{phase_gadgets_00}{%
            \def\dotdist{0.7em}
\begin{ZX}[ampersand replacement=\&]
                               \& \zxN{} \ar[rd, start anchor={[xshift=\dotdist]}] 
    \& \dots                   \& \zxN{} \ar[ld, start anchor={[xshift=-\dotdist]}] \ar[rd, start anchor={[xshift=\dotdist]}] 
    \& \dots                   \& \zxN{} \ar[ld, start anchor={[xshift=-\dotdist]}]          
    \& 
    \\
       \zxN{} \rar             \& \zxN{} \rar 
    \& \zxZ{} \rar             \& \zxX{t\pi} \rar                                   
    \& \zxZ{} \rar             \& \zxN{} \rar             
    \& \zxN{}
    \\
                               \& \zxZ-{\gamma} \rar  
    \& \zxX{\ell\pi} \uar \dar \&                                                   
    \& \zxX{r\pi} \uar \dar    \& \zxZ{\gamma} \lar      
    \& \zxN{}
    \\
       \zxN{} \rar             \& \zxN{} \rar   
    \& \zxZ{} \rar             \& \zxX{b\pi} \rar   
    \& \zxZ{} \rar             \& \zxN{} \rar             
    \& \zxN{}
    \\
                               \& \zxN{} \ar[ru, start anchor={[xshift=\dotdist]}] 
    \& \dots                   \& \zxN{} \ar[lu, start anchor={[xshift=-\dotdist]}] \ar[ru, start anchor={[xshift=\dotdist]}] 
    \& \dots                   \& \zxN{} \ar[lu, start anchor={[xshift=-\dotdist]}]          
    \& 
\end{ZX}

        }
    }}
    &\stackrel{\spiderrule}{=}
    \vcenter{\hbox{
        \externalizezx{phase_gadgets_01}{%
            \begin{ZX}[ampersand replacement=\&]
       \zxN{} \ar[rd]          \& \dots                   
    \& \zxN{} \ar[ld]          \&
    \&                         \&
    \&                         \& 
    \& \zxN{} \ar[rd]          \& \dots                   
    \& \zxN{} \ar[ld]          
    \\
       \zxN{} \rar             \& \zxZ{} \rar
    \& \zxN{} \rar             \& \zxN{} \rar 
    \& \zxZ{} \rar             \& \zxX{t\pi} \rar                                   
    \& \zxZ{} \rar             \& \zxN{} \rar             
    \& \zxN{} \rar             \& \zxZ{} \rar    
    \&
    \\
                               \&
    \&                         \& \zxZ-{\gamma} \rar  
    \& \zxX{\ell\pi} \uar \dar \&                                                   
    \& \zxX{r\pi} \uar \dar    \& \zxZ{\gamma} \lar      
    \& \zxN{}                  \&
    \&
    \\
       \zxN{} \rar             \& \zxZ{} \rar
    \& \zxN{} \rar             \& \zxN{} \rar   
    \& \zxZ{} \rar             \& \zxX{b\pi} \rar   
    \& \zxZ{} \rar             \& \zxN{} \rar             
    \& \zxN{} \rar             \& \zxZ{} \rar
    \&
    \\
       \zxN{} \ar[ru]          \& \dots                   
    \& \zxN{} \ar[lu]          \&
    \&                         \&
    \&                         \& 
    \& \zxN{} \ar[ru]          \& \dots                   
    \& \zxN{} \ar[lu]          
\end{ZX}

        }
    }} \, .
\end{align}
Then, we just consider the inner part
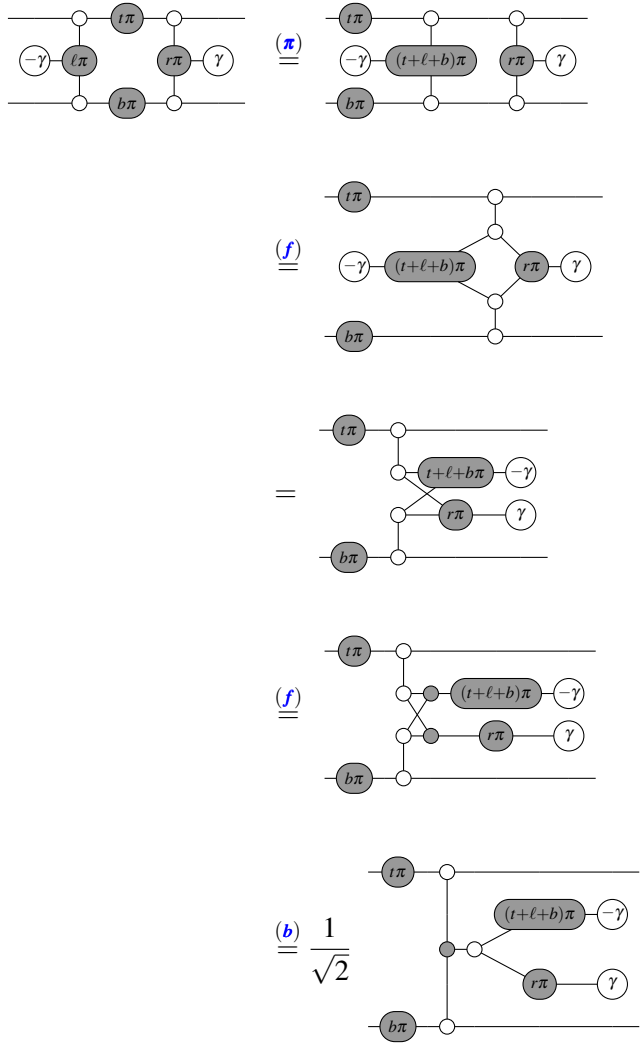
\begin{align}
    \vcenter{\hbox{
        \externalizezx{phase_gadgets_02}{%
            \begin{ZX}[ampersand replacement=\&]
       \zxN{} \rar             \& \zxN{} \rar 
    \& \zxZ{} \rar             \& \zxX{t\pi} \rar                                   
    \& \zxZ{} \rar             \& \zxN{} \rar             
    \& \zxN{}
    \\
                               \& \zxZ-{\gamma} \rar  
    \& \zxX{\ell\pi} \uar \dar \&                                                   
    \& \zxX{r\pi} \uar \dar    \& \zxZ{\gamma} \lar      
    \& \zxN{}
    \\
       \zxN{} \rar             \& \zxN{} \rar   
    \& \zxZ{} \rar             \& \zxX{b\pi} \rar   
    \& \zxZ{} \rar             \& \zxN{} \rar             
    \& \zxN{}
\end{ZX}

        }
    }}
    &\stackrel{\pirule}{=}
    \vcenter{\hbox{
        \externalizezx{phase_gadgets_03}{%
            \begin{ZX}[ampersand replacement=\&]
       \zxN{} \rar             \& \zxX{t\pi} \rar
    \& \zxZ{} \rar             \& \zxN{} \rar                                   
    \& \zxZ{} \rar             \& \zxN{} \rar             
    \& \zxN{}
    \\
                               \& \zxZ-{\gamma} \rar  
    \& \zxX{(t + \ell + b)\pi} \uar \dar \&                                                   
    \& \zxX{r\pi} \uar \dar    \& \zxZ{\gamma} \lar      
    \& \zxN{}
    \\
       \zxN{} \rar             \& \zxX{b\pi} \rar  
    \& \zxZ{} \rar             \& \zxN{} \rar   
    \& \zxZ{} \rar             \& \zxN{} \rar             
    \& \zxN{}
\end{ZX}

        }
    }}
    \\[4ex]
    &\stackrel{\spiderrule}{=}
    \vcenter{\hbox{
        \externalizezx{phase_gadgets_04}{%
            \begin{ZX}[ampersand replacement=\&]
       \zxN{} \rar             \& \zxX{t\pi} \rar
    \& \zxN{} \rar             \& \zxZ{} \rar \dar
    \& \zxN{} \rar             \& \zxN{} \rar             
    \& \zxN{}
    \\
                               \&
    \&                         \& \zxZ{} \ar[ld] \ar[rd]
    \&                         \&
    \&
    \\
                               \& \zxZ-{\gamma} \rar  
    \& \zxX{(t + \ell + b)\pi}   \&                                                   
    \& \zxX{r\pi}              \& \zxZ{\gamma} \lar      
    \& \zxN{}
    \\
                               \&
    \&                         \& \zxZ{} \ar[lu] \ar[ru]
    \&                         \&
    \&
    \\
       \zxN{} \rar             \& \zxX{b\pi} \rar  
    \& \zxN{} \rar             \& \zxZ{} \rar \uar
    \& \zxN{} \rar             \& \zxN{} \rar             
    \& \zxN{}
\end{ZX}

        }
    }}
    \\[4ex]
    &=
    \vcenter{\hbox{
        \externalizezx{phase_gadgets_05}{%
            \begin{ZX}[ampersand replacement=\&]
       \zxN{} \rar             \& \zxX{t\pi} \rar
    \& \zxN{} \rar             \& \zxZ{} \rar \dar
    \& \zxN{} \rar             \& \zxN{} \rar             
    \& \zxN{}
    \\
                               \&
    \&                         \& \zxZ{} \rar \ar[rd]
    \& \zxX{t + \ell + b\pi}   \&  \zxZ-{\gamma} \lar 
    \&
    \\
                               \&
    \&                         \& \zxZ{} \rar \ar[ru]
    \& \zxX{r\pi}              \& \zxZ{\gamma} \lar      
    \&
    \\
       \zxN{} \rar             \& \zxX{b\pi} \rar  
    \& \zxN{} \rar             \& \zxZ{} \rar \uar
    \& \zxN{} \rar             \& \zxN{} \rar             
    \& \zxN{}
\end{ZX}

        }
    }}
    \\[4ex]
    &\stackrel{\spiderrule}{=}
    \vcenter{\hbox{
        \externalizezx{phase_gadgets_06}{%
            \begin{ZX}[ampersand replacement=\&]
       \zxN{} \rar             \& \zxX{t\pi} \rar
    \& \zxN{} \rar             \& \zxZ{} \rar \dar
    \& \zxN{} \rar             \& \zxN{} \rar             
    \& \zxN{} \rar             
    \& \zxN{}
    \\
                               \&
    \&                         \& \zxZ{} \rar \ar[rd]
    \& \zxX{} \rar             
    \& \zxX{(t + \ell + b)\pi}   \&  \zxZ-{\gamma} \lar 
    \&
    \\
                               \&
    \&                         \& \zxZ{} \rar \ar[ru]
    \& \zxX{} \rar             
    \& \zxX{r\pi}              \& \zxZ{\gamma} \lar      
    \&
    \\
       \zxN{} \rar             \& \zxX{b\pi} \rar  
    \& \zxN{} \rar             \& \zxZ{} \rar \uar
    \& \zxN{} \rar             \& \zxN{} \rar             
    \& \zxN{} \rar            
    \& \zxN{}
\end{ZX}

        }
    }}
    \\[4ex]
    &\stackrel{\bialgrule}{=}
    \frac{1}{\sqrt{2}}
    \vcenter{\hbox{
        \externalizezx{phase_gadgets_07}{%
            \begin{ZX}[ampersand replacement=\&]
       \zxN{} \rar             \& \zxX{t\pi} \rar
    \& \zxN{} \rar             \& \zxZ{} \rar \dar
    \& \zxN{} \rar             \& \zxN{} \rar             
    \& \zxN{} \rar             
    \& \zxN{}
    \\
                               \&
    \&                         \& \zxN{} \dar
    \& 
    \& \zxX{(t + \ell + b)\pi}   \&  \zxZ-{\gamma} \lar 
    \&
    \\
                               \&
    \&                         \& \zxX{} \rar 
    \& \zxZ{} \ar[ru] \ar[rd]             
    \&                         \&  
    \&
    \\
                               \&
    \&                         \& \zxN{} \uar
    \& 
    \& \zxX{r\pi}              \& \zxZ{\gamma} \lar      
    \&
    \\
       \zxN{} \rar             \& \zxX{b\pi} \rar  
    \& \zxN{} \rar             \& \zxZ{} \rar \uar
    \& \zxN{} \rar             \& \zxN{} \rar             
    \& \zxN{} \rar            
    \& \zxN{}
\end{ZX}

        }
    }}
    \\[4ex]
    &\stackrel{\pirule \spiderrule}{=}
    \frac{1}{\sqrt{2}}
    \vcenter{\hbox{
        \externalizezx{phase_gadgets_08}{%
            \begin{ZX}[ampersand replacement=\&]
       \zxN{} \rar             \& \zxX{(t+r)\pi} \rar
    \& \zxN{} \rar             \& \zxZ{} \rar \dar
    \& \zxN{} \rar             \& \zxX{r\pi} \rar             
    \& \zxN{} \rar             
    \& \zxN{}
    \\
                               \&
    \&                         \& \zxX{} \rar 
    \& \zxZ{\gamma} \rar             
    \& \zxX{(t + \ell + b + r)\pi}   \&  \zxZ-{\gamma} \lar 
    \&
    \\
       \zxN{} \rar             \& \zxX{b\pi} \rar  
    \& \zxN{} \rar             \& \zxZ{} \rar \uar
    \& \zxN{} \rar             \& \zxN{} \rar             
    \& \zxN{} \rar            
    \& \zxN{}
\end{ZX}

        }
    }} \, . 
\end{align}
If $t+l+b+r$ even, we have
\begin{align}
    \frac{1}{\sqrt{2}}
    \vcenter{\hbox{
        \externalizezx{phase_gadgets_09even}{%
            \begin{ZX}[ampersand replacement=\&]
       \zxN{} \rar             \& \zxX{(t+r)\pi} \rar
    \& \zxN{} \rar             \& \zxZ{} \rar \dar
    \& \zxN{} \rar             \& \zxX{r\pi} \rar             
    \& \zxN{} \rar             
    \& \zxN{}
    \\
                               \&
    \&                         \& \zxX{} \rar 
    \& \zxZ{\gamma} \rar             
    \& \zxX{}   \&  \zxZ-{\gamma} \lar 
    \&
    \\
       \zxN{} \rar             \& \zxX{b\pi} \rar  
    \& \zxN{} \rar             \& \zxZ{} \rar \uar
    \& \zxN{} \rar             \& \zxN{} \rar             
    \& \zxN{} \rar            
    \& \zxN{}
\end{ZX}

        }
    }}
    &\stackrel{\idrule, \spiderrule}{=}
    \frac{1}{\sqrt{2}}
    \vcenter{\hbox{
        \externalizezx{phase_gadgets_10even}{%
            \begin{ZX}[ampersand replacement=\&]
       \zxN{} \rar             \& \zxX{(t+r)\pi} \rar
    \& \zxN{} \rar             \& \zxZ{} \rar \dar
    \& \zxN{} \rar             \& \zxX{r\pi} \rar             
    \& \zxN{} 
    \\
                               \&
    \&                         \& \zxX{} \rar 
    \& \zxZ{} 
    \&                         \& 
    \\
       \zxN{} \rar             \& \zxX{b\pi} \rar  
    \& \zxN{} \rar             \& \zxZ{} \rar \uar
    \& \zxN{} \rar             \& \zxN{} \rar             
    \& \zxN{} 
\end{ZX}

        }
    }}
    \\[4ex]
    &\stackrel{\copyrule}{=}
    \frac{1}{2}
    \vcenter{\hbox{
        \externalizezx{phase_gadgets_11even}{%
            \begin{ZX}[ampersand replacement=\&]
       \zxN{} \rar             \& \zxX{(t+r)\pi} \rar
    \& \zxN{} \rar             \& \zxZ{} \rar \dar
    \& \zxN{} \rar             \& \zxX{r\pi} \rar             
    \& \zxN{} 
    \\
                               \&
    \&                         \& \zxZ{} 
    \& 
    \&                         \& 
    \\
                               \&
    \&                         \& 
    \& 
    \&                         \& 
    \\
                               \&
    \&                         \& \zxZ{} 
    \& 
    \&                         \& 
    \\
       \zxN{} \rar             \& \zxX{b\pi} \rar  
    \& \zxN{} \rar             \& \zxZ{} \rar \uar
    \& \zxN{} \rar             \& \zxN{} \rar             
    \& \zxN{} 
\end{ZX}

        }
    }}
    \\[4ex]
    &\stackrel{\idrule,\spiderrule}{=}
    \frac{1}{2}
    \vcenter{\hbox{
        \externalizezx{phase_gadgets_12even}{%
            \begin{ZX}[ampersand replacement=\&]
       \zxN{} \rar             \& \zxX{t\pi} \rar
    \& \zxN{} 
    \\[\zxwRow, \zxwRow]
       \zxN{} \rar             \& \zxX{b\pi} \rar  
    \& \zxN{} 
\end{ZX}

        }
    }} \,,
\end{align}
else, if $t+l+b+r$ odd, we have
\begin{align}
    \frac{1}{\sqrt{2}}
    \vcenter{\hbox{
        \externalizezx{phase_gadgets_09odd}{%
            \begin{ZX}[ampersand replacement=\&]
       \zxN{} \rar             \& \zxX{(t+r)\pi} \rar
    \& \zxN{} \rar             \& \zxZ{} \rar \dar
    \& \zxN{} \rar             \& \zxX{r\pi} \rar             
    \& \zxN{} \rar             
    \& \zxN{}
    \\
                               \&
    \&                         \& \zxX{} \rar 
    \& \zxZ{\gamma} \rar             
    \& \zxX{\pi}   \&  \zxZ-{\gamma} \lar 
    \&
    \\
       \zxN{} \rar             \& \zxX{b\pi} \rar  
    \& \zxN{} \rar             \& \zxZ{} \rar \uar
    \& \zxN{} \rar             \& \zxN{} \rar             
    \& \zxN{} \rar            
    \& \zxN{}
\end{ZX}

        }
    }}
    &\stackrel{\spiderrule,\pirule}{=}
    \frac{1}{\sqrt{2}} \ee^{\ii \gamma}
    \vcenter{\hbox{
        \externalizezx{phase_gadgets_10odd}{%
            \begin{ZX}[ampersand replacement=\&]
       \zxN{} \rar             \& \zxX{(t+r)\pi} \rar
    \& \zxN{} \rar             \& \zxZ{} \rar \dar
    \& \zxN{} \rar             \& \zxX{r\pi} \rar             
    \& \zxN{} \rar             
    \& \zxN{}
    \\
                               \&
    \&                         \& \zxX{} \rar 
    \& \zxZ-{2\gamma} 
    \&                         \&
    \&
    \\
       \zxN{} \rar             \& \zxX{(b+1)\pi} \rar  
    \& \zxN{} \rar             \& \zxZ{} \rar \uar
    \& \zxN{} \rar             \& \zxX{\pi} \rar             
    \& \zxN{} \rar            
    \& \zxN{}
\end{ZX}

        }
    }} \, ,
\end{align}
which proves \eqref{eqn:conjugates_cancel_phase_gadgets}.
\end{proof}

\subsection{Hardware Efficient Ansatz}%
\label{sec:app_proof_zxidentity_hweffansatz}
\begin{proof}[Proof of \eqref{eqn:zxidentity_hweffansatz}]
\begin{align}
        &
        \vcenter{\hbox{
            \externalizezx{hw_eff_ansatz_side_calculation_00}{
                \begin{ZX}[ampersand replacement=\&]
    \zxX{\ell_1 \pi} \rar 
    \& \zxZ{} \dar \rar \& \zxZ{} \dar \rar \& \zxN{} \rar 
    \& \zxN{} \rar
    \& \zxN{} \rar
    \& \zxN{} \rar \& \zxZ{} \dar \rar \& \zxZ{} \dar \rar 
    \& \zxX{r_1 \pi} 
    \\
    \zxX{\ell_2 \pi} \rar 
    \& \zxX{} \rar \& \zxN{} \dar \rar \& \zxZ{} \dar \rar 
    \& \zxX{m_2 \pi} \rar
    \& \zxZ{m_2 \pi} \rar
    \& \zxZ{} \dar \rar \& \zxN{} \dar \rar \& \zxX{}  \rar 
    \& \zxX{r_2 \pi} 
    \\
    \zxX{\ell_3 \pi} \rar 
    \& \zxN{} \rar \& \zxX{} \rar \& \zxX{} \rar 
    \& \zxX{m_3 \pi} \rar
    \& \zxZ{m_3 \pi} \rar
    \& \zxX{} \rar \& \zxX{} \rar \& \zxN{} \rar 
    \& \zxX{r_3 \pi} 
\end{ZX}

            }
        }}
        \stackrel{\pirule, \spiderrule}{=}
        \vcenter{\hbox{
            \externalizezx{hw_eff_ansatz_side_calculation_01}{
                \begin{ZX}[ampersand replacement=\&]
    \zxX{\ell_1 \pi} \rar 
    \& \zxZ{} \dar \rar \& \zxZ{} \dar \rar \& \zxN{} \rar 
    \& \zxN{} \rar
    \& \zxN{} \rar \& \zxZ{} \dar \rar \& \zxZ{} \dar \rar 
    \& \zxZ{m_2 \pi} \rar
    \& \zxX{r_1 \pi} 
    \\
    \zxX{\ell_2 \pi} \rar 
    \& \zxX{} \rar \& \zxN{} \dar \rar \& \zxZ{} \dar \rar 
    \& \zxX{m_2 \pi} \rar
    \& \zxZ{} \dar \rar \& \zxN{} \dar \rar \& \zxX{}  \rar 
    \& \zxZ{(m_2+m_3) \pi} \rar
    \& \zxX{r_2 \pi} 
    \\
    \zxX{\ell_3 \pi} \rar 
    \& \zxN{} \rar \& \zxX{} \rar \& \zxX{} \rar 
    \& \zxX{m_3 \pi} \rar
    \& \zxX{} \rar \& \zxX{} \rar \& \zxN{} \rar 
    \& \zxZ{m_3 \pi} \rar
    \& \zxX{r_3 \pi} 
\end{ZX}

            }
        }}
        \\[4ex]
        &
        \stackrel{\copyrule, \pirule, \spiderrule}{=}
        \underbrace{(-1)^{m_2 r_1 + (m_2 \oplus m_3) r_2 + m_3 r_3}}_{=:f_{m_2 m_3}^{r_1 r_2 r_3}}
        \vcenter{\hbox{
            \externalizezx{hw_eff_ansatz_side_calculation_02}{
                \begin{ZX}[ampersand replacement=\&]
    \zxX{\ell_1 \pi} \rar 
    \& \zxZ{} \dar \rar \& \zxZ{} \dar \rar \& \zxN{} \rar 
    \& \zxN{} \rar
    \& \zxN{} \rar \& \zxZ{} \dar \rar \& \zxZ{} \dar \rar 
    \& \zxX{r_1 \pi} 
    \\
    \zxX{\ell_2 \pi} \rar 
    \& \zxX{} \rar \& \zxN{} \dar \rar \& \zxZ{} \dar \rar 
    \& \zxX{m_2 \pi} \rar
    \& \zxZ{} \dar \rar \& \zxN{} \dar \rar \& \zxX{}  \rar 
    \& \zxX{r_2 \pi} 
    \\
    \zxX{\ell_3 \pi} \rar 
    \& \zxN{} \rar \& \zxX{} \rar \& \zxX{} \rar 
    \& \zxX{m_3 \pi} \rar
    \& \zxX{} \rar \& \zxX{} \rar \& \zxN{} \rar 
    \& \zxX{r_3 \pi} 
\end{ZX}

            }
        }}
        \\[4ex]
        &
        \stackrel{\spiderrule}{=}
        f_{m_2 m_3}^{r_1 r_2 r_3}
        \vcenter{\hbox{
            \externalizezx{hw_eff_ansatz_side_calculation_03}{
                \begin{ZX}[ampersand replacement=\&]
    \zxX{\ell_1 \pi} \rar 
    \& \zxN{} \rar \& \zxZ{} \dar \rar \& \zxN{} \rar 
    \& \zxN{} \rar
    \& \zxN{} \rar \& \zxZ{} \dar \rar \& \zxN{} \rar 
    \& \zxX{r_1 \pi} 
    \\
    \zxX{\ell_2 \pi} \rar 
    \& \zxX{} \rar \ar[ru] \& \zxN{} \dar \rar \& \zxZ{} \ar[ld] \rar 
    \& \zxX{m_2 \pi} \rar
    \& \zxZ{} \ar[rd] \rar \& \zxN{} \dar \rar \& \zxX{} \ar[lu] \rar 
    \& \zxX{r_2 \pi} 
    \\
    \zxX{\ell_3 \pi} \rar 
    \& \zxN{} \rar \& \zxX{} \rar \& \zxN{} \rar 
    \& \zxX{m_3 \pi} \rar
    \& \zxN{} \rar \& \zxX{} \rar \& \zxN{} \rar 
    \& \zxX{r_3 \pi} 
\end{ZX}

            }
        }}
        \stackrel{\spiderrule}{=}
        f_{m_2 m_3}^{r_1 r_2 r_3}
        \vcenter{\hbox{
            \externalizezx{hw_eff_ansatz_side_calculation_04}{
                \begin{ZX}[ampersand replacement=\&]
    \zxX{\ell_1 \pi} \rar 
    \& \zxN{} \rar 
    \& \zxZ[a=zt]{} \dar \rar 
    \& \zxN{} \rar 
    \& \zxX{r_1 \pi} 
    \\
    \&
    \& \zxX{(\ell_3 + m_3 + r_3) \pi} \ar[ld] \ar[rd] 
    \&
    \&
    \\
    \zxX[a=l3]{\ell_2 \pi} \rar 
    \& \zxZ{} \rar
    \& \zxX{m_2 \pi} \rar
    \& \zxZ{} \rar
    \& \zxX[a=r3]{r_2 \pi} 
    \ar[from=l3, to=zt, )]
    \ar[from=r3, to=zt, (]
\end{ZX}

            }
        }}
        \\[4ex]
        &
        \stackrel{\pirule, \spiderrule}{=}
        f_{m_2 m_3}^{r_1 r_2 r_3}
        \vcenter{\hbox{
            \externalizezx{hw_eff_ansatz_side_calculation_05}{
                \begin{ZX}[ampersand replacement=\&]
    \zxX{\ell_1 \pi} \rar 
    \& \zxN{} \rar 
    \& \zxZ[a=zt]{} \dar \rar 
    \& \zxN{} \rar 
    \& \zxX{r_1 \pi} 
    \\
    \&
    \& \zxX{(m_2 + \ell_3 + m_3 + r_3) \pi} \ar[ld] \ar[rd] 
    \&
    \&
    \\
    \zxX[a=l3]{(\ell_2 + m_2) \pi} \rar 
    \& \zxZ{} \rar
    \& \zxN{} \rar
    \& \zxZ{} \rar
    \& \zxX[a=r3]{r_2 \pi} 
    \ar[from=zt, to=l3, (.=20]
    \ar[from=zt, to=r3, (.=-20]
\end{ZX}

            }
        }}
        \\[4ex]
        &
        \stackrel{\spiderrule}{=}
        f_{m_2 m_3}^{r_1 r_2 r_3}
        \vcenter{\hbox{
            \externalizezx{hw_eff_ansatz_side_calculation_06}{
                \begin{ZX}[ampersand replacement=\&]
    \zxX{\ell_1 \pi} \rar 
    \& \zxN{} \rar 
    \& \zxZ[a=zt]{} \dar \rar 
    \& \zxN{} \rar 
    \& \zxX{r_1 \pi} 
    \\[2ex]
    \zxX[a=l3]{(\ell_2 + m_2) \pi} 
    \&
    \& \zxX{(m_2 + \ell_3 + m_3 + r_3) \pi} \dar
    \&
    \& \zxX[a=r3]{r_2 \pi} 
    \\
    \&
    \& \zxX{} \ar[d,)] \ar[d,(] 
    \&
    \&
    \\
    \& 
    \& \zxZ[a=zb]{} 
    \& 
    \&
    \ar[from=zt, to=l3]
    \ar[from=zt, to=r3]
    \ar[from=zb, to=l3]
    \ar[from=zb, to=r3]
\end{ZX}

            }
        }}
        \\[4ex]
        &
        \stackrel{\hopfrule, \spiderrule}{=}
        \frac{f_{m_2 m_3}^{r_1 r_2 r_3}}{2}
        \vcenter{\hbox{
            \externalizezx{hw_eff_ansatz_side_calculation_07}{
                \begin{ZX}[ampersand replacement=\&]
    \zxX{\ell_1 \pi} \ar[rd]
    \& 
    \& \zxX{r_1 \pi} \ar[ld]
    \\
    \& \zxZ[a=zt]{} 
    \&
    \\
    \zxX[a=l3]{(\ell_2 + m_2 + r_2) \pi} \ar[ru, o-] \ar[ru, o']
    \&
    \& \zxX{(m_2 + \ell_3 + m_3 + r_3) \pi} \ar[lu]
\end{ZX}

            }
        }}
        \\[4ex]
        &
        \stackrel{\hopfrule, \spiderrule}{=}
        \frac{f_{m_2 m_3}^{r_1 r_2 r_3}}{2^2}
        \vcenter{\hbox{
            \externalizezx{hw_eff_ansatz_side_calculation_08}{
                \begin{ZX}[ampersand replacement=\&]
    \zxX{\ell_1 \pi} \ar[rd]
    \& 
    \& \zxX{r_1 \pi} \ar[ld]
    \\
    \& \zxZ[a=zt]{} 
    \&
    \\
    \zxX[a=l3]{(\ell_2 + m_2 + r_2) \pi}
    \&
    \& \zxX{(m_2 + \ell_3 + m_3 + r_3) \pi} \ar[lu]
\end{ZX}

            }
        }}
        \\[4ex]
        &
        \stackrel{\copyrule, \spiderrule}{=}
        \frac{f_{m_2 m_3}^{r_1 r_2 r_3}}{2^3}
        \vcenter{\hbox{
            \externalizezx{hw_eff_ansatz_side_calculation_09}{
                \begin{ZX}[ampersand replacement=\&]
    \zxX{(\ell_1 + m_2 + \ell_3 + m_3 + r_3)\pi}
    \& 
    \& \zxX{(r_1 + m_2 + \ell_3 + m_3 + r_3)\pi}
    \\[\zxwRow, \zxwRow]
    \zxX[a=l3]{(\ell_2 + m_2 + r_2) \pi}
    \&
    \& 
\end{ZX}

            }
        }}
\end{align}
\end{proof}

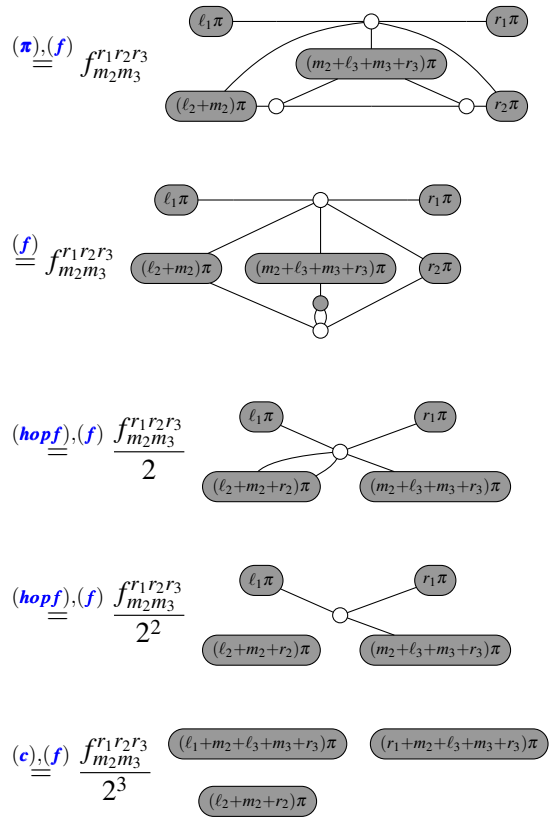
\end{document}